%% file: preprint.tex
\documentclass{preprint}
\usepackage[backend=biber, style=numeric-comp, sorting=none, maxbibnames=20, giveninits=true]{biblatex}%
\AtEveryBibitem{\iffieldundef{doi}{}{\clearfield{url}\clearfield{urlyear}\clearfield{urlmonth}\clearfield{urlday}}}
\DeclareNameAlias{default}{family-given}

\DeclareFieldFormat*{title}{#1}
\DeclareFieldFormat[article]{volume}{#1}
\DeclareFieldFormat{pages}{#1}
\renewbibmacro{in:}{}
\DeclareSourcemap{\maps[datatype=bibtex]{\map{\step[fieldset=language, null]}}}

\usepackage{amsmath}
\usepackage{booktabs}
\usepackage{subcaption}
\usepackage{multirow}
\usepackage{svg}
\usepackage{xcolor}

\makeatletter
\newcommand{\conflict}[1]{\section*{Declaration of competing interests}\if@anonymous Removed for anonymity \else #1 \fi}
\makeatother
\renewcommand{\ack}[1]{\section*{Declaration of generative AI use}#1}

\begin{document}

\articletype{Article type} %

\title{Low-regret pathways for the energy transition under uncertainty}

\author{Gabriel Wiest$^1$\orcid{0000-0001-7508-872X}, Evren Mert Turan$^1$\orcid{0000-0002-2662-1978}, André Bardow$^1$\orcid{0000-0002-3831-0691} and Stefano Moret$^{2,*}$\orcid{0000-0001-7716-3084}}

\affil{$^1$Energy and Process Systems Engineering, Department of Mechanical and Process Engineering, ETH Zürich, Tannenstrasse 3, Zürich, 8092, Switzerland}

\affil{$^2$Department of Management, Economics and Industrial Engineering, Politecnico di Milano, Via Lambruschini 4/b, 20156, Milan, Italy}

\affil{$^*$Author to whom any correspondence should be addressed.}

\email{stefano.moret@polimi.it}


\begin{abstract}
Long-term decisions for the energy transition must be made before future costs and demands are known. Energy system modelers often address this uncertainty by generating many scenarios, each yielding a cost-optimal transition pathway. However, these pathway ensembles are not actionable on their own. They show which transition is optimal in each scenario, but not the risk of committing to a decision when the future unfolds differently than anticipated. We propose a method that condenses large pathway ensembles into a few representative states of the energy system, defines strategies as sequences of these states, and evaluates their regret over time. By fixing only selected states along the transition, we can assess lock-in effects of near-term decisions and path dependencies toward target states. We apply this method to Switzerland's transition to net zero by 2050, generating 1000 scenarios by sampling six uncertain macro drivers (population, economy, technology, imports, behavior, and climate) and eight weather years. We reduce these 1000 pathways to six representative states, which combine into 20 candidate strategies. Focusing on three representative strategies, we identify under which conditions each strategy has the lowest regret. We show that the preferred strategy across uncertain futures depends on the decision-maker's risk aversion. Furthermore, we find that the optimal net-zero state in 2050 depends on the path followed up to 2035, showing that a low-regret transition requires choosing the right path, not just the right target.
\end{abstract}

\section{Introduction}

Energy systems, including electricity, heating, industry, transportation, and building sectors, contribute the most to global greenhouse gas emissions \cite{ritchie_breakdown_2020}. Rapidly transforming these systems is essential to mitigate climate change.
Researchers use energy system optimization models to support decision-makers in exploring future pathways for the energy system \cite{ruiz_manuel_optimising_2026}.
Pathway models optimize energy system investments and operations over several decades. These investments are often long-lived and partly irreversible; thus, decisions early in the transition can strongly influence the later stages (see e.g., carbon lock-in \cite{unruh_understanding_2000, seto_carbon_2016}). The success of the energy transition therefore depends on the entire transition pathway, which determines cumulative emissions and cost, not just on the system's final state \cite{victoria_early_2020}.

However, planning long-term energy infrastructure, such as deciding whether to build a hydrogen grid or not, involves large uncertainties, including future resource prices and technology costs. This uncertainty can cause decision-makers to postpone investments \cite{pindyck_irreversibility_1991}.
Hence, to avoid delays in the energy transition, decision-makers need robust strategies that perform well across a wide range of plausible futures. %
We refer to these strategies as \emph{low-regret} strategies \cite{wiest_low-regret_2026}.

Regret is a natural metric for evaluating the performance of decisions under uncertainty. As introduced by Savage \cite{savage_theory_1951}, regret measures the additional cost of a decision compared to the optimal decision in hindsight, once uncertainty is realized. Regret is particularly well-suited as a decision metric in energy planning, where total system costs are often dominated by external factors such as resource prices, on which a single country's strategy has limited influence. 
Measuring cost relative to each scenario's optimum separates the cost of a decision from the cost level of the scenario.
Minimax regret was early on applied to a greenhouse gas abatement model \cite{loulou_minimax_1999}. More recently, regret has been used to improve robustness in strategic planning for the Spanish energy system \cite{rodriguez-matas_improving_2024}, to evaluate urban energy system designs for Accra, Ghana \cite{yazdanie_strengthening_2025}, and to identify low-regret strategies for biomass use in Switzerland \cite{wiest_low-regret_2026}.

Traditional approaches to address uncertainty in transition pathways fall into two groups: The first group incorporates uncertainty into the optimization itself, through two-stage \cite{usher_critical_2012} or multi-stage stochastic programming \cite{mcdonald_sequential_2024, guevara_uncertainty_2024}, or through robust optimization \cite{bertsimas_price_2004, patankar_using_2022}. Stochastic formulations require probability distributions that are difficult to determine for decades ahead \cite{moret_characterization_2017}, and their problem size grows quickly with the number of scenarios and decision stages. Robust formulations can be overly conservative, and the parameters controlling their conservatism are difficult to interpret \cite{moret_decision_2020}. Moreover, both formulations return a single optimal decision, which may provoke reactance and single-option aversion \cite{brehm_psychological_1981, mochon_single-option_2013}, rather than supporting decision-makers in exploring trade-offs \cite{marchau_decision_2019}. 

The second group of approaches runs deterministic optimization models many times, either to obtain a variety of optimal solutions from sampled input parameters in Monte Carlo analyses \cite{pye_integrated_2015} or to explore near-optimal solutions with modeling to generate alternatives (MGA) \cite{decarolis_modelling_2016, sinha_diverse_2024, van_greevenbroek_little_2025}. Applied to pathways, these methods produce ensembles of hundreds to thousands of transitions, e.g., 1100 near-optimal pathways for the United States \cite{sinha_diverse_2024}, which differ in both their target system and the timing of investments. Presenting so many options to decision-makers can lead to choice overload \cite{chernev_choice_2015}.

Therefore, neither a single plan nor hundreds of pathways are well suited for decision support. Effective decision support should instead simplify complex model outcomes into interpretable insights that are accessible to non-experts, including policymakers. Several approaches aim to do so, e.g., clustering algorithms group similar pathways into a few representative pathways \cite{li_using_2020}, and decision trees identify key trade-offs and the conditions under which decisions are made \cite{baader_seeing_2026, wiegner_exploring_2026}. Other studies reduce the complexity of uncertainty studies by quantifying technology interdependencies \cite{pye_technology_2019} and cost thresholds at which technology choices tip \cite{muller_probabilistic_2026}. Beyond energy system models, approaches such as modernized stabilization wedges convert large solution spaces into a list of options without prescribing a single route \cite{johnson_democratizing_2026}.
While all of these methods indicate which options appear in the ensemble of solutions, none of them assesses how an option performs if future conditions differ from expectations.

Regret provides exactly this assessment; however, few studies have evaluated regret or related robustness metrics on transition pathways. Schricker et~al.~\cite{schricker_exploring_2026} explore transition pathways for an industrial site in Germany whose minimax regret stays below a tolerance, Ganter et~al.~\cite{ganter_minimum-regret_2025} investigate minimum-regret strategies for the European hydrogen and CO$_{2}$ infrastructure rollout, Kumar et~al.~\cite{kumar_resilient_2026} cluster district heating pathways in Sweden and assess their robustness, and Fitzgibbon et~al.~\cite{fitzgibbon_planning_2026} evaluate predefined social and environmental policies relative to business-as-usual planning for the power system of Yukon, Canada.

We see two gaps in the existing literature.
First, regret-based studies of transition pathways typically evaluate predefined strategies and report only aggregated regret. They reveal neither where regret arises, i.e., in which phase of the transition and in which technologies or resources, nor which uncertain parameters drive it.
Second, methods that reduce the complexity of pathway ensembles typically cluster them into representative pathways. Each representative pathway is fully fixed over the entire transition. 
Fixing the entire transition path, however, limits the analysis of path dependencies and lock-in effects.

In this work, we build on the low-regret framework by Wiest et~al.~\cite{wiest_low-regret_2026} and extend it to energy system transition pathways. To do so, we optimize the energy transition across a large set of scenarios sampled from uncertain macro drivers. We cluster the states that the pathways pass through into a few representative states that can occur in different years of the transition. We define transition strategies as sequences of these states and measure their regret by evaluating their performance across all scenarios. 

Our contributions are in three parts.
First, we shift the perspective from clustering entire pathways to clustering states of the energy system. Defining strategies as sequences of representative states accounts for the fact that similar states can occur at different stages of the transition. This approach reduces a large ensemble of pathways to a handful of states and strategies. %
Second, we develop a regret analysis tailored to transition pathways, decomposing regret by transition phase, technology, and resource and identifying the conditions under which each strategy has the lowest regret. %
Third, by defining strategies as sequences of states, we can also prescribe strategies partially, e.g., by fixing only the final state or the path up to an intermediate year. This definition separates the target state regret from that of the near-term path, and allows the analysis of path dependency and lock-in effects. %
We demonstrate our method for the transition of the Swiss energy system to net zero by 2050, using a whole-energy system pathway model.

\section{Methods}
To provide decision-makers with actionable insights for the energy transition under uncertainty, we propose a multi-step process:
By optimizing the energy transition using an energy system model for a large set of plausible scenarios, we obtain an ensemble of possible pathways. From this ensemble, we identify representative states of the energy system, i.e., states that the energy system typically passes through during the transition. Connecting these states temporally over the transition results in a network of sequences of states which can be evaluated as candidate transition strategies. We analyze the performance of these strategies using regret and visualize the results in an interpretable form for decision-makers.

\subsection{Energy system model and case study}

In our study, we use EnergyScope Pathway \cite{limpens_energyscope_2024}, an energy system transition model formulated as a linear program that minimizes total transition cost, subject to constraints such as energy balances. EnergyScope Pathway is an open-source multi-sector model covering electricity, heat (decentralized, district heating, industrial), mobility (private, public, freight, aviation), and non-energy (methanol, ammonia, high-value chemicals (HVC)) demands.

We make two key modifications to the model.
First, we revise the initialization of the existing technology stock.
In the original model, all pre-existing capacity is built immediately before the optimization start year, which delays the transition in early years to avoid sunk costs. We replace this formulation so that the initial technology stock at the start of the transition follows a user-defined age distribution (or a uniform distribution if none is specified; see Note~S1.1).
Second, we add multi-phase construction to the model.
This change ensures that technologies with construction times longer than the model's 5-year phases, such as new nuclear power plants \cite{portugal-pereira_better_2018}, become operational only after construction is finished (see Note~S1.2).

We model Switzerland's energy transition from 2025 to net zero in 2050, consistent with the Paris Agreement. The transition is modeled in 5-year steps, resulting in six \emph{representative years} $y\in \mathcal{Y} = \{2025, 2030, 2035, 2040, 2045, 2050\}$, assuming perfect foresight within each scenario over the whole transition. Each representative year is resolved using 12 typical days \cite{limpens_energyscope_2019} at an hourly resolution. 
The time span between two representative years is the operational \emph{phase} $p\in \mathcal{P}$, named $y_{\text{start}}$--$y_{\text{end}}$, e.g., the phase between 2025 and 2030 is called $2025$--$2030$. We model the activities within a phase as a linear interpolation between the two representative years (compare \cite{ruiz_manuel_optimising_2026}).

Furthermore, we adapt the model input data to reflect the Swiss energy system. We adjust the hourly demand profiles (heating, transport, etc.) and capacity factors (PV, Hydro, Wind) to match Swiss conditions. We update technologies with outdated parameters using Switzerland-specific values where available.
We initialize the Swiss energy system using the existing technology stock as of 2025 (see Note~S2.7).
Following Swiss legislation, we enforce renewable electricity generation targets, excluding hydropower, of 35\,TWh in 2035 and 45\,TWh in 2050 \cite{bundesgesetz2023} and allow no new installation of resistive heating in the building sector \cite{endk_muken_2025}. Based on Swiss emission reduction targets, we derive a CO$_{2}$ budget of 468.7\,Mt\,CO$_{2,\mathrm{eq}}$ for the entire transition until 2050 (see Note~S2.6). Finally, we enforce a net-zero operational constraint for Switzerland in 2050.

\subsection{Scenario generation}

Large uncertainties in population growth, economic development, behavioral change, climate change, technology costs, and resource prices and availability impact the optimal pathway of the energy transition. 
Parameter correlations significantly influence energy system outcomes \cite{rodriguez-matas_how_2025}. Ideally, we would specify a joint probability distribution to capture these interdependencies, but such distributions are generally unknown, and reliable data to estimate them is lacking.
To address this data limitation, we account for parameter interdependencies by evaluating a large ensemble of plausible future scenarios structured around six uncertain \emph{macro drivers}: population \cite{kanyako_compounding_2024}, eco-friendly behavior \cite{eerma_potential_2022}, economy \cite{kanyako_compounding_2024, christensen_uncertainty_2018}, technology cost \cite{fais_impact_2016, vatankhah_ghadim_are_2025}, import cost \cite{mersch_impact_2023}, and climate change \cite{yalew_impacts_2020, schoniger_climate_2026}, each identified in previous work as an influential and uncertain factor for energy systems.

To explore the scenario space, we generate $N=1000$ scenarios $i \in \mathcal{I} = \{1, \dots, N\}$ using Latin Hypercube Sampling \cite{mckay_comparison_1979} over the six macro drivers. Additionally, we capture weather uncertainty by independently sampling a historical weather year from 2016--2023 for each representative year in a scenario. %

For each scenario $i$, we transform the sampled macro drivers and weather years into model input parameters $\theta_i$ through a three-step process (see Note~S3 for details):
First, we use the realized macro drivers to derive macroeconomic parameters of the system, e.g., population size, gross domestic product, number of households.
Second, we scale input parameters with the macroeconomic parameters to which they are sensitive, e.g., we scale the residential heating demand with the number of households.
Third, we adjust the input parameters according to the remaining realized macro drivers and weather uncertainties, e.g., we increase the heating demand because of low eco-friendly behavior and a cold weather year. 

This approach implies that parameters sensitive to a single macro driver are perfectly correlated. For example, all technology investment costs move together. We investigate relaxing this full-correlation assumption in Note~S4. Each macro driver is sampled only once per scenario and applied in all years. A scenario thus describes a persistent trend, e.g., a scenario with high population growth remains at the upper end of the range until 2050. 

\subsection{Clustering of energy system states and identification of strategies}

Optimizing the energy transition for each of the $N$ sampled scenarios yields a large ensemble of pathways that is difficult for decision-makers to navigate. To reduce this complexity, we view the energy transition as a transition between energy system states. Our premise is that comparing individual states is more comprehensible than comparing full pathways. Furthermore, representing the energy transition as a sequence of individual states lets us define strategies more flexibly, e.g., fixing states only in a subset of years, while leaving the other years unconstrained (Section \ref{sec:partial_strategies}). 

To find representative states, we cluster the model outputs as follows: We categorize all technologies $t \in \mathcal{T}$ into six sectors $s$, several of which are further divided into subsectors $ss$.
We distinguish capacity-based sectors (Electricity, Storage, Fuels \& feedstocks), for which the installed capacity of each technology is the relevant quantity, and share-based sectors (Building, Industry, Transport), for which a shift is naturally expressed as a change in the share of the subsector (Table~\ref{tab:subsectors}). Thus, we characterize the energy system with the features:

\begin{equation}
    x_{i,y,t} = \begin{cases} F_{i,y,t} & \text{if } s(t) \in \{\text{Electricity, Storage, Fuels \& feedstocks}\} \\ \phi_{i,y,t} & \text{if } s(t) \in \{\text{Building, Industry, Transport}\} \end{cases} 
    \label{eq:features}
\end{equation}
where $F_{i,y,t}$ is the installed capacity of technology $t$ in scenario $i$ and year $y$, and $\phi_{i,y,t}$ is the technology's annual production share within its subsector.

\begin{table}[h]
\caption{Hierarchical structuring of the energy system into sectors and subsectors.} 
\centering
\begin{tabular}{l l}%
\toprule
Sector & Subsectors\\
\midrule
Electricity     & \\
Industry        & \\
Building        & Decentralized heating, District heating\\
Storage         & Short-term, Seasonal\\
Transport       & Private, Public, Freight, Aviation\\
Fuels \& feedstocks     & Oil, Gas, H$_2$, Methanol, Ammonia, HVC, Captured CO$_2$\\
\bottomrule
\end{tabular}
\label{tab:subsectors}
\end{table}

To ensure that sectors and subsectors of differing scale and number of constituent technologies contribute comparably to the clustering, we standardize features, adapting the sectoral standardization of Li~et~al.~\cite{li_using_2020} to our hierarchical sector–subsector–technology structure.
Each technology feature $x_{i,y,t}$ within subsector $ss(t)$, itself within sector $s(t)$, is standardized using the average standard deviation across all technologies within its
subsector,
\begin{equation}
    \hat{x}_{i,y,t} = \frac{x_{i,y,t} - \overline{x}_t}{\overline{\sigma}_{ss(t)} \sqrt{n_{s(t)}}},
    \label{eq:subsector_std}
\end{equation}
where $x_{i,y,t}$ is the feature value of technology $t$ in year $y$ of scenario $i$, $\overline{x}_t$ its mean across all years and scenarios, and $\overline{\sigma}_{ss(t)}$ the average of the individual standard deviations of all technologies belonging to subsector $ss(t)$. The correction factor $1/\sqrt{n_{s(t)}}$, where $n_{s(t)}$ is the number of subsectors in the technology's parent sector $s(t)$, ensures comparable weighting across sectors. For sectors without subsectors, the sector itself is treated as a single subsector, i.e., $\overline{\sigma}_{ss(t)}$ is the average standard deviation over all technologies of the sector and $n_{s(t)}=1$.
The transformation in Eq. \eqref{eq:subsector_std} gives metrics with relatively larger variance a stronger influence on the clustering, emphasizing technology substitution within a subsector, while avoiding disproportionately amplifying features with very small absolute variance.
Unlike Li et~al.~\cite{li_using_2020}, who standardize per representative year for clustering pathways, we standardize consistently across years, making states from different years directly comparable.

We apply $k$-means clustering to the standardized features $\hat{x}_{i,y,t}$ to identify $K$ representative states $\mathcal{K} = \{\kappa_0, \dots, \kappa_{K-1}\}$ of the energy system. We treat each scenario--year combination as an independent observation. The clustering is thus independent of the year, and the same state can occur in different years of the transition. We choose the number of states $K$ with the elbow method (Note~S5), trading off within-cluster variance against interpretability.

We define a \emph{strategy} $\pi$ as a sequence of state--year pairs ordered by year,
\begin{equation}
    \pi = \left((\kappa, 2025), \dots, (\kappa', 2050)\right), \qquad \kappa, \dots, \kappa' \in \mathcal{K} \cup \{\mathrm{X}\},
\end{equation}
where selecting ``X'' leaves the state in a year unconstrained. The first state is always the initial state $\kappa_0$, as 2025 is fixed by the initialization. We write a strategy compactly by concatenating its state labels, e.g., \texttt{0-1-2-2-X-4} prescribes $\kappa_1$ in 2030, $\kappa_2$ in both 2035 and 2040, and $\kappa_4$ in 2050, while leaving 2045 unconstrained. 

Denoting by $\mathcal{I}_{\kappa,y}$ the subset of scenarios whose energy system is in state $\kappa$ in year $y$, we prescribe a state in a given year $y$ by the mean over its members in that year,
\begin{equation}
    \overline{F}_{\kappa,y,t} = \frac{1}{|\mathcal{I}_{\kappa,y}|} \sum_{i \in \mathcal{I}_{\kappa,y}} F_{i,y,t},
    \qquad
    \overline{\phi}_{\kappa,y,t} = \frac{1}{|\mathcal{I}_{\kappa,y}|} \sum_{i \in \mathcal{I}_{\kappa,y}} \phi_{i,y,t}.
    \label{eq:state_year_mean}
\end{equation}

To enforce a strategy, we bound every technology for each prescribed pair $(\kappa, y) \in \pi$, $\kappa \neq \mathrm{X}$, around the mean of that state in that year (Eq.~\ref{eq:state_year_mean}):
\begin{equation}
    F^{\min}_{y,t}, F^{\max}_{y,t} = (1 \pm \alpha)\, \overline{F}_{\kappa,y,t}, \qquad
    \phi^{\min}_{y,t}, \phi^{\max}_{y,t} = \overline{\phi}_{\kappa,y,t} \pm \alpha.
    \label{eq:state_year_bound}
\end{equation}
The bandwidth $\alpha$ defines how much flexibility the model retains to adapt to a scenario. A too small $\alpha$ confines the energy system to a narrow corridor around the state means, which inflates regret with artifacts of the strategy definition rather than of the decisions it represents. A too large $\alpha$ makes strategies blurry, no longer representing distinct decisions, and their regret approaches zero. In our case study, we select $\alpha = 0.05$, i.e., $\pm5\%$ of the capacity and $\pm5$ percentage points of the share. We explore the sensitivity of our results to $\alpha$ in Note~S6. All bounds are clipped to valid values (i.e., $F^{\min}_{y,t}, \phi^{\min}_{y,t} \geq 0$, $\phi^{\max}_{y,t} \leq 1$) and to the model's original bounds on the technology. %

\subsection{Regret calculation}

The energy system model minimizes the total transition cost. For scenario $i$ and strategy $\pi$, we can write this objective as the sum of cost contributions of technologies $t \in \mathcal{T}$ and resources $r \in \mathcal{R}$ over the phases $p \in \mathcal{P}$ of the transition,

\begin{equation}
    C_{i,\pi} = C^{\text{init}}_{i,\pi} +\sum_{p \in \mathcal{P}} \Big( \sum_{t \in \mathcal{T}} C_{i,\pi,p,t} + \sum_{r \in \mathcal{R}} C_{i,\pi,p,r} \Big),
    \label{eq:transition_cost}
\end{equation}

where $C^{\mathrm{init}}_{i,\pi}$ is the operating and investment cost in the initial year, $C_{i,\pi,p,t}$ includes the investment and maintenance cost of technology $t$ minus its salvage value at the end of the transition, and $C_{i,\pi,p,r}$ describes the cost of resource $r$. All costs are discounted to the base year 2024.

We define the regret of strategy $\pi$ in scenario $i$ as the additional transition cost of following $\pi$ compared with the unconstrained optimum of the same scenario, $C^{\mathrm{opt}}_{i} = C_{i, \texttt{X-X-X-X-X-X}}$,
\begin{equation}
    R_{i,\pi} = C_{i,\pi} - C^{\mathrm{opt}}_{i} \geq 0.
    \label{eq:regret}
\end{equation}
The regret is non-negative, as enforcing a strategy adds constraints to the optimization.

We can decompose the regret into the contributions of individual technologies and resources in each phase,
\begin{equation}
\begin{aligned}[t]
    R_{i,\pi} &= R^{\mathrm{init}}_{i,\pi} + \sum_{p \in \mathcal{P}} \Bigl( \sum_{t \in \mathcal{T}} R_{i,\pi,p,t} + \sum_{r \in \mathcal{R}} R_{i,\pi,p,r} \Bigr), \\
    \text{where} \quad 
    R^{\mathrm{init}}_{i,\pi} &= C^{\mathrm{init}}_{i,\pi}-C^{\mathrm{init, opt}}_{i},\\
    R_{i,\pi,p,t} &= C_{i,\pi,p,t} - C^{\mathrm{opt}}_{i,p,t},\\
    R_{i,\pi,p,r} &= C_{i,\pi,p,r} - C^{\mathrm{opt}}_{i,p,r}.
\end{aligned}
\label{eq:regret_breakdown}
\end{equation}
Even with a fixed initial state, the initial cost can differ slightly across scenarios and strategies, since the cumulative emission budget couples the optimal dispatch in the initial year to the subsequent pathway.
The regret components show when and where a strategy causes additional cost. Unlike total regret, individual regret components can be negative, e.g., when a strategy over-invests in one technology but saves on resource costs, or when a strategy has high costs early in the transition but saves costs in later years. The regret breakdown therefore shows by which technologies and resources, and in which phases of the transition, a strategy deviates from the optimal pathways and quantifies how much each deviation costs.

\section{Results}
\label{sec:results}

We optimize Switzerland's energy transition from 2025 to 2050 across 1000 scenarios constructed from six uncertain macro drivers. In the following, we first condense the resulting ensemble of pathways into a small set of energy system states.
We then select three of these sequences as strategies for the entire transition and analyze their regret across all scenarios.
Finally, to reveal path dependencies and lock-in effects in the energy transition, we investigate partial strategies, i.e., strategies that fix the near-term decisions up to 2035, the target state of the energy system in 2050, or both.

\subsection{Similar energy system states occur at different stages of the transition}
\label{sec:states}
\begin{figure}[h]
\centering
\includegraphics[width=1\linewidth]{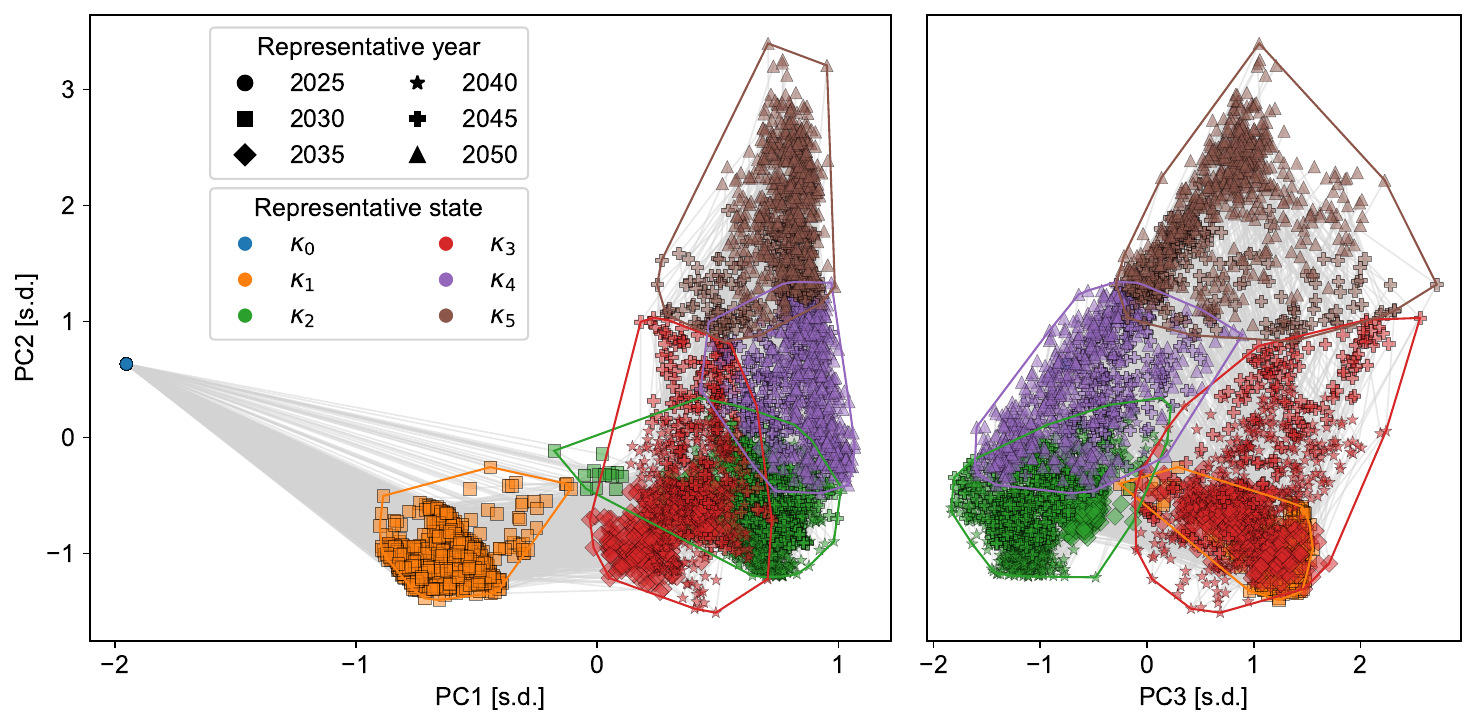}
\caption{Energy transition pathways of all 1000 scenarios in the space of the first three principal components of the sectoral standardized outputs, in standard deviations (s.d.). PC1, PC2, and PC3 explain 63\%, 16\%, and 8\% of the variance. Each grey line represents a pathway. Symbols along a pathway indicate the states in representative years, and their color indicates the representative state of the energy system. The outlines are convex hulls of all members of a representative state.}
\label{fig:PCA}
\end{figure}

With six representative years per pathway and 1000 scenarios, the pathway ensemble contains 6000 energy system states. Each state is described by 72 features: installed capacities for production, conversion, and storage technologies, and shares for sectoral output. To visualize this large output space, we project the states onto their first three principal components (PC1--PC3), which explain 63\%, 16\%, and 8\% of the total variance in the data, respectively (Fig.~\ref{fig:PCA}). 
Tables~S18 and S19 in Note S5 characterize each principal component by its loadings.

We use k-means clustering to reduce the 6000 energy system states to six representative states $\kappa_0$--$\kappa_5$ (Table~\ref{tab:state_means}). %
The first two states are tied to a single year: $\kappa_0$ represents the initial system of 2025, and $\kappa_1$ contains the states of 2030 (99\% of scenarios). The four remaining states are not tied to a single year, but each state contains systems from several years. %
The same representative state can thus be reached earlier in some pathways and later in others.
The four later states play different roles in the transition: $\kappa_2$ and $\kappa_3$ are predominantly intermediate states, whereas $\kappa_4$ and $\kappa_5$ are target states of the energy transition.

\begin{table}[h]
\caption{Representative states characterized by selected technologies: mean installed capacity, or share of the subsector output, over all scenario–years of a state. CCGT: combined-cycle gas turbine; Dec.: decentralized; DH: district heating; CHP: combined heat and power; CC: carbon capture.}

\centering
\input{tables/state_means}
\label{tab:state_means}
\end{table}

The general direction of the energy transition is captured by the first principal component (PC1).
Because the model is initialized with today's energy system, all pathways start from the same point in 2025 (blue circle in Fig.~\ref{fig:PCA}). All pathways share key developments in the transition, which can therefore be seen as robust to the uncertainties considered. Across all scenarios, we observe a large build-out of photovoltaics (PV) from today's 8.2\,GW to 32\,GW on average in 2050, dominated by open-field PV (24.7\,GW). Wind power is expanded to its maximum potential of 4.6\,GW in most pathways (88\% of the scenarios in 2050), and nuclear power is always fully phased out by 2040. The transport sector rapidly electrifies by 2035, with a complete shift to battery-electric cars in private transport and freight transport split between electric trucks (60\%) and trains (40\%). Heat pumps become the dominant supplier of district heating (DH), with an average share of 76\% in 2050. In industry, waste boilers and, where possible, high-temperature heat pumps replace gas boilers as the main sources for process heat. 

The pathways diverge more in the later years of the transition, i.e., the further the transition proceeds, the greater the spread in possible energy systems. This spread is captured by PC2 and PC3. PC3 trades off alpine against open-field PV, combined with a slight shift from gas boilers to decentralized heat pumps. This dimension separates the two intermediate states. State $\kappa_2$ expands open-field PV (20\,GW) and keeps gas boilers for decentralized heat (86\% on average). State $\kappa_3$ instead builds alpine PV early (9.2\,GW) to complement open-field PV (10.3\,GW) and begins replacing gas boilers with heat pumps, reducing the gas boiler share to 70\%.
PC2 separates states that heat with heat pumps from those that heat with gas boilers, and therefore distinguishes the two target states. State $\kappa_4$ keeps a largely gas-based heating system (72\% gas boilers) and offsets its emissions with direct air capture (DAC, 1.1\,kt\,CO$_2$/h), together with some alpine PV (5.6\,GW). State $\kappa_5$ instead electrifies decentralized heating with heat pumps (83\%) and substantially expands PV (24\,GW open-field, 8.7\,GW alpine), complemented by backup gas power plants (1.4\,GW). Thus, $\kappa_4$ and $\kappa_5$ represent two different ways to reach net zero: compensating gas-heating emissions with CO$_2$ removal, or avoiding them in the first place through electrification. We discuss the sensitivity of the gas-based target system to our data assumptions in Section~\ref{sec:limitations}.

\subsection{A few sequences of states describe most transition pathways}
\label{sec:sequences}

\begin{figure}[h]
\centering
\includegraphics[width=1\linewidth]{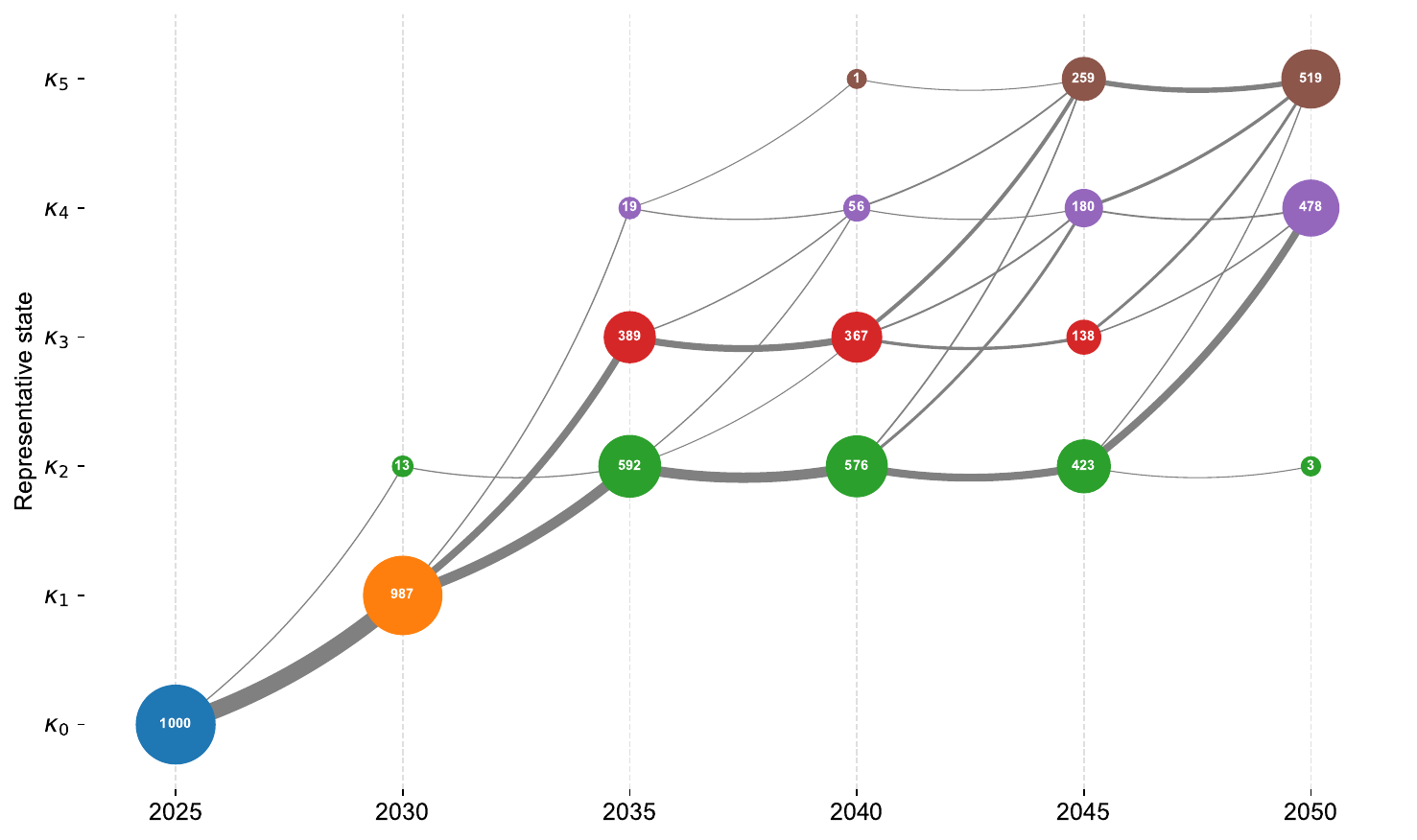}
\caption{Network of transitions between representative states. Mapping the representative states to their year of occurrence transforms the 1000 transition pathways into 20 distinct sequences. Each circle shows a representative energy system state in a given year. The color of a circle indicates the representative state it belongs to. The size of each circle and the number within it indicate how often the state occurs in that year. The thickness of the connecting lines illustrates the transition frequency between two states from one representative year to the next.}
\label{fig:state_transitions}
\end{figure}

Mapping each representative state of the pathway ensemble to the years in which it occurs turns 1000 pathways into a network of transitions between states (Fig.~\ref{fig:state_transitions}). The six representative states are connected to 20 unique sequences, each of which is a candidate transition strategy.

\begin{table}[h]
\caption{The five most frequent sequences of states and their share of the 1000 scenarios. Together, they cover 79\% of the scenarios.}
\centering
\input{tables/sequences}
\label{tab:sequences}
\end{table}

Until 2030, the pathways largely coincide: all start from $\kappa_0$ in 2025, and 99\% move to $\kappa_1$ in 2030. The first major branching occurs in 2035, when 59\% of the scenarios are in the gas-based state with open-field PV ($\kappa_2$) and 39\% in the state with early alpine PV ($\kappa_3$). Most pathways remain in their 2035 state until 2040. By 2050, the pathways end in either $\kappa_4$ (48\%) or $\kappa_5$ (52\%), with only 0.3\% remaining in $\kappa_2$. Although 20 different sequences occur, the five most frequent ones cover 79\% of the scenarios (Table~\ref{tab:sequences}).

From these five sequences, we select three strategies that are diverse in the timing and the direction of the energy transition:
\begin{itemize}
\item \emph{Gas \& CO$_2$ removal} (\texttt{0-1-2-2-2-4}), the most frequent sequence (39\% of the scenarios), keeps gas boilers in decentralized heating throughout the transition and captures and exports the remaining emissions in 2050.
\item \emph{Early electrification} (\texttt{0-1-3-3-5-5}), the second most frequent sequence (17\%), builds alpine PV from 2035 and reaches the electrified state $\kappa_5$ already in 2045.
\item \emph{Late switch} (\texttt{0-1-2-2-4-5}), occurring in 7\% of scenarios, follows \emph{Gas \& CO$_2$ removal} until 2040 and then switches to the electrified state $\kappa_5$ by 2050. The strategy thus shares its near-term decisions with the first strategy and its target state with the second, thereby isolating the effect of postponing electrification.
\end{itemize}

When enforcing a strategy, we prescribe the year-specific mean of the state's members (Eq.~\ref{eq:state_year_mean}) rather than the mean over all years in which a state occurs (Table~\ref{tab:state_means}). For example, when \emph{Late switch} passes through $\kappa_4$ in 2045, no direct air capture is installed, whereas when \emph{Gas \& CO$_2$ removal} reaches $\kappa_4$ in 2050, 1.7\,kt\,CO$_2$/h of direct air capture is built. The year-specific state means are given in Tables~S20 and S21 in Note~S6.

\subsection{The choice of low-regret strategy depends on the  decision-maker's risk aversion}
\label{sec:regret_curves}

\begin{figure}[h]
\centering
\includegraphics[width=0.8\linewidth]{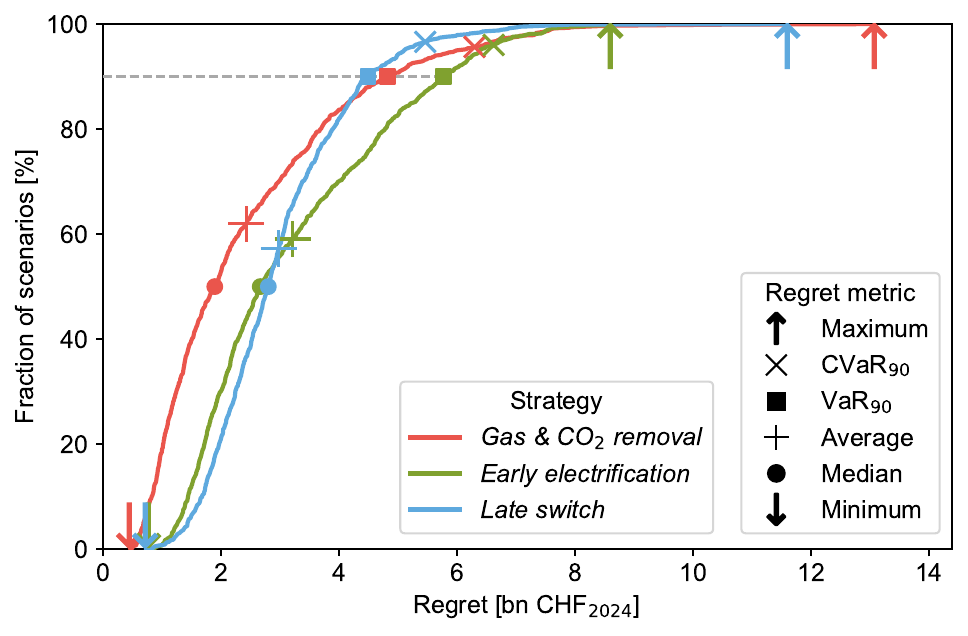}
\caption{The cumulative regret curves display the regret of the three strategies across all 1000 scenarios. Downward arrows mark the minimum regret for each strategy, while upward arrows mark their maximum regret. Circles represent the median and plus signs represent the average regret across all scenarios. The value-at-risk VaR$_{\beta}$ for a strategy is determined by selecting a confidence level $\beta\in[0,100]$ on the y-axis and identifying the corresponding regret value on the x-axis. This regret value is exceeded in (100-$\beta$)\% of scenarios. As an example, the VaR at a confidence level of 90\% is indicated (square). For the same confidence level, the conditional value-at-risk CVaR$_{90}$, i.e., the average regret of the 10\% worst scenarios, is shown by an $\times$.}
\label{fig:regret_curves_full_pathway}
\end{figure}

We enforce each of the three strategies on all 1000 scenarios and compute each strategy's regret as the difference between the scenario's optimal pathway transition cost and the strategy's transition cost (Eq. \eqref{eq:regret}). The cumulative regret curves (Fig.~\ref{fig:regret_curves_full_pathway}) show the regret distribution across all scenarios and allow at-a-glance comparison of the strategies under several ensemble-based decision criteria, including minimum, median, average, and maximum regret, value-at-risk (VaR), and conditional value-at-risk (CVaR) (Table~\ref{tab:regret_metrics}). Using the regret curves, a decision-maker can choose a strategy based on these various metrics, depending on their personal risk aversion \cite{mavromatidis_comparison_2018}.

\begin{table}[h]
\caption{Regret metrics of the three strategies across all 1000 scenarios, in billion CHF$_{2024}$. Within each metric, the lowest value among the strategies is highlighted in \textbf{bold}.}
\centering
\input{tables/regret_metrics}
\label{tab:regret_metrics}
\end{table}

The average regret of the three strategies ranges from 2.4 to 3.2\,billion CHF$_{2024}$, i.e., 0.5--0.6\% of the average total transition cost of 505\,billion CHF$_{2024}$ from 2025 to 2050. Thus, relative to the total cost, the regret is small. In absolute terms, it amounts to about 100--130 million CHF$_{2024}$ per year of the transition.

None of the strategies stochastically dominates \cite{levy_stochastic_2016}, i.e., no strategy has a lower regret at every percentile, and the cumulative regret curves cross. 
\emph{Gas \& CO$_2$ removal} has the lowest minimum, median, and average regret and is the lowest-regret strategy in 54\% of the 1000 scenarios. At the same time, it has the heaviest tail, with a maximum regret of 13.1\,billion CHF$_{2024}$. \emph{Early electrification} shows the opposite profile. The strategy has the highest average regret (3.2\,billion) and CVaR$_{90}$ (6.6\,billion), but the lowest maximum regret (8.6\,billion), and is the lowest-regret strategy in 39\% of scenarios. \emph{Late switch} is the lowest-regret strategy in only 7\% of scenarios but has the lowest VaR$_{90}$ (4.5\,billion) and CVaR$_{90}$ (5.5\,billion). Although rarely the best choice, it limits the regret in the worst 10\% of the scenarios. Consequently, the choice of strategy depends on the decision-maker's risk attitude. A risk-neutral decision-maker might choose \emph{Gas \& CO$_2$ removal}, a decision-maker averse to tail risk \emph{Late switch}, and a decision-maker minimizing the maximum regret \emph{Early electrification}.

\subsection{Strategies trade off domestic investment against resource import costs}
\label{sec:regret_decomposition}

\begin{figure}[h]
\centering
\includegraphics[width=0.99\linewidth]{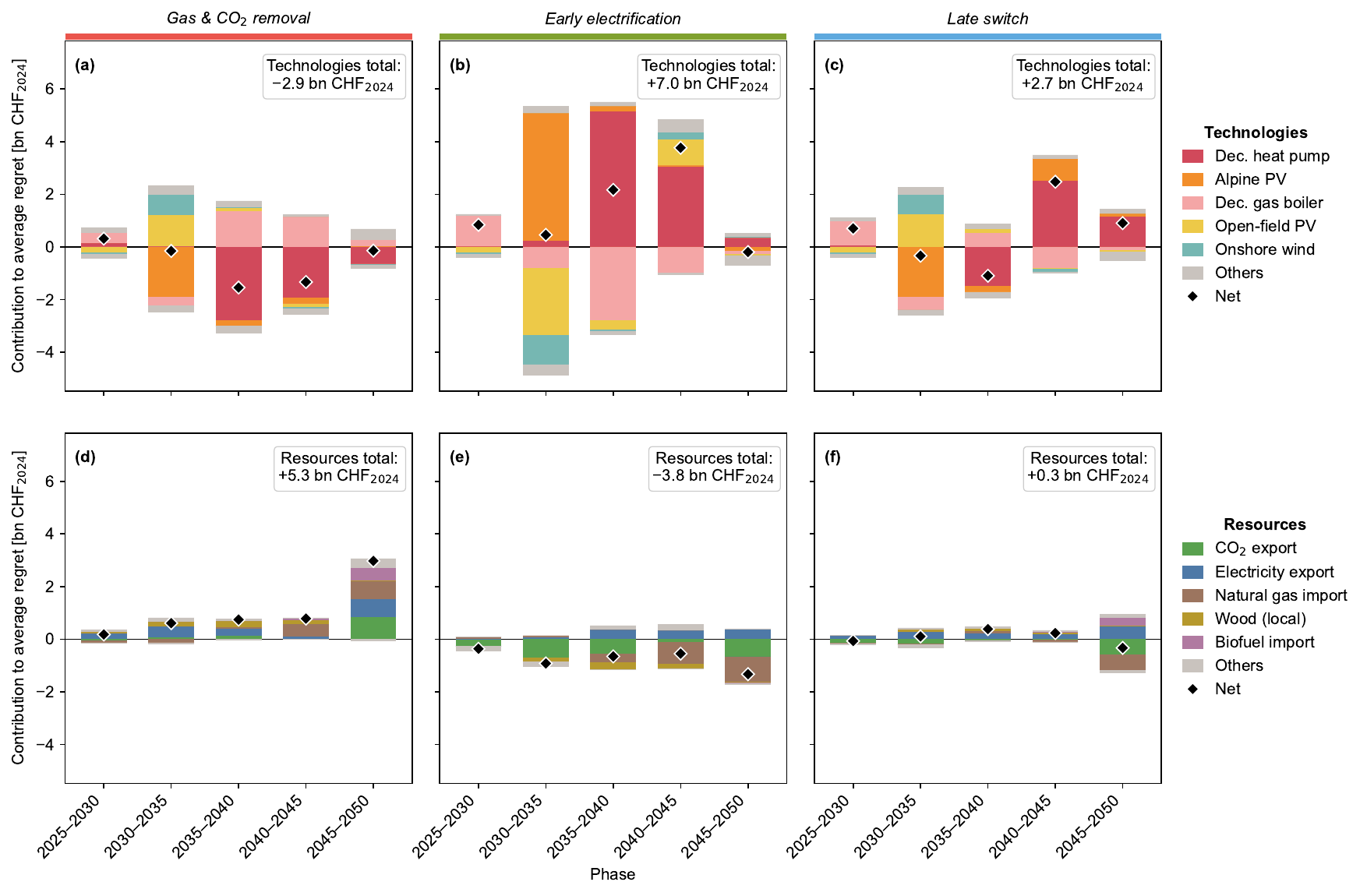}
\caption{Average regret components of the three strategies across all scenarios, decomposed by phase of the transition and by technologies (a--c) and resources (d--f). Stacked bars show the five components with the largest absolute contribution over the entire transition. The remaining components are summed in ``Others''. Positive values represent additional costs compared to the optimal pathways, while negative values represent savings. Diamonds mark the net regret per phase, and the box gives the sum over all phases.}
\label{fig:regret_breakdown}
\end{figure}

Beyond overall performance, decision-makers need to understand what causes the regret of each strategy. We therefore decompose each strategy's average regret by technology, resource, and over the phases of the transition (Fig.~\ref{fig:regret_breakdown}). By construction, a strategy's total regret in a scenario is non-negative, since enforcing a strategy adds constraints to the cost minimization (Eq. \eqref{eq:regret}). The regret components, however, can be negative, as a strategy can save costs in one technology, resource, or phase at the expense of others (Eq. \eqref{eq:regret_breakdown}). For example, a hypothetical strategy that keeps oil boilers saves investments in heat pumps and thus has a negative regret component for heat pumps. However, this alone does not mean it is a good decision, as in turn, it would have positive regret components for oil imports and, later in the transition, for the CO$_2$ removal needed to offset the additional emissions.

The phases in which regret accumulates reflect the timing of the strategies. \emph{Gas \& CO$_2$ removal} accumulates most of its regret at the end of the transition (2.8\,billion CHF$_{2024}$ in 2045--2050), when net zero must be reached with a gas-based heating system. \emph{Early electrification} accumulates regret in the middle of the transition (2035--2045), when it invests earlier than most optimal pathways, but recovers part of the regret in 2045--2050. \emph{Late switch} experiences most of its regret in 2040--2045, when switching from gas-based to electrified decentralized heating.

The time decomposition also explains why \emph{Early electrification} and \emph{Late switch} differ strongly in their regret for decentralized heat pumps (+8.8 vs.\ +2.3\,billion), although both reach almost the same heat pump capacity in 2050 (14.1 vs.\ 13.7\,GW). About half the difference stems from the residual value at the end of the horizon, which is higher for the younger heat pumps of \emph{Late switch} and thus lowers regret. About two fifths come from investing earlier, when unit costs are higher, and discounting is weaker. The remainder comes from accumulated maintenance costs over the years. However, as earlier heat pumps replace gas boilers and save resources, the total regret of the two strategies remains close (3.2 vs.\ 3.0\,billion). Individual technology components must therefore be read together with the resource components they offset.

The breakdown by technologies and resources reveals a trade-off between domestic investment and the cost of imported resources. \emph{Gas \& CO$_2$ removal} saves on technologies ($-$2.9\,billion), mainly decentralized heat pumps ($-$5.2\,billion) and alpine PV ($-$2.3\,billion), but spends more on resources (+5.3\,billion). The system exports less electricity and spends more on imported natural gas and biofuels, and on CO$_2$ export. The savings in domestic investment, thus, come at the cost of a higher dependence on other countries. \emph{Early electrification} shows the opposite behavior. The strategy's additional investments (+7.0\,billion), dominated by decentralized heat pumps (+8.8\,billion) and alpine PV (+5.0\,billion), reduce its resource costs ($-$3.8\,billion), mainly by requiring less CO$_2$ export and natural gas. \emph{Late switch} lies in between. The strategy's technology regret (+2.7\,billion) comes from installing decentralized heat pumps when switching to $\kappa_5$, while its overall resource regret is small (+0.3\,billion).

\subsection{Technology cost decides between electrified and gas-based heating}

\label{sec:decision_map}
\begin{figure}[h]
\centering
\includegraphics[width=\linewidth]{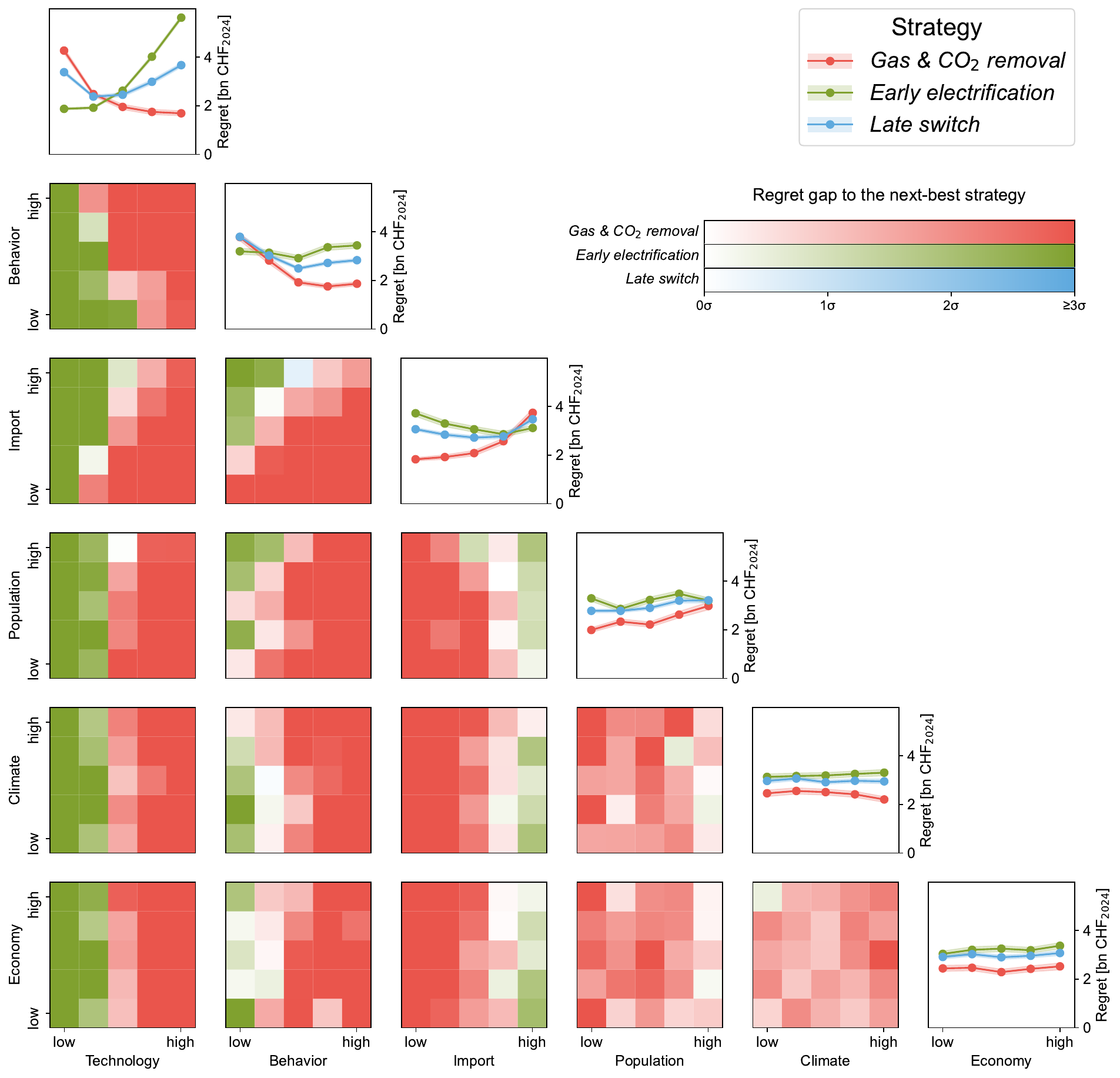}
\caption{Decision maps show the dependence of regret of the three strategies on the six uncertain macro drivers. The diagonal panels show the regret of each strategy vs.\ each driver. The drivers are sorted by the mean absolute Pearson correlation with the regret of the three strategies. The uncertainty range of every driver is divided into $n_\text{b} = 5$ bins of equal width, each containing 200 scenarios. In the diagonal panels, the average regret (line) and the standard error of the average (shaded area) are shown for each bin. The lower triangular panels present decision maps for pairwise driver combinations, with $5\times5=25$ pixels per map and thus on average 40 scenarios per pixel. The color of a pixel shows the strategy with the lowest average regret under the given conditions. The stronger the color, the larger a strategy's regret gap, in standard errors of the average regret.}
\label{fig:decision_maps}
\end{figure}

To identify under which conditions each strategy performs best, we examine how the strategies' regret depends on the six sampled macro drivers (Fig.~\ref{fig:decision_maps}). The panels on the diagonal show how each strategy's average regret moves with an individual macro driver. The drivers are sorted from left to right by decreasing influence on regret. The lower triangular panels are decision maps. For each pair of drivers, a pixel's color indicates the strategy with the lowest average regret under these conditions, and the color's intensity indicates the \emph{regret gap} (Note~S7) to the next-best strategy.

The diagonal panels show that of all drivers, technology cost has the greatest impact on the strategies' regret. As technology costs increase, the average regret of \emph{Early electrification} rises from about 1.9 to 5.6\,billion CHF$_{2024}$, while the regret of \emph{Gas \& CO$_2$ removal} falls from about 4.3 to 1.7\,billion. These dependencies reflect the regret breakdown (Fig.~\ref{fig:regret_breakdown}), where \emph{Early electrification} exhibits the largest investments, while \emph{Gas \& CO$_2$ removal} reduces investment costs and instead relies on resource imports. \emph{Late switch}, which balances investments and resources, varies only between 2.4 and 3.7\,billion across the technology-cost range. Overall, the strategy is the least sensitive to changes in the macro drivers.
Eco-friendliness of behavior is the second most influential driver; in particular, \emph{Gas \& CO$_2$ removal} depends on eco-friendly behavior to keep its regret low.
Less eco-friendly behavior raises the demand for energy services. In a gas-based heating system, higher demand means more residual emissions, which must be offset elsewhere. \emph{Gas \& CO$_2$ removal} does so mainly by replacing fossil jet fuel with biofuels in aviation, by importing more and exporting less electricity, and by paying for CO$_2$ compensation abroad.
Import costs are the third most influential driver. 
With high import costs, the optimal pathways shift towards domestic electricity generation. Because \emph{Gas \& CO$_2$ removal} invests the least in domestic generation, its regret increases. \emph{Early electrification} is closer to the optimal pathways when imports are expensive. Hence, low import costs favor \emph{Gas \& CO$_2$ removal}, while very high import costs make \emph{Early electrification} the lowest-regret strategy.
Population growth, climate change, and economic development influence regret less.

The lower triangular panels help to navigate dependencies between drivers when choosing a strategy.
\emph{Late switch} is the lowest-regret strategy in only 3 of 375 pixels, all with a small regret gap, underlining that the strategy hedges against the tail of the regret distribution rather than being the best choice for a particular future.
At low technology costs, \emph{Early electrification} has the lowest regret, while at high technology costs, \emph{Gas \& CO$_2$ removal} does. At medium technology costs, however, a second driver decides. With little eco-friendly behavior, \emph{Early electrification} remains the lowest-regret strategy up to the middle of the technology-cost range, whereas with high eco-friendly behavior or low import costs, \emph{Gas \& CO$_2$ removal} already performs best at the second-lowest technology cost bin.
For driver combinations with little influence, e.g., climate change together with economic development, the map shows \emph{Gas \& CO$_2$ removal} (the strategy with the overall lowest average regret) in pale color, i.e., with a small regret gap. 

\clearpage

\subsection{Near-term decisions create lock-in effects; fixing target states reveals path dependencies}
\label{sec:partial_strategies}

Full-pathway strategies fix the state of the energy system in each representative year and thus accumulate regret over the entire transition. In the following, we separate near-term decision regret from target-state regret by defining partial strategies (summarized in Table~\ref{tab:alternatives_design}). Fixing only the target state shows whether a final state is desirable on its own, similar to a greenfield analysis but starting from today's energy system. Fixing only the path up to 2035 shows the regret of near-term decisions while keeping later decisions flexible. Fixing both shows how the regret of reaching a target state depends on the path towards it.

\begin{table}[h]
\caption{The path up to 2035 (rows) is combined with the target state in 2050 (columns). ``X'' leaves the respective years unconstrained. The upper-left cell is the unconstrained optimization, i.e., the optimal base run for the regret evaluation. The combinations are visualized in Fig.~\ref{fig:alternatives}a.}
\centering
\input{tables/alternatives_design}
\label{tab:alternatives_design}
\end{table}

\begin{figure}[h]
\centering
\includegraphics[width=0.7\linewidth]{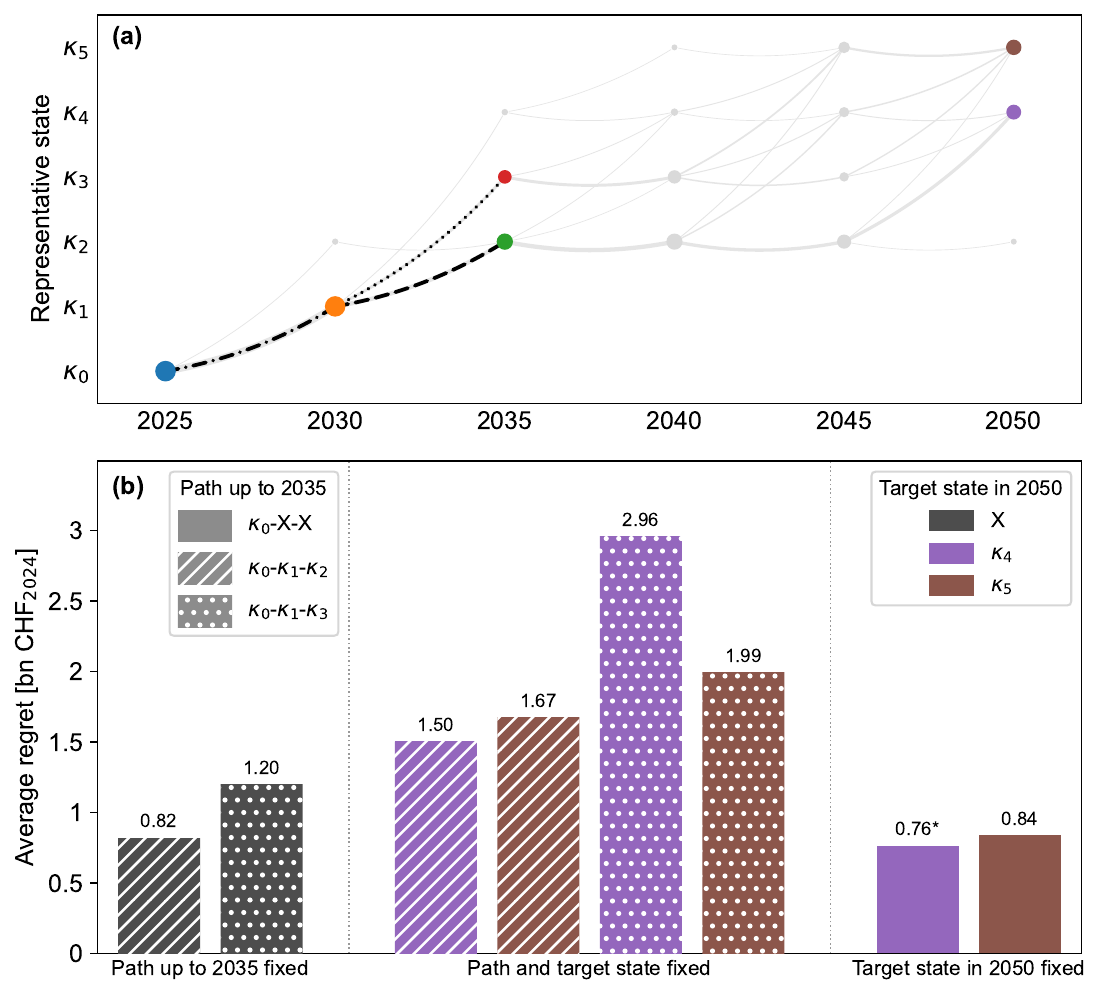}
\caption{Partial strategies: (a) The paths up to 2035 ($\kappa_0$-$\kappa_1$-$\kappa_2$, dashed; $\kappa_0$-$\kappa_1$-$\kappa_3$, dotted) and the target states in 2050 ($\kappa_4$, $\kappa_5$) in the state transition diagram of Fig.~\ref{fig:state_transitions}. (b) Average regret of the partial strategies that fix only the path up to 2035, both the path and the target state, or only the target state. Hatching indicates the path, and color indicates the target state.}
\label{fig:alternatives}
{\raggedright\par\footnotesize $^*$Upper bound: The strategy \texttt{0-X-X-X-X-4} encounters numerical trouble in 10 out of 1000 scenarios. We bound the regret of these scenarios from above by using the regret of the more constrained strategy \texttt{0-1-X-X-X-4}.\par}
\end{figure}

Fixing the target state alone causes little regret (Fig.~\ref{fig:alternatives}b), i.e., enforcing $\kappa_4$ in 2050 results in an average regret of about 0.76\,billion CHF$_{2024}$ and $\kappa_5$ in 0.84\,billion. Both ways of reaching net zero are thus nearly equivalent if the path towards them is left unconstrained. The near-term decisions matter more. Fixing the path up to 2035 causes an average regret of 0.82\,billion for the sequence $\kappa_0$-$\kappa_1$-$\kappa_2$ and 1.20\,billion for $\kappa_0$-$\kappa_1$-$\kappa_3$. The near-term decisions of the next decade accumulate as much regret as the choice of the target state, or more.

Evaluating the near-term path and target together shows that the regret of a target state depends on the path taken to reach it. After following $\kappa_0$-$\kappa_1$-$\kappa_2$, the two target states remain close in regret, with an average regret of 1.50\,billion for $\kappa_4$ and 1.67\,billion for $\kappa_5$. However, after $\kappa_0$-$\kappa_1$-$\kappa_3$, which invests in alpine PV early, $\kappa_5$ is clearly preferable (1.99\,billion) to $\kappa_4$ (2.96\,billion). For three of the four path--target combinations, the regret is close to the sum of the regrets of path and target alone. However, reaching $\kappa_4$ after $\kappa_0$-$\kappa_1$-$\kappa_3$ results in an average regret of 2.96\,billion, i.e., 1\,billion more than the sum of the parts of the strategy (i.e., $\kappa_0$-$\kappa_1$-$\kappa_3$: 1.20\,billion + $\kappa_4$: 0.76\,billion). Thus, early commitment to alpine PV and electrified heating makes the gas-based target state costly and locks the transition in on the electrified target state $\kappa_5$. Near-term decisions therefore can restrict which target states remain low-regret.

Finally, comparing the regret of the partial strategies with the full strategies shows the cost of additionally fixing the intermediate years 2040 and 2045. With the same path and target state, fixing also 2040 and 2045 raises the average regret from 1.50 to 2.43\,billion for \emph{Gas \& CO$_2$ removal}, from 1.99 to 3.21\,billion for \emph{Early electrification}, and from 1.67 to 2.97\,billion for \emph{Late switch}. About 40\% of the regret of the full strategies thus results from the intermediate states, i.e., from the timing of the transition rather than from its start or its end. Keeping intermediate decisions flexible and revisiting them as uncertainty resolves can therefore substantially reduce regret.

\clearpage
\section{Discussion}

\subsection{A state-based perspective on regret in energy transition pathways}

In this work, we introduce a state-based perspective to the analysis of pathway ensembles. Previous studies reduce pathway ensembles by clustering entire pathways, either by combining the technology metrics of selected years into one vector per pathway, which compares pathways year by year \cite{li_using_2020}, or by treating each pathway as a time series and measuring similarity with dynamic time warping, which tolerates shifts in timing \cite{kumar_resilient_2026}. In both approaches, a representative pathway is defined by the mean or medoid of a cluster. In contrast, we cluster the energy system states of all scenarios and years together (Fig. \ref{fig:PCA}), meaning that states are defined independent of the year. Thus, similar system designs that are reached earlier in some pathways and later in others are described by the same state (e.g., $\kappa_2$ occurs from 2030 to 2050). The timing of the transition is then explicitly described as the sequence of states. 
Csereklyei et al.~\cite{csereklyei_energy_2017} apply the idea of clustering energy system states to historical data. They cluster the energy mixes of EU member states over 1971--2010 as country-year observations and trace the movement of each country between the clusters as its energy pathway. We transfer this perspective from historical data to an ensemble of optimized future pathways. This perspective turns the pathway ensemble into a network of transitions between states (Fig.~\ref{fig:state_transitions}), in which the branching points of the transition become visible. 

Strategies follow directly from the data as sequences of states and can be enforced in the model bounded around the state means.
Describing strategies as sequences of states also allows for defining partial strategies, which we use to analyze path dependencies (Fig. \ref{fig:alternatives}). These path dependencies relate to the literature on carbon lock-in \cite{unruh_understanding_2000, seto_carbon_2016}, in which long-lived fossil infrastructure constrains later decarbonization. In our results, however, we observe that early investments in low-carbon technologies, such as heat pumps and alpine PV in $\kappa_3$, narrow the later options as well, i.e., we see an \emph{electricity lock-in}.

Our state-based perspective resembles Dynamic Adaptive Policy Pathways (DAPP) \cite{haasnoot_dynamic_2013, kwakkel_developing_2015}, in which a plan is a sequence of actions over time. For each action, DAPP determines an adaptation tipping point, i.e., the condition under which the action no longer meets its objectives, and monitors signposts that indicate when a switch to another pathway becomes necessary. Our decision maps (Fig.~\ref{fig:decision_maps}) are comparable to adaptation tipping points, as they show the conditions under which a strategy is the lowest-regret option (Note~S8).

The regret of a strategy depends on how tightly the strategy is defined by the bandwidth $\alpha$ (sensitivity analysis in Note~S6). Too tight bounds can enforce unrealistic system configurations and cause high regret, whereas too loose bounds make the strategies indistinguishable and thus meaningless. In future work, the bounds could instead be derived directly from the data, e.g., from quantiles of the members of a state.

\subsection{Comparison with other studies of the Swiss energy transition}

We compare our results for 2050 with three studies of the Swiss energy transition: the Energy Perspectives 2050+ (EP2050+, scenario ZERO Basis) \cite{prognos_ag_energieperspektiven_2021}, Panos et al.~\cite{panos_assessment_2023}, and the Energiezukunft 2050 update of the Swiss electricity industry association (VSE) \cite{vse_energiezukunft_2025}.
All studies agree on the main elements of the transition.
Nuclear power is phased out. While the other studies impose the phase-out, with fixed lifetimes for the existing plants and no construction of new plants, our model could build new nuclear power plants but never does so.
Road transport is electrified, with battery-electric cars dominating in all studies. Large heat pumps are the main source of district heat in our results and in the VSE study. Finally, all studies rely on CO$_2$ storage abroad to reach net zero.

The studies differ slightly in the installed capacity for renewable energy generation.
While the total PV capacity is similar (32\,GW on average in our results, compared with 38\,GW in the EP2050+, 21--47\,GW in Panos et al., and 30\,GW in the VSE study), our model builds mainly open-field (25\,GW) and alpine PV (7\,GW), whereas the other studies invest mostly in rooftop PV.
In our results, wind power is expanded to its maximum potential of 4.6\,GW in most scenarios by 2050, about twice the 1.9--2.6\,GW of the other studies.
The difference is explained by the assumed maximum potential, which is 8.4\,TWh/yr in our study, compared with 4.3\,TWh/yr in the EP2050+ and in Panos et al., and 3.8\,TWh/yr in the VSE study, where the potential is limited for reasons of social acceptance.
However, in a variant of the VSE study that allows the full sustainable potential of about 30\,TWh/yr \cite{meyer_windpotenzial_2022}, wind power is expanded to about 13\,GW.
Thus, all studies agree on the benefit of expanding wind power for the Swiss energy system.

A further difference lies in the treatment of international aviation. Unlike our study, the EP2050+ and Panos et al.\ exclude international aviation from the net-zero target, and the VSE study does not address aviation. The EP2050+ assumes 97\% of fuel for international aviation is synthetic kerosene imported from 2045 onwards, while Panos et al.\ do not report the fuel mix of international aviation. In contrast, aviation fuel remains mostly fossil in our results. In 2050, on average, about two-thirds of the 22\,TWh is fossil jet fuel. Most of the remainder is biofuel produced from pyrolysis. Brazzola et~al.~\cite{brazzola_role_2025} show that fossil jet fuel compensated by direct air capture plus storage is cheaper when only CO$_2$ is considered, whereas synthetic aviation fuels can become cheaper when non-CO$_2$ effects such as contrails are included. Since our model considers only the direct CO$_2$ emissions of aviation, it corresponds to the first case.

Residual emissions are offset by capturing CO$_2$ and exporting it for storage abroad. The CO$_2$ is captured at point sources, such as waste incineration, cement production, biomethanation, and pyrolysis, and directly from the air.
In 2050, our model exports on average 21\,Mt\,CO$_2$/yr (25\,Mt/yr in $\kappa_4$ and 18\,Mt/yr in $\kappa_5$), compared with 7--12\,Mt/yr of CO$_2$ capture and removal in the EP2050+ and Panos et al. The residual emissions from waste incineration, cement production, and agriculture are similar to those in the EP2050+ (about 12\,Mt\,CO$_{2,\mathrm{eq}}$/yr; Panos et al.\ exclude non-CO$_2$ emissions from agriculture). The higher export in our study has three main causes. First, unlike the other studies that restrict emissions per year, we impose a cumulative CO$_2$ budget of 468.7\,Mt\,CO$_{2,\mathrm{eq}}$ for the entire transition. The budget makes the system net negative in 2050 in 28\% of scenarios, adding 2\,Mt/yr CO$_2$ removal on average. Second, our net-zero target includes the 3.8\,Mt\,CO$_2$/yr from fossil jet fuel for international aviation, which the other studies do not include. Third, decentralized gas boilers still emit 7.0\,Mt\,CO$_2$/yr in $\kappa_4$, and 0.8\,Mt/yr in $\kappa_5$.

The persistence of decentralized gas boilers in about half of the scenarios ($\kappa_4$: 78\% decentralized gas boilers in 2050) is the largest difference from the other studies. Heat pumps are the main heating technology in the VSE study, supply 85\% of decentralized heat in Panos et al., and 88\% in the EP2050+. These shares are consistent with our target state $\kappa_5$, in which heat pumps provide 90\% of decentralized heat. Our model's preference for gas boilers in $\kappa_4$ stems from model assumptions, which we discuss next.

\subsection{Limitations}
\label{sec:limitations}

Our results are subject to limitations in the model and in the generation of scenarios. 
Two model assumptions are favorable for decentralized gas boilers. First, the model does not represent the cost of distribution grids, despite their impact on technology choice in the energy system \cite{schnidrig_role_2023}. Neglecting the maintenance and renewal costs of the existing gas distribution grid underestimates the cost of decentralized gas heating. While the electricity distribution grid is also not represented, its cost per unit of heat is lower for heat pumps due to their high efficiency. Second, captured CO$_2$ can be exported for storage abroad at a fixed cost of 100\,CHF$_{2024}$/t, with no limit on the exported amount. In practice, CO$_2$ transport and storage infrastructure does not exist yet, and slow deployment may limit the availability of CO$_2$ export \cite{burger_beyond_2026}.
With these two assumptions, the levelized costs of heat in 2050 are similar for heat pumps and gas boilers, at about 80--90\,CHF$_{2024}$/MWh.
Thus, small changes in the assumptions can tip the balance between the two technologies. A sensitivity analysis around the central scenario (Note~S9) shows that gas boilers disappear from decentralized heating by 2050 if the investment cost of heat pumps is 10\% lower, if gas boilers pay an additional distribution grid fee of about 3.4\,CHF$_{2024}$/MWh of gas, if CO$_2$ export costs 150 instead of 100\,CHF$_{2024}$/t, or if CO$_2$ export is limited to 12\,Mt/yr in 2050. Thus, the gas-based target state $\kappa_4$, and with it the strategy \emph{Gas \& CO$_2$ removal}, is only competitive under favorable conditions for CO$_2$ storage abroad and without accounting for grid costs.

Further limitations concern scenario generation. Because macro drivers describe persistent trends, i.e., are fully autocorrelated over the transition, we do not generate scenarios that are, for example, at the upper end of the range in 2035 and at the lower end in 2050. Guevara et al.~\cite{guevara_uncertainty_2024} analyze the temporal dependence of uncertain parameters with autoregressive processes, ranging from fully persistent trends, as in our work, to fully independent parameters across years. They show that temporal correlation allows investment decisions to adapt to the observed development, whereas independent values lead to overcapacities that protect against unlikely changes. Our scenarios do not contain abrupt changes after a decision has been made, such as disruptions of energy imports, a loss of social acceptance, or technologies that fail to mature \cite{coppitters_identifying_2026}. As each pathway is optimized with perfect foresight, our regret values do not reflect the cost of adapting to such unexpected events.

Furthermore, we model the uncertain parameters as fully correlated with their macro driver. For instance, a high value of the technology cost driver implies that PV, wind, heat pumps, and electric vehicles are all expensive at the same time. While common drivers such as material and labor costs justify some correlation, full correlation is a strong assumption.
Relaxing this assumption (Note~S4) leaves the average energy transition pathway almost unchanged. The individual energy system states spread more widely, mainly in 2035 and 2040, while the spread of total cost decreases. Full correlation with the drivers thus narrows the range of energy system states and widens the range of costs, but does not shift the average choice of technologies substantially.

\section{Conclusion}

Energy system optimization models can support decisions on the energy transition under uncertainty, but only if their results are translated into terms that decision-makers can act on. We propose a method that reduces a large ensemble of cost-optimal transition pathways to a small number of representative energy system states and transition strategies, and evaluates these strategies by their regret.

Applied to the Swiss energy transition, the method reduces 1000 pathways to six representative states that connect to a network of 20 unique sequences of states. All pathways share a large build-out of PV (mostly open-field), the electrification of road transport, heat pumps in district heating, and the phase-out of nuclear power. They differ mainly in decentralized heating, where gas boilers compete with heat pumps, and in the type of PV, open-field or alpine. The state network branches first in 2035, and the five most frequent sequences of states cover 79\% of the scenarios. 

We define three strategies directly from the data: \emph{Gas \& CO$_2$ removal}, \emph{Early electrification}, and \emph{Late switch}. No strategy stochastically dominates the others. Which strategy has the lowest regret depends on the decision criterion and thus on the risk attitude of the decision-maker. The strategies differ in how they trade off investments against resource costs. \emph{Gas \& CO$_2$ removal} saves investments but pays for natural gas imports and CO$_2$ exports. \emph{Early electrification}, in contrast, invests early to reduce resource costs. \emph{Late switch} lies between the two. Technology cost uncertainty largely determines which strategy is best. Low technology costs favor \emph{Early electrification}, while medium to high costs favor \emph{Gas \& CO$_2$ removal}. Fixing the target state in 2050 causes less regret than fixing the path up to 2035. A low-regret energy transition thus depends not only on choosing the correct net-zero target system but also on finding the right path toward it.

The competitiveness of gas-based heating relies on favorable assumptions on grid costs and on CO$_2$ export, whose infrastructure has yet to be built. Future work should therefore represent the rollout of CO$_2$ transport and storage and extend the evaluation beyond cost, e.g., to energy security and acceptance. Combining the representative states with signposts, as in adaptive pathways, could turn the strategies into plans that are adjusted as the uncertainty unfolds.

\data{The data that support the findings of this study are openly available. The model results are available at \url{https://doi.org/10.5281/zenodo.24363646}, and the code and model input data at \url{https://gitlab.ethz.ch/epse/systems-design-public/low-regret-pathways/-/tree/v1}.}

\roles{Conceptualization, G.W., E.M.T., A.B., and S.M.; Methodology, G.W., E.M.T., and S.M.; Investigation, G.W.; Writing – Original Draft, G.W.; Writing – Review \& Editing, G.W., E.M.T., A.B., and S.M.; Funding Acquisition, A.B. and S.M.; Resources, A.B. and S.M.; Supervision, E.M.T. and S.M.}

\funding{S.M. and G.W. acknowledge support from the Swiss National Science Foundation (SNSF) under Grant No. PZ00P2\_202117. S.M. acknowledges support from the European Union’s European Research Council (ERC), grant project no. 101164071 (UNITES). E.M.T. acknowledges support from TotalEnergies OneTech, France, for partial funding through project CT0072000.}

\conflict{A.B. served on review committees for research and development at ExxonMobil and TotalEnergies. A.B. and S.M. have ownership interests in firms that render services to industry, some of which may provide energy planning services. All other authors have no competing financial interests.}

\ack{During the preparation of this manuscript, the authors used Claude Opus 5.5 (Anthropic) to support improving the code and the manuscript's readability. The authors reviewed, validated and critically assessed all AI-assisted outputs and take full responsibility for the content of the manuscript.}

\clearpage
\printbibliography

\end{document}


\maketitle
\tableofcontents

\clearpage
\section*{Supplemental Notes}

\sinote{model}{Model modifications}

\sisubnote{vintage}{Technology stock initialization}

In the original formulation of EnergyScope Pathway \cite{limpens_energyscope_2024}, the existing technology stock at the start of a transition pathway is instantiated as having been entirely constructed during the phase immediately preceding the start year, i.e., when starting in 2025, all existing technologies are assumed to have been built during the phase $2020$--$2025$.
This assumption implies that, at the start year, every technology is at the beginning of its technical lifetime. Since the replacement of the existing stock determines the speed of the technological change \cite{fleiter_methodology_2018}, this assumption biases the model toward a delayed transition.
Switching away from an existing technology early (e.g., replacing an oil boiler with a heat pump shortly after the start year) requires decommissioning an asset long before the end of its technical lifetime, resulting in stranded assets.

To remove this bias, we reformulate the technology stock initialization to follow a user-defined age distribution at the start of the transition. When no distribution is given, the model assumes a uniform deployment of the technology before the start year, i.e., the existing stock of a technology is uniformly distributed across its lifetime. The detailed changes to the governing constraints are given below in pseudo-AMPL code, following the convention of the original EnergyScope Pathway model formulation \cite{limpens_energyscope_2024}.

\paragraph*{Initialize $\mathbf{F}_{new}$ for past phases with known age distribution}
\begin{equation}
\mathbf{F}_{new}(p, i) = \mathbf{F}(y_{initial}, i) \cdot \operatorname{age\_distribution}(p, i)
\end{equation}
\begin{equation*}
\forall\ i \in TECH_{age\_distrib},\ p \in PHASE\_PAST
\end{equation*}

\paragraph*{Initialize $\mathbf{F}_{new}$ for past phases with uniform build assumption}
\begin{equation}
\mathbf{F}_{new}(p, i) =
\begin{cases}
\dfrac{\mathbf{F}(y_{initial}, i) \cdot t_{phase}}{l(y_{initial}, i) - \big(l(y_{initial}, i) \bmod t_{phase}\big)}
& \text{if } y_{initial} - y_{stop} \leq l(y_{initial}, i) - t_{phase} \\[2ex]
0 & \text{otherwise}
\end{cases}
\end{equation}
\begin{equation*}
\forall\ i \in TECH \setminus TECH_{age\_distrib},\ p \in PHASE\_PAST,\ y_{stop} \in Y\_STOP(p)
\end{equation*}

\sisubnote{multiphase}{Multi-phase construction}

In the original model formulation, a technology planned for construction during phase $p=y_{\text{start}}$--$y_{\text{end}}$ already becomes fully operational in year $y_{\text{end}}$. This approximation is valid only if the technology's construction time does not exceed the length of a single phase (5 years in our case study). However, the construction time of new nuclear power plants typically exceeds 5 years \cite{portugal-pereira_better_2018}. To avoid misrepresenting such technologies, we introduce multi-phase construction. For example, a new nuclear power plant with 15 years of construction time that is initiated in phase $2025$--$2030$ is commissioned only in phase $2040$--$2045$. Capacity under construction can, in principle, be canceled in any later phase. The changes to the governing constraints are given below in pseudo-AMPL code.

\paragraph*{Number of construction phases}
\begin{equation}
n_{constr}(y,i) = \max\left(1, \left\lceil \frac{t_{constr}(y,i) + t_{phase}/2}{t_{phase}} \right\rceil \right)
\qquad \forall y \in YEARS,\ i \in TECH
\end{equation}

\paragraph*{Initial construction pipeline is empty}
\begin{equation}
\mathbf{F}_{constr}(p, p_{init}, i, k) = 0
\end{equation}
\begin{equation*}
\begin{aligned}
\forall\ & p \in PHASE\_WND,\ p_{init} \in PHASE\_WND,\ y_{start} \in Y\_START(p_{init}), \\
& i \in TECH,\ k \in \{2,\dots,n_{constr}(y_{start},i)\} : \operatorname{ord}(p) = 1
\end{aligned}
\end{equation*}

\paragraph*{No construction before the initial phase}
\begin{equation}
\mathbf{F}_{constr}(p, p_{init}, i, k) = 0
\end{equation}
\begin{equation*}
\begin{aligned}
\forall\ & p, p_{init} \in PHASE\_WND,\ y_{start} \in Y\_START(p_{init}), \\
& i \in TECH,\ k \in \{1,\dots,n_{constr}(y_{start},i)\} : \operatorname{ord}(p) < \operatorname{ord}(p_{init})
\end{aligned}
\end{equation*}

\paragraph*{No construction after the final phase}
\begin{equation}
\mathbf{F}_{constr}(p, p_{init}, i, k) = 0
\end{equation}
\begin{equation*}
\begin{aligned}
\forall\ & p, p_{init} \in PHASE\_WND,\ y_{start} \in Y\_START(p_{init}), \\
& i \in TECH,\ k \in \{1,\dots,n_{constr}(y_{start},i)\} : \operatorname{ord}(p) > \operatorname{ord}(p_{init}) + n_{constr}(y_{start},i) - 1
\end{aligned}
\end{equation*}

\paragraph*{No construction/cancellation beyond the required number of phases}
\begin{equation}
\mathbf{F}_{constr}(p, p_{init}, i, k) = 0
\end{equation}
\begin{equation*}
\begin{aligned}
\forall\ & p, p_{init} \in PHASE\_WND,\ y_{start} \in Y\_START(p_{init}), \\
& i \in TECH,\ k \in MAX\_CONSTR\_PHASES : k > n_{constr}(y_{start},i)
\end{aligned}
\end{equation*}

\paragraph*{Construction pipeline recursion}
\begin{equation}
\mathbf{F}_{constr}(p, p_{init}, i, k) = \mathbf{F}_{constr}\big(\operatorname{prev}(p), p_{init}, i, k-1\big) - \mathbf{F}_{cancel}(p, p_{init}, i, y_{start}, k)
\end{equation}
\begin{equation*}
\begin{aligned}
\forall\ & p, p_{init} \in PHASE\_WND,\ y_{start} \in Y\_START(p_{init}), \\
& i \in TECH,\ k \in \{2,\dots,n_{constr}(y_{start},i)\} : \operatorname{ord}(p) > 1
\end{aligned}
\end{equation*}

\paragraph*{Commissioning --- linking new capacity to the completed pipeline}
\begin{equation}
\mathbf{F}_{new}(p,i) = \sum_{\substack{p_{init} \in PHASE\_WND \\ y_{start} \in Y\_START(p_{init})}} \mathbf{F}_{constr}\big(p, p_{init}, i, n_{constr}(y_{start},i)\big)
\qquad \forall p \in PHASE\_WND,\ i \in TECH
\end{equation}

\clearpage

\sinote{data}{Input data for the Swiss energy system}

EnergyScope Pathway \cite{limpens_energyscope_2024} input data are used unless stated otherwise, with costs converted to CHF$_{2024}$. This note documents the data that are new or changed for the Swiss case study.

\sisubnote{conventions}{Conventions and currency conversion}

We convert costs by first inflating them in the original currency, then converting at the 2024 exchange rate, using World Bank consumer price inflation \cite{world_bank_inflation_nodate} and official exchange rates \cite{world_bank_official_nodate} (Tab.~\ref{tab:currency_factors}). Prices given in CHF$_{2025}$ are taken as CHF$_{2024}$, as the Swiss consumer price index changed by $+0.2\,\%$ between the two years. We assume a real social discount rate of 1.5\,\% over the whole transition, close to the about 1.4\,\% implied by the Stern Review \cite{stern_economics_2007}, which uses a near-zero pure rate of time preference to avoid burden shifting to future generations. %

\begin{table}[h]
\caption{Currency conversion factors to CHF$_{2024}$}
\label{tab:currency_factors}
\centering
\input{tables/currency_factors}
\end{table}

\clearpage
\sisubnote{demands}{Energy service demands}

All end-use demands are replaced by Swiss data. We use the CROSS end-use energy demand, release 2025-06-06 \cite{sweet_cross_end-use_2025}, reference variant with climate scenario RCP4.5, per sector and end-use type (Tab.~\ref{tab:demands}). The modeled passenger demand includes personal cars, passenger rail, buses, and trams, which together account for about 90\,\% of the total passenger-kilometers. Process heat is split into a medium- (below 200\,°C) and a high-temperature share, with the share of process heat from Energieperspektiven 2050+ \cite{prognos_ag_energieperspektiven_2021} (Fig.~129, scenario ZERO Basis: 29\,\% in 2025 decreasing to 23\,\% in 2050), applied to industry and services alike. Non-energy demand is held constant at 2118\,GWh (split into high-value chemicals 85\,\%, methanol 10\,\%, ammonia 5\,\%) \cite{wiest_low-regret_2026}.  %

\begin{table}[h]
\caption{End-use demands of the reference scenario.}
\label{tab:demands}
\centering
\small
\input{tables/demands}
\end{table}

\clearpage
\sisubnote{resources}{Resources: potentials and prices}

Tab.~\ref{tab:resources} lists the resource parameters that are new or changed compared with EnergyScope Pathway \cite{limpens_energyscope_2024}. All domestic waste must be used within the country, as the model cannot store or export it. Electricity is traded at the same price for import and export.

The model accounts for the emissions by tracking which resources enter or leave Switzerland. Additionally, it accounts for local agricultural emissions and the geogenic process emissions of cement production. While agricultural emissions are considered decentralized and thus are directly emitted to the atmosphere, cement emissions are modeled as point-source emissions and can be captured. Captured CO$_2$ can then either be utilized or exported for storage abroad. Remaining residual emissions can be compensated abroad by paying for direct air capture with storage. Both CO$_2$ export and CO$_2$ compensation count negatively towards the budget.

\begin{table}[h]
\caption{Resource parameters that are new or changed with respect to EnergyScope Pathway \cite{limpens_energyscope_2024}. %
}
\label{tab:resources}
\centering
\small
\input{tables/resources}
\end{table}

\clearpage
\sisubnote{technologies}{Technologies: costs and performance}

Tables~\ref{tab:technologies}--\ref{tab:technologies_heat} document new or changed technology parameters. The table notes give the source of each value, and all other technologies keep their EnergyScope Pathway data. The investment costs of PV, onshore wind, run-of-river, CCGT, heat pumps, boilers, solar thermal, district heating networks, electrolyzers and Li-ion batteries are taken from the CROSS data platform, release 2026-02-24 \cite{sweet_cross_investment_2026}
for the size classes of Tab.~\ref{tab:cross_mapping}. Because CROSS reports no operation \& maintenance (O\&M) costs, we keep the O\&M costs of EnergyScope Pathway. We update the costs for buses, trucks, Fischer-Tropsch fuel production, and decentralized direct electric heating using technology data from the Danish Energy Agency \cite{danish_energy_agency_technology_nodate, danish_energy_agency_technology_nodate-1, danish_energy_agency_technology_nodate-2}. High-temperature heat pumps use the original EnergyScope Pathway cost of the decentralized heat pump as a proxy.

The potentials of rooftop PV, alpine PV and onshore wind are the mean of the High and Low potential of Tab.~12 in the CROSS scenario protocol \cite{marcucci_cross_2025} (40, 15 and 8.35\,TWh/yr), divided by the mean full-load hours of the weather years 2016--2023, which gives 43.8, 9.3 and 4.6\,GW.
For the open-field PV, we assume the full potential of free areas along roads (24.7\,TWh/yr) and motorway embankments (5.6\,TWh/yr) \cite{remund_schweizer_2019}, as compiled in Tab.~4 of \cite{hugi_meta-study_2025}, i.e., 30.3\,TWh/yr or 25.9\,GW. 
Hydropower is not expanded beyond its 2024 capacities, except for the reservoir storage, which can increase from the 6.5\,TWh used today to 8.9\,TWh in 2050 through dam heightening \cite{boes_swiss_2021}.

\begin{table}[h]
\caption{Mapping of CROSS technology classes to model technologies.}
\label{tab:cross_mapping}
\centering
\small
\input{tables/cross_mapping}
\end{table}

\begin{table}[p]
\caption{Technology parameters that are new or changed with respect to EnergyScope Pathway
\cite{limpens_energyscope_2024}. Costs are given per unit of the main output (per kWh of storage capacity for storage).
}
\label{tab:technologies}
\centering
\small
\input{tables/technologies_electricity}
\end{table}

\begin{table}[p]
\caption{Continued from Tab.~\ref{tab:technologies}: transport and fuel production. Vehicle costs are given per pkm/h or tkm/h of capacity.}
\label{tab:technologies_transport}
\centering
\small
\input{tables/technologies_transport}
\end{table}

\begin{table}[H]
\caption{Continued from Tab.~\ref{tab:technologies}: heating technologies in buildings and industry.}
\label{tab:technologies_heat}
\centering
\small
\input{tables/technologies_heat}
\end{table}

\paragraph*{Other model parameters}
The electricity import and export capacities are the sum of the net transfer capacities between Switzerland and its neighbors from CROSS \cite{sweet_cross_net_2025}, based on ENTSO-E TYNDP 2024 (historical values for 2020, the Reference Grid scenario for 2030 and the Distributed Energy scenario from 2035 on): 11.4 and 11.9\,GW in 2025, rising to 17.3 and 17.0\,GW in 2050. District heating supplies at least its 2025 share of the low-temperature heat, 9.9\,\% (the share of the population heated by district heating in 2024 \cite{bundesamt_fur_statistik_bfs_bevolkerung_2025}), and at most 15\,\% in 2030, rising by 5 percentage points per phase to 30\,\% in 2045 (own assumption). Public transport covers at most 23.6\,\% of the passenger-kilometers, its 2025 share in the CROSS passenger demand \cite{sweet_cross_end-use_2025}; eco-friendly behavior raises this bound (Note~\ref{note:sampling}). Rail can supply 38--40\,\% of the freight transport, around the 39\,\% presented in the CROSS freight demand \cite{sweet_cross_end-use_2025}. As already implemented for other sectors in the EnergyScope Pathway, we restrict changes in industrial heat supply to a maximum of 50\,\% per phase.

\clearpage
\sisubnote{timeseries}{Time series and typical days}

All hourly series are rebuilt from Swiss data for the eight weather years 2016--2023 (Tab.~\ref{tab:timeseries}). 
Electricity demand, run-of-river, rooftop PV, wind power \cite{fraunhofer_ise_net_nodate}, and reservoir inflow \cite{fraunhofer_ise_storage_nodate} are constructed from Energy-Charts. Observed metering outages (rooftop PV, 20--26 February 2018; onshore wind, April 2021) are filled with the same hours of the previous year. 
Open-field and alpine PV capacity factors are constructed using PVGIS \cite{european_commission_joint_research_centre_jrc_nodate} (14\,\% system losses, averaging across eight hypothetical open-field sites and the locations of the three alpine projects: SedrunSolar, MadrisaSolar and Hohsaas). Solar thermal capacity factors are calculated from the PV series using ambient temperature data from PVGIS and fixed collector parameters. 
Space heating time series for households and services come from the ERA5-based dataset of Antonini et al.~\cite{antonini_weather-_2024}. The time series for hot water and high-temperature heat are built using JASM data \cite{panos_demand_2019}. The passenger mobility profile is constructed from FSO pocket statistics \cite{bundesamt_fur_statistik_bfs_mobility_2025} for the day types Monday--Friday, Saturday, and Sunday. 
In leap years, 29 February is dropped, and the series are renormalized to the 365 modeled days. 
Each weather year is reduced to 12 typical days by the MILP k-medoids clustering of EnergyScope TD \cite{limpens_energyscope_2019} with the series weights of Tab.~\ref{tab:timeseries}, each day being assigned to its nearest typical day. 
Once the typical days have been selected, the hourly electricity price profile is added from the Nexus-e projection \cite{gjorgiev_nexus-e_2022}. 
Because clustering removes extremes from the heating time series, the heating capacity is sized using a peak factor per weather year (Tab.~\ref{tab:peak_factor}). The model adds the three low-temperature demands before it sizes the heating technologies. The factor is the ratio of the peaks of that summed profile.

\begin{table}[h]
\caption{Hourly time series and their weight in the selection of the typical days.}
\label{tab:timeseries}
\centering
\small
\input{tables/timeseries_weights}
\end{table}

\begin{table}[h]
\caption{Peak low-temperature heat demand per weather year: day, hour, and annual peak demand compared with the highest demand retained by the 12 typical days. Both refer to the summed low-temperature demand that the model sizes the heating against (sum of space heating of households, of services and industry, and hot water), scaled with the reference demand of 2025 (66.1\,TWh). The last column gives the ratio between annual peak and typical-day peak.}
\label{tab:peak_factor}
\centering
\input{tables/peak_factor}
\end{table}

\clearpage
\sisubnote{co2budget}{CO$_2$ budget}

The emission budget replaces the EnergyScope Pathway target with the Swiss one.

\paragraph*{Historic emissions}
We calculate the Swiss historic emissions from the total greenhouse gas emissions of the CO$_2$ statistics \cite{bundesamt_fur_umwelt_bafu_co2-statistik_2026}, including the balance of land use, land-use change and forestry (LULUCF), and add the emissions of international aviation from the national greenhouse gas inventory \cite{bundesamt_fur_umwelt_bafu_treibhausgasinventar_2026}, which the official Swiss targets up to 2050 do not cover, but our model does. This results in:
\begin{equation}
E_{1990} = 53.17 + 3.09 = 56.26\,\mathrm{MtCO}_{2,\text{eq}},
\end{equation}
\begin{equation}
E_{2024} = 41.30 + 5.53 = 46.83\,\mathrm{MtCO}_{2,\text{eq}}.
\end{equation}

\paragraph*{Target emission levels (as shares of 1990)}

We calculate the emission levels in the future based on the Swiss emission reduction goals: $-$50\,\% in 2030 (CO$_2$ Act \cite{schweizerische_eidgenossenschaft_bundesgesetz_2011}, revision of 15 March 2024), $-$65\,\% in 2035 \cite{swiss_confederation_switzerlands_2025}, net zero in 2050 \cite{schweizerische_eidgenossenschaft_bundesgesetz_2022}, i.e.,

\begin{equation}
E_{2030} = 0.50 \cdot E_{1990} \qquad (-50\%)
\end{equation}
\begin{equation}
E_{2035} = 0.35 \cdot E_{1990} \qquad (-65\%)
\end{equation}
\begin{equation}
E_{2050} = 0.00 \cdot E_{1990} \qquad (-100\%).
\label{eq:net_zero}
\end{equation}

\paragraph*{Linear interpolation between anchor years}
For each interval $[y_a, y_b]$ with known start level $E_{y_a}$ and target level $E_{y_b}$, the yearly emissions path is defined by linear interpolation:
\begin{equation}
E_y = E_{y_a} + \frac{y - y_a}{y_b - y_a}\left(E_{y_b} - E_{y_a}\right), \qquad y = y_a, y_a+1, \dots, y_b
\end{equation}

This is applied over the three anchor pairs $(y_a, y_b) \in \{(2024, 2030),\ (2030, 2035),\ (2035, 2050)\}$, using the corresponding $E_{y_a}, E_{y_b}$ defined above.

\paragraph*{Period budgets}
For each interval, the budget is the cumulative sum of the interpolated path, less the starting year's value (to avoid double-counting across adjacent periods):
\begin{equation}
B_{y_a\text{-}y_b} = \sum_{y=y_a}^{y_b} E_y - E_{y_a}, \qquad (y_a, y_b) \in \{(2024,2030),\ (2030,2035),\ (2035,2050)\}
\end{equation}

giving $B_{2025\text{-}2030}$, $B_{2030\text{-}2035}$, and $B_{2035\text{-}2050}$.

\paragraph*{Total emission budget}

Summing the emission budgets from 2025 until 2050 gives:

\begin{equation}
B_{total} = B_{2025\text{-}2030} + B_{2030\text{-}2035} + B_{2035\text{-}2050} = 468.7\,\mathrm{MtCO}_{2,\text{eq}}.
\label{eq:total_budget}
\end{equation}

In the model, we prescribe only the total budget (Eq.~\ref{eq:total_budget}) and the net-zero state in 2050 (Eq.~\ref{eq:net_zero}) and leave the reduction trajectory free for the optimizer.

\clearpage
\sisubnote{init2025}{Initialization of the 2025 energy system}

To ensure that differences between scenarios arise only from the transition and not from different starting points, we fix the initial energy system and always use the same weather year (2023) in 2025. However, energy statistics data provide only the production of the main power plants and the shares of the technologies in mobility and heating, not the capacities of all technologies in the model. Fixing only the reported values would let the model build the remaining 2025 capacities freely in each scenario. To avoid this flexibility in the initial state, we calibrate the complete 2025 system once, and fix the result across all runs.

As calibration data (Tab.~\ref{tab:init_2025}), we use 2024 values unless stated otherwise: the electricity production of nuclear, wind, and PV and the biofuel import are taken from the Swiss overall energy statistics \cite{bundesamt_fur_energie_bfe_gesamtenergiestatistik_2025}, the installed capacities from Energy-Charts \cite{fraunhofer_ise_installed_nodate}. We fix the shares of cars, buses, and trucks in the Swiss vehicle stock \cite{bundesamt_fur_statistik_bfs_stock_2025, bundesamt_fur_statistik_bfs_mobility_2025}, the heating systems of the population, including district heating \cite{bundesamt_fur_statistik_bfs_bevolkerung_2025}, and the energy use of industry by source \cite{bundesamt_fur_energie_bfe_energieverbrauch_2025}. For the source mix of district heating, we use data for the year 2020 \cite{verband_fernwarme_schweiz_vfs_fernwarme-statistik_2021}.

The calibration run solves the central scenario, with all macro drivers and parameters at their median ($u = 0.5$) and the weather years 2023, 2023, 2022, 2023, 2020 and 2022 for 2025--2050, the same as in Note~\ref{note:sensitivity}. In the calibration run, the initial investment is weighted by a factor of 10 in the objective to prevent the perfect foresight model from building overcapacities that are needed only later. The resulting 2025 fleet (Tab.~\ref{tab:init_2025_fleet}) is fixed across all runs, along with the constraints in Tab.~\ref{tab:init_2025}. Capacities below $10^{-4}$ in their respective model unit (GW, GWh, Mpkm/h or Mtkm/h) are treated as numerical artifacts and set to zero.

\begin{table}[h]
\caption{Values of the 2025 energy system taken from statistics. Electricity production and net biofuel import are fixed; technology shares are lower bounds on the output of their subsector unless marked otherwise.}
\label{tab:init_2025}
\centering
\small
\input{tables/init_2025_targets}

\end{table}

\begin{table}[h]
\caption{Installed capacities of the 2025 energy system fixed in all runs.}
\label{tab:init_2025_fleet}
\centering
\small
\input{tables/init_2025_fleet}
\end{table}

\clearpage

\sinote{sampling}{Sampling design}

To evaluate system performance under uncertainty, we generate $N=1000$ scenarios by sampling six macro drivers $d \in D =\{\text{population, economy, technology, imports, behavior, climate}\}$ via Latin Hypercube Sampling as uniform random variables $u \sim \mathcal{U}(0,1)$. Each driver is sampled only once per scenario, with the resulting value $u$ held fixed for all years across the transition.
Additionally, we sample a weather year from the eight historical years 2016--2023, independently for each representative year from 2030 onward. For 2025, we use the weather year 2023 in all scenarios to ensure the same starting point across pathways.

To generate input parameters for the energy system model, we follow a three-step procedure: First, we compute trajectories of relevant system macro parameters (population size, GDP, number of households) over the entire transition. Second, we scale model input parameters (e.g., electricity and heating demand) according to the macro parameters. Third, we adjust the scaled model input parameters according to the macro drivers (e.g., higher eco-friendly behavior reduces demand for aviation, a colder weather year increases heating demand).

Each uncertainty range is characterized by a Low, High, and Reference scenario. We interpret the Low scenario as the lower bound $a$, the High scenario as the upper bound $b$, and the Reference as the median $m$. Following the principle of maximum entropy, the uncertainty range is described by a two-piece uniform distribution. A uniform random variable $u \sim \mathcal{U}(0,1)$ is transformed to the two-piece uniform distribution via:

\begin{align}
    f(u;\, a, m, b) = \begin{cases} a + 2u(m - a) & \text{if } u < 0.5 \\ m + 2(u - 0.5)(b - m) & \text{if } u \geq 0.5 \end{cases}.
    \label{two-piece uniform}
\end{align}

\subsubsection*{Macro parameters}

We calculate the macro parameters population size $P$, number of households $H$ and gross domestic product $GDP$ from the sampled macro drivers using Eq. \eqref{two-piece uniform}. The macro parameters are calculated as follows:

\begin{align}
    P &= f_{\text{population size}}\big(u_{\mathrm{population}}\big) \\
    H &= P \cdot f_{\text{households per capita}}\big(1 - u_{\mathrm{behavior}}\big) \\
    GDP &= P \cdot f_{\text{share working pop.}}\big(u_{\mathrm{economy}}\big) \cdot f_{\text{GDP per working pop.}}\big(u_{\mathrm{economy}}\big),
\end{align}

where we write $f_{\text{parameter}}(u_{d}) = f(u_{d};\, a_{\text{parameter}}, m_{\text{parameter}}, b_{\text{parameter}})$ for brevity, with the Low/Ref/High values for each parameter given in Tab.~\ref{tab:macro_parameters}. 

\begin{table}[h]
\caption{Low, reference and high trajectories of the macro parameters. The population and the reference trajectories are the CROSS macro-variables, release 2025-06 \cite{sweet_cross_macroeconomic_2025}. The low and high trajectories of the other three parameters deviate from the reference by 1\,\% in 2030, rising by 1 percentage point per period to 5\,\% in 2050 (2\,\% to 10\,\% for GDP per worker), and are based on our own assumptions. Households per capita decrease with the \emph{behavior} driver, i.e., the Low trajectory applies at $u_{\mathrm{behavior}} = 1$ and the high trajectory at $u_{\mathrm{behavior}} = 0$.}
\label{tab:macro_parameters}
\centering
\small
\input{tables/macro_parameters}
\end{table}
\newpage
\subsubsection*{Scaling model inputs with macro parameters}

Given the macro parameters population size $P$, number of households $H$, and gross domestic product $GDP$, we scale the relative model input parameters $\tilde{\theta}$ given in Tab.~\ref{tab:parameter_scaling} by multiplying each by its corresponding macro parameter $M \in \{P, H, GDP\}$, as indicated in the ``Scaled with'' column of Tab.~\ref{tab:parameter_scaling}:

\begin{equation}
    \theta = \tilde{\theta} \cdot M, \qquad M \in \{P, H, GDP\}.
    \label{eq:scaling}
\end{equation}

For example, the annual residential heating demand is then obtained by multiplying the per-household residential heating demand $\tilde{\theta}_{\text{heat, res}}$ by the number of households $H$:
\begin{equation}
    \theta_{\text{heat, res}} = \tilde{\theta}_{\text{heat, res}} \cdot H.
\end{equation}

\begin{table}[h]
\caption{Model inputs relative to the macro parameters (reference values): the end-use demands of Note~\ref{note:demands} and the CROSS biomass potentials (reference scenario) \cite{sweet_cross_biomass_2025} are divided by the reference trajectories of population, households, or GDP of Tab.~\ref{tab:macro_parameters}. GDP-related intensities are given per CHF$_{2017}$, the price basis of the CROSS GDP.}
\label{tab:parameter_scaling}
\centering
\resizebox{\textwidth}{!}{\input{tables/parameter_scaling}}

\end{table}

\subsubsection*{Adjusting model input parameters}

In the last step, the scaled model input parameters $\theta$ (Eq.~\eqref{eq:scaling}) are further adjusted according to the sampled macro drivers, to account for driver-specific effects on demands, resource costs and availabilities, and technology investment costs. For each parameter $\theta$ and each driver $d$ affecting it, Tables~\ref{tab:uncertainties_demand}--\ref{tab:uncertainties_tech} give a Low and High adjustment multiplier for each representative year. Together with a multiplier of $1$ in the reference scenario (i.e., no adjustment), the Low and High adjustment multipliers give a two-piece uniform transform analogous to Eq.~\eqref{two-piece uniform}:
\begin{equation}
    g_\theta(u_d) = f(u_d;\, a_{\theta,d},\, 1,\, b_{\theta,d}),
    \label{eq:adjustment}
\end{equation}
where $a_{\theta,d}$ and $b_{\theta,d}$ are the Low and High multipliers for parameter $\theta$ under driver $d$, and $u_d$ is the sampled value of driver $d$.

The weather year adjustment accounts for interannual variability in weather-sensitive parameters (e.g., heating demand, hydro reservoir inflow) by drawing, for each scenario and each representative year from 2030 onward, one historical weather year $w \in \{2016, \dots, 2023\}$ independently and uniformly at random (Tab.~\ref{tab:weather_year_adjustment}). Each weather-sensitive parameter $\theta$ is then adjusted by the corresponding column entry $h_\theta(w)$, the historical realization of $\theta$. For hydro dam inflow and run-of-river generation, $h_\theta(w)$ is the weather year value relative to the mean of 2016–2023 \cite{fraunhofer_ise_storage_nodate, fraunhofer_ise_net_nodate}. Since space heating demand systematically shifts over time (e.g., warmer climate, renovated buildings, ...), we measure the weather year adjustment relative to the linear trend of the demand from 2014–2023 in CROSS \cite{sweet_cross_end-use_2025}. 
Because a weather year $w$ is drawn per representative year $y$, we write $w_y$ for the weather year realized in year $y$, and the adjusted parameter value becomes
\begin{equation}
    \theta^{\mathrm{adj}}_y = \theta_y \cdot h_\theta(w_y) \cdot \prod_{d \,\in\, D_\theta} g_\theta(u_d),
    \label{eq:adjustment_total_weather}
\end{equation}

where $\theta_y$ denotes the scaled parameter value in representative year $y$, and $D_\theta$ is the subset of drivers affecting parameter $\theta$.%

For example, the residential low-temperature heat demand in year $y$ is adjusted for both behavioral and climate effects as well as the sampled weather year $w_y$:

\begin{equation}
\theta_{\text{heat, res}, y}^{\mathrm{adj}} = \theta_{\text{heat, res}, y} \cdot h_{\text{heat, res}}(w_y) \cdot g_{\text{heat, res}}(u_{\text{behavior}}) \cdot g_{\text{heat, res}}(u_{\text{climate}})
\end{equation}

Because CROSS \cite{sweet_cross_import_2025} gives uncertainty ranges only for oil, gas, and renewable hydrogen, the remaining carriers follow the closest one when the import-price uncertainty is sampled: biofuel, coal, and uranium follow oil; hydrogen, ammonia, and methanol follow gas; and the renewable methane, ammonia, and methanol follow renewable hydrogen.

\begin{table}[h]
\caption{Factors of the weather-dependent model inputs per weather year, relative to their reference value.}
\label{tab:weather_year_adjustment}
\centering
\small
\input{tables/weather_scaling}

\end{table}

\begin{table}[h]
\caption{Multipliers of the model inputs at the low/high end of their macro driver: demands and other parameters.}
\label{tab:uncertainties_demand}
\centering
\small
\input{tables/drivers_demands}
\end{table}

\begin{table}[h]
\caption{Multipliers of the model inputs at the low/high end of their macro driver: resources. Import prices follow the CROSS price scenarios.}
\label{tab:uncertainties_resources}
\centering
\small
\input{tables/drivers_resources}
\end{table}

\begin{table}[h]
\caption{Multipliers of the model inputs at the low/high end of their macro driver: investment costs of technologies.}
\label{tab:uncertainties_tech}
\centering
\small
\input{tables/drivers_technologies}
\end{table}

\clearpage

\sinote{correlation}{Relaxation of correlation with macro drivers in sampling}

In the main paper, we assume that parameters are fully correlated with the macro drivers. For example, in a scenario where the macro driver \emph{technology} is high, the investment costs for all technologies will be high. More realistically, parameters should only be partially correlated and have an independent component of uncertainty.
To model a relaxed correlation between drivers and parameters, we apply a Gaussian copula transformation.

\subsubsection*{Mathematical formulation}

Let $u_{\theta} \sim \mathcal{U}(0,1)$ represent the independent uniform sample of a parameter $\theta$, and $u_{d} \sim \mathcal{U}(0,1)$ represent the Latin Hypercube Sample (LHS) of its corresponding macro driver $d$, and $\omega_d \in [0,1]$ the weight assigned to the macro driver's random variable in the blend, controlling the degree of correlation between $\theta$ and $d$. 

We transform the uniform realizations into standard normal space using the inverse cumulative distribution function of the standard normal distribution $\Phi^{-1}$:
\begin{equation}
    z_{\theta} = \Phi^{-1}(u_{\theta}), \quad z_{d} = \Phi^{-1}(u_{d}).
\end{equation}

Then, we calculate the blended standard normal quantile $z_{\text{blended}}$ via a variance-preserving linear combination:
\begin{equation}
    z_{\text{blended}} = \sqrt{\omega_d} \, z_{d} + \sqrt{1 - \omega_d} \, z_{\theta}.
\end{equation}

Because $z_d, z_{\theta}\sim \mathcal{N}(0,1)$ are independent, the linear combination preserves unit variance,
\begin{equation}
    \operatorname{Var}(z_{\text{blended}}) = (\sqrt{\omega_d})^2 + (\sqrt{1-\omega_d})^2 = 1,
\end{equation}

ensuring that $z_{\text{blended}}$ again follows a standard normal distribution $\mathcal{N}(0,1)$ with $\operatorname{Corr}(z_{\text{blended}}, z_d) = \sqrt{\omega_d}$.

Last, we map the blended Gaussian variable back to the uniform domain $\mathcal{U}(0,1)$ using the standard normal cumulative distribution function $\Phi$:
\begin{equation}
    u_{\text{blended}} = \Phi(z_{\text{blended}}) \sim \mathcal{U}(0,1)
\end{equation}

For $\omega_d=1$, as used in the main paper, we obtain $u_{\text{blended}}=u_{d}$, i.e., the parameter is fully correlated with the macro driver. Conversely, for $\omega_d=0$, we get $u_{\text{blended}}=u_{\theta}$, i.e., each parameter is sampled independently. For $0<\omega_d<1$, a parameter is partially influenced by its specific random variable $u_{\theta}$ and partially by the random variable of the macro driver $u_{d}$.

Note that the weight $\omega_d$ does not directly correspond to the Spearman correlation coefficient $\rho_s$ between $u_{\text{blended}}$ and $u_d$, but as shown by Moran \cite{moran_rank_1948} it can be calculated as:

\begin{equation}
    \rho_s = \frac{6}{\pi} \arcsin\left(\frac{\sqrt{\omega_d}}{2}\right).
    \label{eq:moran}
\end{equation}

\subsubsection*{Effect on parameter correlation}

\begin{figure}[h]
    \centering
    \includegraphics[width=0.75\linewidth]{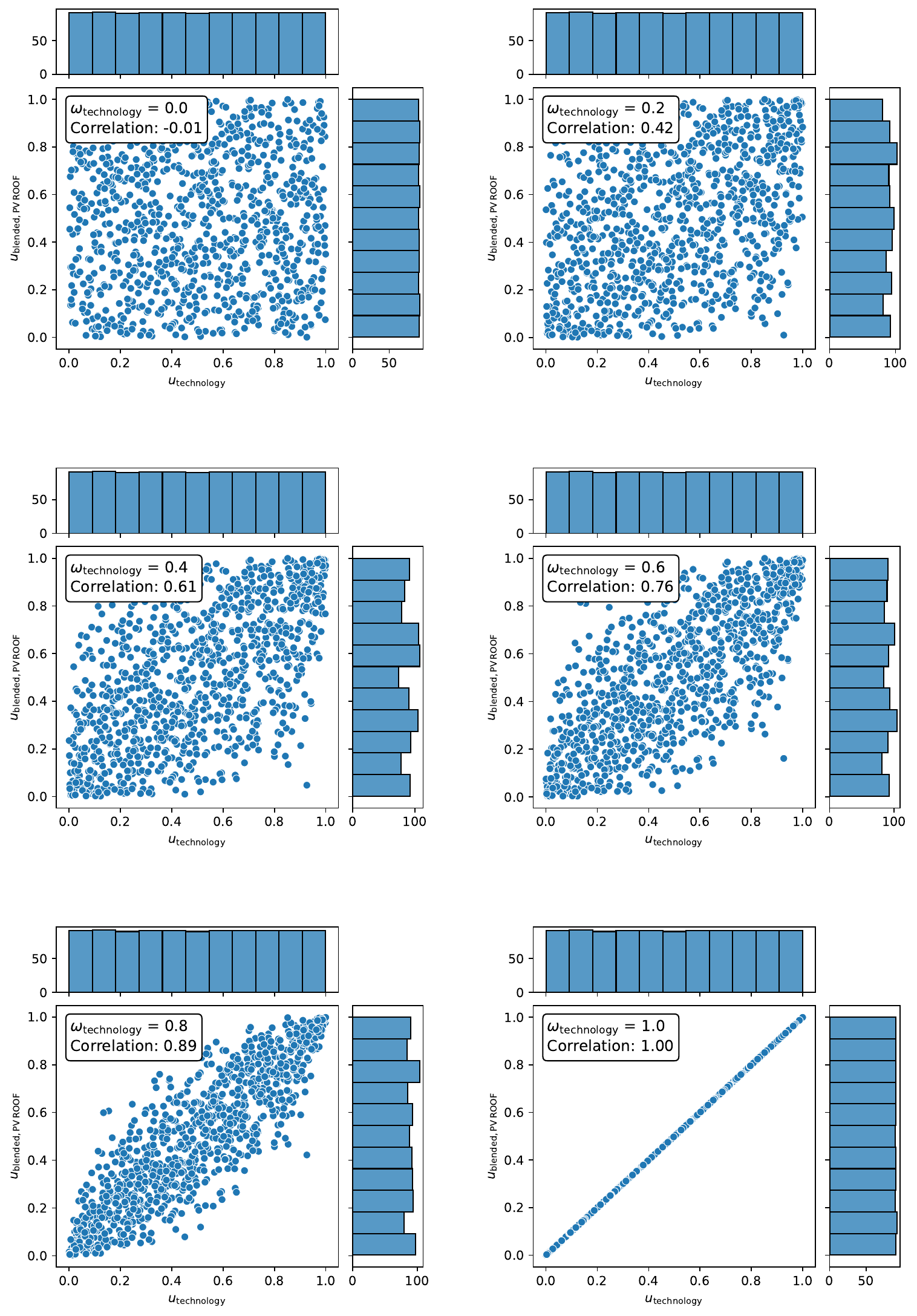}
    \caption{Effect of the blending weight $\omega_d$ on the joint distribution of the sampled parameter uncertainty $u_{\mathrm{blended}}$ for the investment cost of rooftop PV (PV\_ROOF) and the corresponding \emph{technology} driver uncertainty $u_{\mathrm{technology}}$ across $N=1000$ scenarios. Histograms (top, right) show the marginal distributions of $u_{\mathrm{blended}}$ and $u_{d}$. The scatter plots show the corresponding Spearman correlation $\rho_s$ increasing from $\approx 0$ (independent, $\omega_d=0$) to $1$ (fully correlated, $\omega_d=1$).}
    \label{fig:correlation}
\end{figure}

Fig.~\ref{fig:correlation} illustrates the effect of the blending weight $\omega_d$ on the joint distribution of the sampled parameter uncertainty $u_{\mathrm{blended}}$ and the driver uncertainty $u_{\mathrm{driver}}$, for the example of investment cost of rooftop PV (PV\_ROOF) under the technology macro driver, i.e., $u_{\mathrm{technology}}$. As $\omega_d$ increases from 0 to 1, the scatter of realizations tightens from an uncorrelated cloud ($\rho_s \approx 0$) to full alignment along the diagonal ($\rho_s = 1$), with the empirical Spearman correlations closely matching those predicted by Eq.~\eqref{eq:moran}.

Notably, the blending procedure preserves the marginal distribution of $u_{\mathrm{blended}}$ (histograms on the right-hand side in Fig.~\ref{fig:correlation} remain uniform for all $\omega_d$). However, for $0 < \omega_d < 1$, the resulting distribution of $u_{\mathrm{blended}}$ becomes visibly noisier. This effect occurs because the blend of two stratified Latin Hypercube columns is itself no longer stratified. 

\clearpage

\subsubsection*{Effect on energy transition pathways}

To test how the assumption of full correlation affects the pathways, we repeat the optimization of all 1000 scenarios with the driver weights $\omega_d = 0.5$ and $\omega_d = 0$ for all parameters. The macro drivers and weather years are the same as in the reference case ($\omega_d = 1$).

\begin{figure}[h]
\centering
\includegraphics[width=0.55\linewidth]{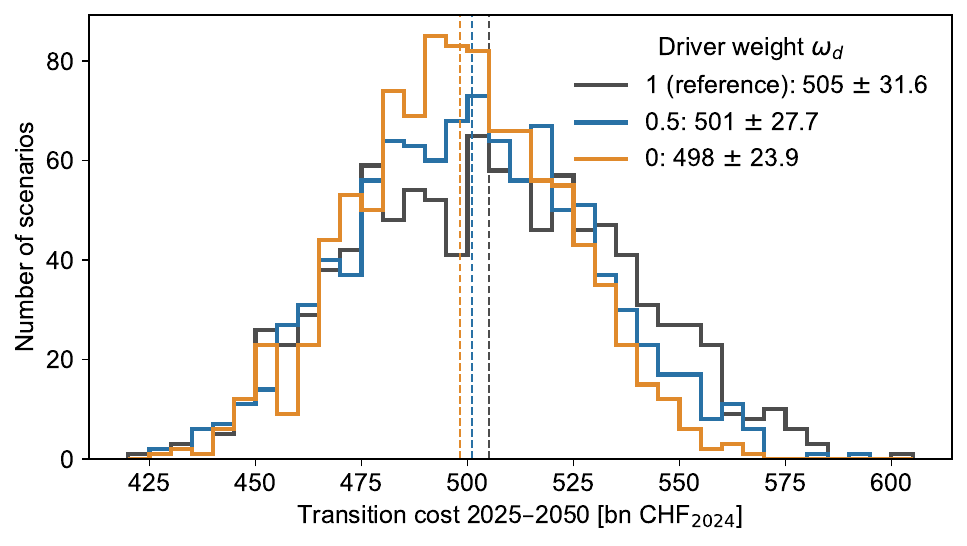}
\caption{Distribution of the transition cost 2025--2050 (without initialization cost) over the 1000 scenarios for the driver weights $\omega_d = 1$ (reference case), $0.5$ and $0$, in bins of 5\,billion CHF$_{2024}$. Dashed lines show the means. The legend gives mean $\pm$ standard deviation.}
\label{fig:weights_cost}
\end{figure}

\begin{figure}[h]
\centering
\includegraphics[width=0.85\linewidth]{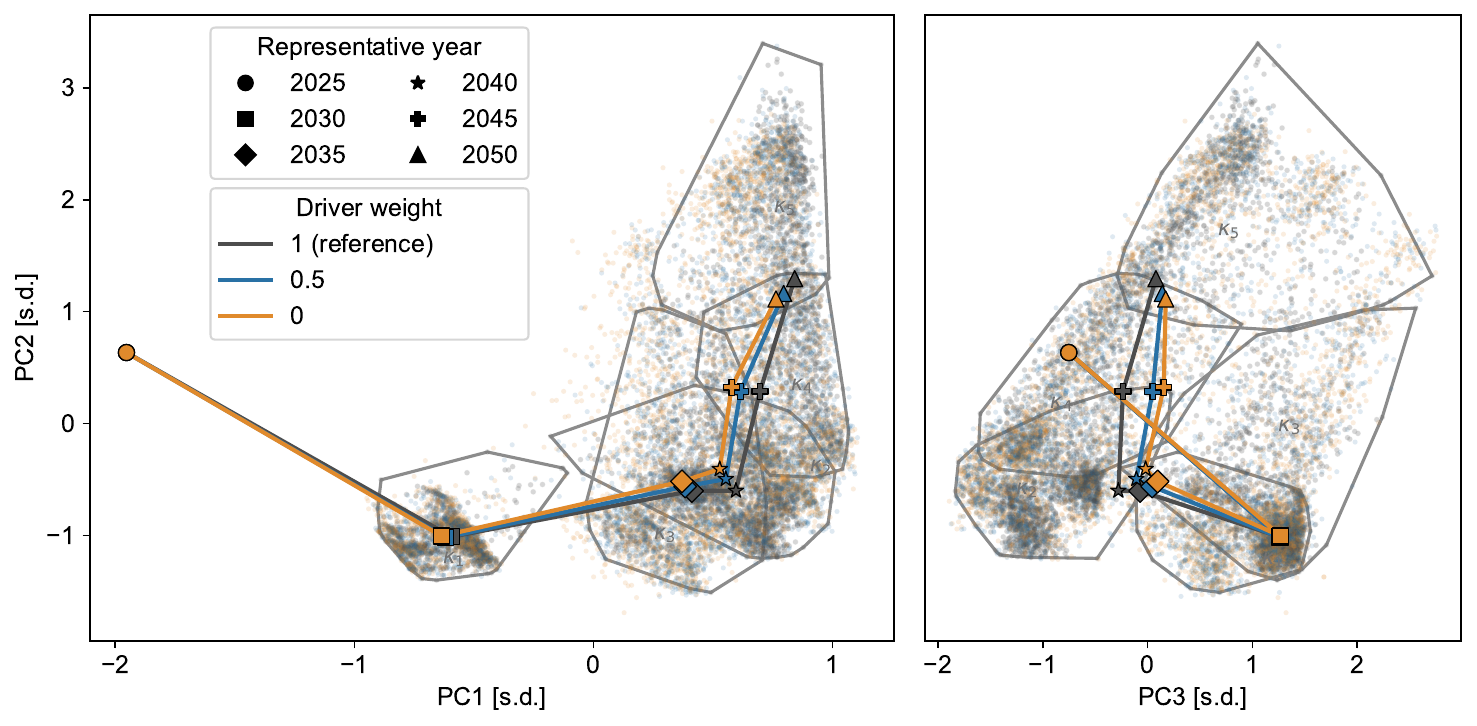}
\caption{Optimal pathways for the driver weights $\omega_d = 1$ (reference case), $0.5$ and $0$ in the PCA space of the reference case. Points show the individual scenario--year pairs, lines connect the mean per year, and grey outlines are the states of the reference case.}

\label{fig:weights_pca}
\end{figure}

The scenario--year pairs of the relaxed cases fall into similar regions of the PCA space as those of the reference case, and their mean trajectories only slightly shift (Fig.~\ref{fig:weights_pca}). However, the individual scenario--year pairs spread more widely as the correlation is relaxed, mainly in 2035 and 2040, since the parameters no longer deviate jointly with their macro drivers. As the correlation is relaxed, the mean transition cost (excluding initialization) decreases slightly, from 505 to 498\,billion CHF$_{2024}$, and its standard deviation from 31.6 to 23.9\,billion, as independent deviations of the parameters partly cancel out (Fig.~\ref{fig:weights_cost}). The narrowing happens almost entirely at the upper end of the distribution: the 5th percentile hardly changes (455 to 458\,billion), while the 95th percentile decreases from 557 to 538 and the maximum from 604 to 566\,billion. The most expensive scenarios of the reference case require all cost-increasing parameters to deviate in the same direction, which becomes unlikely when they are drawn independently.

The relaxed cases occupy the same regions of the PCA space as the states of the main case, so the states, and the strategies built from them, remain meaningful under relaxed correlation. How often each state and path occurs may change slightly. Evaluating the regret of the strategies under relaxed correlation would require strategy runs for these cases, which we leave to future work.

\clearpage

\sinote{clusters}{Clustering of energy system states}
\label{note:clustering}

\subsubsection*{Number of states}

\begin{figure}[h]
\centering
\includegraphics[width=0.55\linewidth]{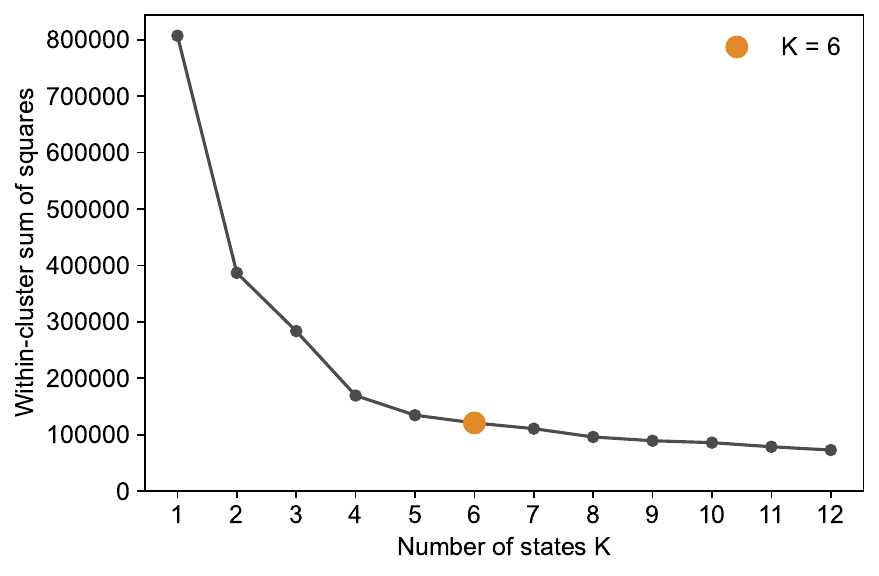}
\caption{Within-cluster sum of squares of the $k$-means clustering of the 6000 scenario--year pairs over the number of states $K$. The chosen $K = 6$ is highlighted.}
\label{fig:elbow}
\end{figure}

We cluster the 6000 scenario--year pairs with $k$-means for $K = 1, \dots, 12$ states, with the same features and random seed as in the main text. The within-cluster sum of squares drops steeply up to $K = 4$ and flattens beyond (Fig.~\ref{fig:elbow}): $K = 5$ and $K = 6$ reduce it by another 4 and 2\,\% of its value at $K = 1$, every further state by less than 2\,\%.

\begin{figure}[h]
\centering
\includegraphics[width=0.85\linewidth]{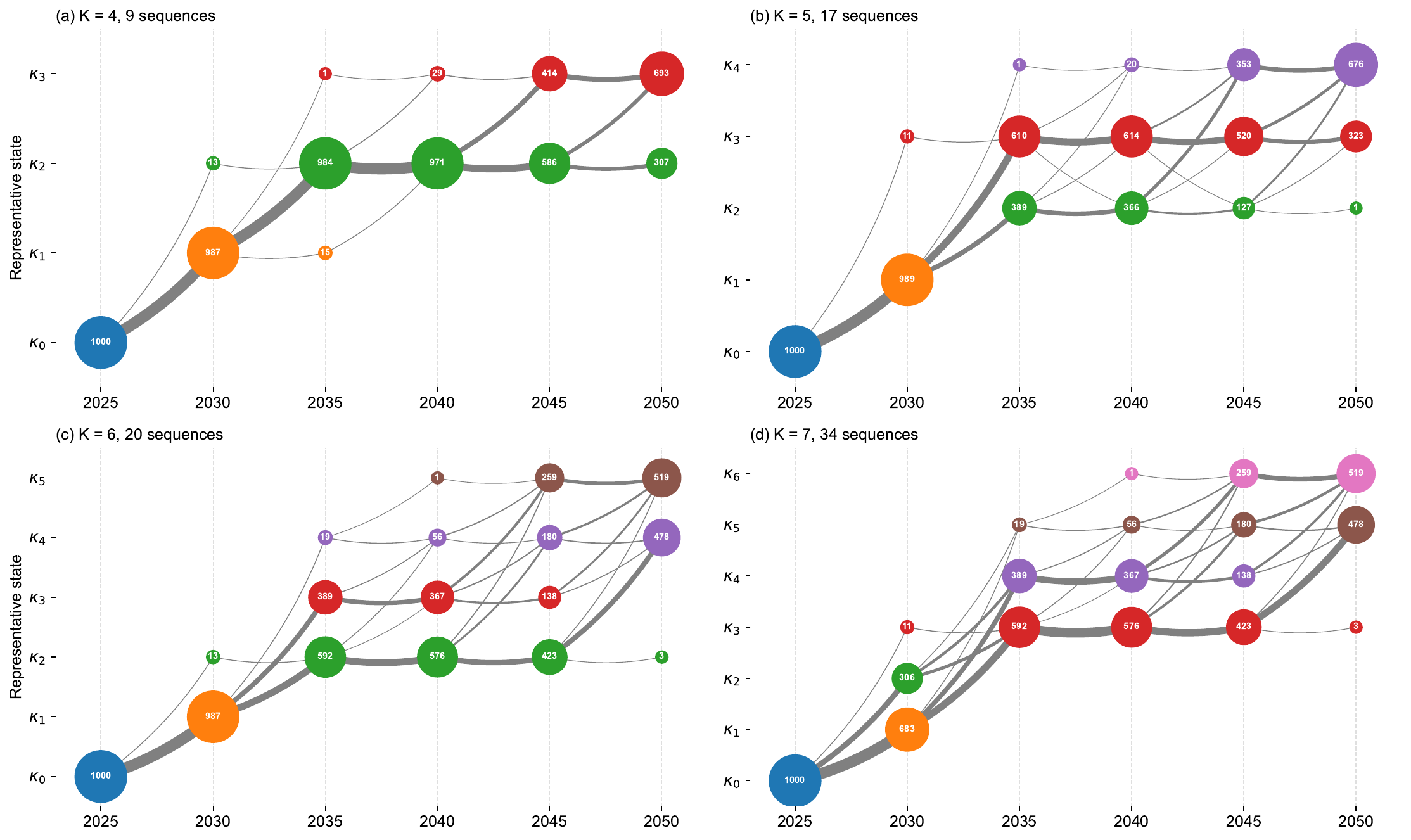}
\caption{Networks of states for $K = 4$ to 7, clustered with the same features and random seed as in the main text. Panel (c) corresponds to Fig.~2. Node sizes and numbers give the scenarios per state and year, line widths the number of transitions. State numbers differ between panels; the text refers to the states of $K = 6$.}
\label{fig:state_networks}
\end{figure}

Since the elbow alone does not single out one number, we compare the states for $K = 4, \dots, 7$ (Fig.~\ref{fig:state_networks}). As state numbers differ between panels, in the following we refer to the state numbers of $K = 6$ (panel c). With $K = 4$, the gas-based and the electrified intermediate states $\kappa_2$ and $\kappa_3$ are not separated, and with $K = 5$, most of the gas-based target states in $\kappa_4$ are combined with the intermediate states $\kappa_2$. Both would hide the decisions on decentralized heating and PV that distinguish the paths. With $K = 7$, the additional state splits the 2030 state $\kappa_1$. We choose $K = 6$, the smallest number of states that separates both intermediate and both target states.

\subsubsection*{Principal components}

\begin{figure}[h]
\centering
\includegraphics[width=0.55\linewidth]{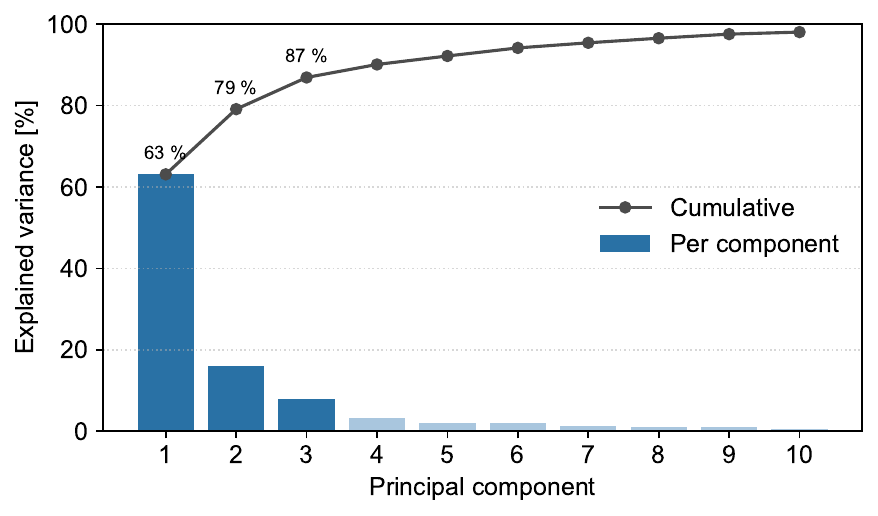}
\caption{Share of the variance of the 6000 scenario--year pairs explained by the first ten principal components (bars) and cumulative share (line). The first three components, shown in the main text, are highlighted.}
\label{fig:pca_variance}
\end{figure}

To visualize the energy system states, we project the 72-dimensional feature vectors of the 6000 scenario--years onto their principal components. The features are the same standardized installed capacities and production shares used for the clustering, excluding technologies with a total capacity or share below $10^{-4}$ (in the respective model unit) in all scenario--years. The first three principal components explain 63\,\%, 16\,\% and 8\,\% of the variance, together 87\,\% (Fig.~\ref{fig:pca_variance}); further components each add about 3\,\% or less. The two-dimensional projections in the main text (PC1--PC2, PC3--PC2) therefore capture most of the differences between the states. The sign of each component is arbitrary. We orient the components so that the states of 2050 have positive mean scores, i.e., positive values point towards the end of the transition. Tables \ref{tab:pca_loadings_all} and \ref{tab:pca_loadings_all_2} characterize the components by all technologies used as features, showing the change in capacity or share per standard deviation along each component.

\begin{table}[H]%
\caption{Characterization of the first three principal components of the energy system states by all technologies. Values give the change in installed capacity or production share for an increase of one standard deviation along each component; the share of explained variance is given in brackets. The components are oriented so that the states of 2050 have positive mean scores. pp: percentage points of the subsector output.}
\label{tab:pca_loadings_all}
\centering
\small
\input{tables/pca_loadings_all_1}
\end{table}

\begin{table}[p]
\caption{Continued from Tab.~\ref{tab:pca_loadings_all}.}
\label{tab:pca_loadings_all_2}
\centering
\small
\input{tables/pca_loadings_all_2}
\end{table}

\clearpage

\sinote{strategies}{Implementation of strategies and bandwidth $\alpha$}
\label{note:bandwidth}

\subsubsection*{From states to strategies}

A strategy prescribes a state in each of its years, and each prescribed state is enforced through bounds on the technologies around the mean of the state's members in that year (Eq.~(4) in the main text). The quantity that defines a state differs by sector: in the electricity, storage, and fuel \& feedstocks sectors, it is the installed capacity $F$, bounded to $(1 \pm \alpha)\overline{F}$; in buildings, industry, and transport, it is the share $\phi$ of a technology in the output of its subsector, bounded to $\overline{\phi} \pm \alpha$. In detail, the bounds are implemented as follows:
\begin{itemize}
\item Technologies with a mean capacity below $10^{-4}$ in their respective model unit (GW, GWh,
Mpkm/h or Mtkm/h) are treated as numerical noise and are fixed to zero.
\item All bounds are clipped to $[0, \infty)$ for capacities and to $[0, 1]$ for shares. Additionally, the original lower and upper bounds of the technology in the model are also enforced, so that a strategy never allows what the model does not.
\item For freight transport and industrial heat, the model defines shares relative to narrower totals than the overall subsector output: In freight, the rail share is relative to its maximum share and the road technologies relative to the freight not covered by rail. In process heat, industrial heat pumps are relative to medium-temperature heat demand, and the other technologies are relative to the heat output not covered by these heat pumps. The bounds are converted to these totals.
\item The year 2025 is never bounded, as it is fixed by the initialization (Note~\ref{note:init2025}).
\end{itemize}

Tables~\ref{tab:state_year_means} and~\ref{tab:state_year_means_2} list the mean capacities and shares for all state--year pairs prescribed by the three strategies analyzed in the main text. These year-specific means can differ considerably from the pooled state means in the main text (Tab.~2), as a state can span several years.

\begin{table}[h]
\caption{Means of the state--year pairs prescribed by the strategies \emph{Gas \& CO$_2$ removal}, \emph{Early electrification}, and \emph{Late switch}, for all technologies used as features. Installed capacity, or share of the subsector output, averaged over the members of a state in the respective year.}
\label{tab:state_year_means}
\centering
\small
\input{tables/state_year_means_1}
\end{table}

\begin{table}[h]
\caption{Continued from Tab.~\ref{tab:state_year_means}.}
\label{tab:state_year_means_2}
\centering
\small
\input{tables/state_year_means_2}
\end{table}

\clearpage
\subsubsection*{Sensitivity to the bandwidth $\alpha$}

\begin{figure}[h]
\centering
\includegraphics[width=\linewidth]{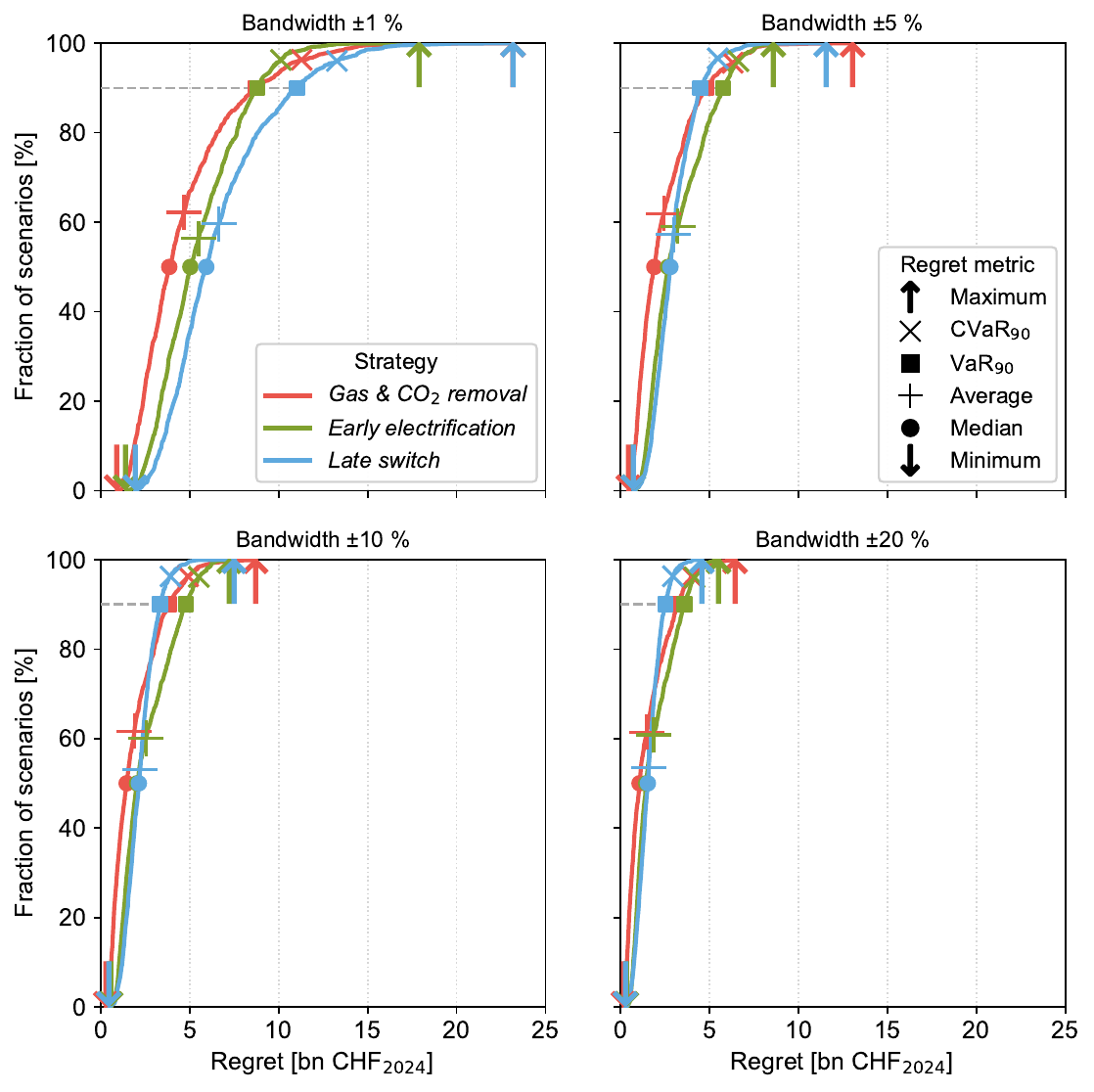}
\caption{Regret curves of the three full-pathway strategies \emph{Gas \& CO$_2$ removal}, \emph{Early electrification}, and \emph{Late switch} for the bandwidths $\alpha = 1, 5, 10$ and $20\,\%$, with the minimum, average, median, VaR$_{90}$, CVaR$_{90}$ and maximum regret marked as in Fig.~3.}

\label{fig:bandwidth_regret}
\end{figure}

We repeat the three full-pathway strategies \emph{Gas \& CO$_2$ removal}, \emph{Early electrification}, and \emph{Late switch} of the main text with $\alpha \in \{1, 5, 10, 20\,\%\}$. The regret decreases with the bandwidth (Fig.~\ref{fig:bandwidth_regret}, Tab.~\ref{tab:bandwidth}): from $\alpha = 1\,\%$ to $5\,\%$, the average regret roughly halves, while it decreases by roughly another 40\,\% from 5 to 20\,\%. %
The ranking of the strategies is robust for $\alpha \geq 5\,\%$: \emph{Gas \& CO$_2$ removal} has the lowest minimum, average and median regret, and \emph{Late switch} the lowest VaR$_{90}$ and CVaR$_{90}$. Only the maximum regret changes its ranking, from \emph{Early electrification} at $\alpha \leq 10\,\%$ to \emph{Late switch} at $\alpha = 20\,\%$. At $\alpha = 1\,\%$, \emph{Gas \& CO$_2$ removal} also has the lowest VaR$_{90}$ and \emph{Early electrification} the lowest CVaR$_{90}$. We choose $\alpha = 5\,\%$, the smallest bandwidth at which the ranking of the strategies no longer changes, except for the maximum regret. At $\alpha = 1\,\%$, much of the regret stems from the narrow corridor rather than from the decisions the strategies represent. Wider bandwidths make the strategies less specific, e.g., at $\alpha = 20\,\%$, the share of a technology may vary by 40 percentage points. While the absolute regret values depend on $\alpha$, the ranking of the strategies remains similar as $\alpha$ increases.

\begin{table}[h]
\caption{Regret metrics of the full-pathway strategies \emph{Gas \& CO$_2$ removal}, \emph{Early electrification}, and \emph{Late switch} for different bandwidths $\alpha$ in billion CHF$_{2024}$. The rank of each strategy within a metric and bandwidth is given in brackets.}
\label{tab:bandwidth}
\centering
\input{tables/bandwidth}
\end{table}

\clearpage
\sinote{gap}{Regret gap}

The decision maps (Fig.~5) color each pixel by the strategy with the lowest average regret and shade it by the regret gap to the next-best strategy. We measure this regret gap in standard errors of the average regret. The range of each macro driver is divided into $n_{\text{b}} = 5$ bins of equal width. A pixel of the decision map of two drivers $d$ and $d'$ contains the scenarios $\mathcal{I}_{b,b'} = \{\, i \in \mathcal{I} \mid u_{i,d} \in B_{b},\ u_{i,d'} \in B_{b'} \,\}$ whose driver values $u_{i,d}$ and $u_{i,d'}$ fall into the bins $B_b$ and $B_{b'}$. All quantities below refer to the scenarios within a pixel. For brevity, we omit the pixel index $(b,b')$. For each strategy $\pi$, we compute the average regret $\overline{R}_{\pi}$ over these scenarios and its standard error $\mathrm{SE}_{\pi} = \mathrm{SD}_{\pi}/\sqrt{|\mathcal{I}_{b,b'}|}$, with $\mathrm{SD}_{\pi}$ the sample standard deviation of the regret. The regret gap of the lowest-regret strategy $\pi^{*} = \arg\min_{\pi} \overline{R}_{\pi}$ over its closest competitor, in units of the combined standard error, is
\begin{equation}
z = \min_{\pi \neq \pi^{*}} \frac{\overline{R}_{\pi} - \overline{R}_{\pi^{*}}}{\sqrt{\mathrm{SE}_{\pi}^{2} + \mathrm{SE}_{\pi^{*}}^{2}}}.
\label{eq:regret_gap}
\end{equation}
The color of a pixel indicates the lowest-regret strategy $\pi^{*}$. The pixel intensity shows the regret gap $z$. Each strategy has its own color scale, from white ($z = 0$, no regret gap) to its full color ($z \geq 3$, i.e., a regret gap of at least three standard errors).

\clearpage
\sinote{partial}{Partial strategies}

Section~3.6 of the main text reports the average regret of each partial strategy (Tab.~5). Here, we break it down by technology, resource, and phase (Figs.~\ref{fig:alternatives_breakdown}--\ref{fig:alternatives_breakdown_3}), show under which conditions each strategy is preferable (Figs.~\ref{fig:dm_path}--\ref{fig:dm_fixed_3}), and estimate the value of keeping the target state open.

For 10 of the 1000 scenarios, the run that fixes only the target state $\kappa_4$ did not converge: 9 runs did not solve to optimality, and in one run the regret exceeded that of a more constrained strategy. For these scenarios, we bound the regret from below by zero and from above by the smallest regret of the strategies that fix $\kappa_4$ together with a path, which are more constrained. The average regret of this strategy is therefore given as a range (0.75--0.76\,billion CHF$_{2024}$).

\paragraph*{Which path up to 2035?} 

The regret of the path up to 2035 mainly reflects which renewables are expanded: open-field PV and wind on the path $\kappa_0$-$\kappa_1$-$\kappa_2$, alpine PV on the path $\kappa_0$-$\kappa_1$-$\kappa_3$ (Figs.~\ref{fig:alternatives_breakdown_2} and \ref{fig:alternatives_breakdown_3}). The choice between the two paths depends almost only on technology costs (Fig.~\ref{fig:dm_path}). The path via $\kappa_3$ has the lower regret in scenarios with low technology costs, whereas the path via $\kappa_2$ is favored at high technology costs.

\paragraph*{Which target state in 2050?}

If the path to the target state is left unconstrained, $\kappa_4$ and $\kappa_5$ yield similar average regret (0.76 and 0.84\,billion CHF$_{2024}$). Their regret stems from the decentralized heating technologies, i.e., investments in gas boilers ($\kappa_4$) or heat pumps ($\kappa_5$), and the resulting demand for natural gas imports and CO$_2$ export (Fig.~\ref{fig:alternatives_breakdown}). Technology costs again determine the preferable target: $\kappa_5$ at low and $\kappa_4$ at high technology costs (Fig.~\ref{fig:dm_target}). Beyond technology costs, high import prices favor $\kappa_5$, while high eco-friendly behavior favors $\kappa_4$. The other drivers are less influential.

\paragraph*{Which target state in 2050 after a given path up to 2035?} 

After $\kappa_0$-$\kappa_1$-$\kappa_2$, fixing $\kappa_4$ or $\kappa_5$ in 2050 causes an average regret of 1.50 and 1.67\,billion CHF$_{2024}$, and each target is preferable in about half of the scenarios. Choosing the better target per scenario reduces the average regret to 1.12\,billion, i.e., by about 0.4\,billion compared with the better fixed target. The decision on the target state thus remains open after this path. After $\kappa_0$-$\kappa_1$-$\kappa_3$, state $\kappa_5$ (1.99\,billion) is preferable to $\kappa_4$ (2.96\,billion) in 87\,\% of the scenarios, and choosing per scenario reduces the average regret by only 0.05\,billion. Thus, keeping the target state unconstrained on this path adds less benefit. Since the target is chosen with all scenario parameters known, these reductions are upper bounds on the value of keeping the target state open.
Following $\kappa_0$-$\kappa_1$-$\kappa_3$, large alpine PV capacities are built by 2035, exceeding those used under $\kappa_4$. Reaching $\kappa_4$ thus requires decommissioning part of this capacity before the end of its lifetime, which reduces its salvage value. Since the salvage value is accounted for in the phase of investment, this as a higher regret of alpine PV in 2030--2035 (Fig.~\ref{fig:alternatives_breakdown_3}).

The decision maps of the targets after a given path (Figs.~\ref{fig:dm_fixed_2} and \ref{fig:dm_fixed_3}) can be read as signposts for an adaptive transition, as in dynamic adaptive policy pathways \cite{haasnoot_dynamic_2013}. After $\kappa_0$-$\kappa_1$-$\kappa_2$, the transition should move towards $\kappa_5$ if technology costs turn out low, import costs high, or eco-friendly behavior weak, and continue towards $\kappa_4$ otherwise (Fig.~\ref{fig:dm_fixed_2}). After $\kappa_0$-$\kappa_1$-$\kappa_3$, state $\kappa_4$ has a lower regret only where very high technology costs coincide with very low import costs or very strong eco-friendly behavior (Fig.~\ref{fig:dm_fixed_3}).

\begin{figure}[p]
\centering
\includegraphics[width=0.9\linewidth]{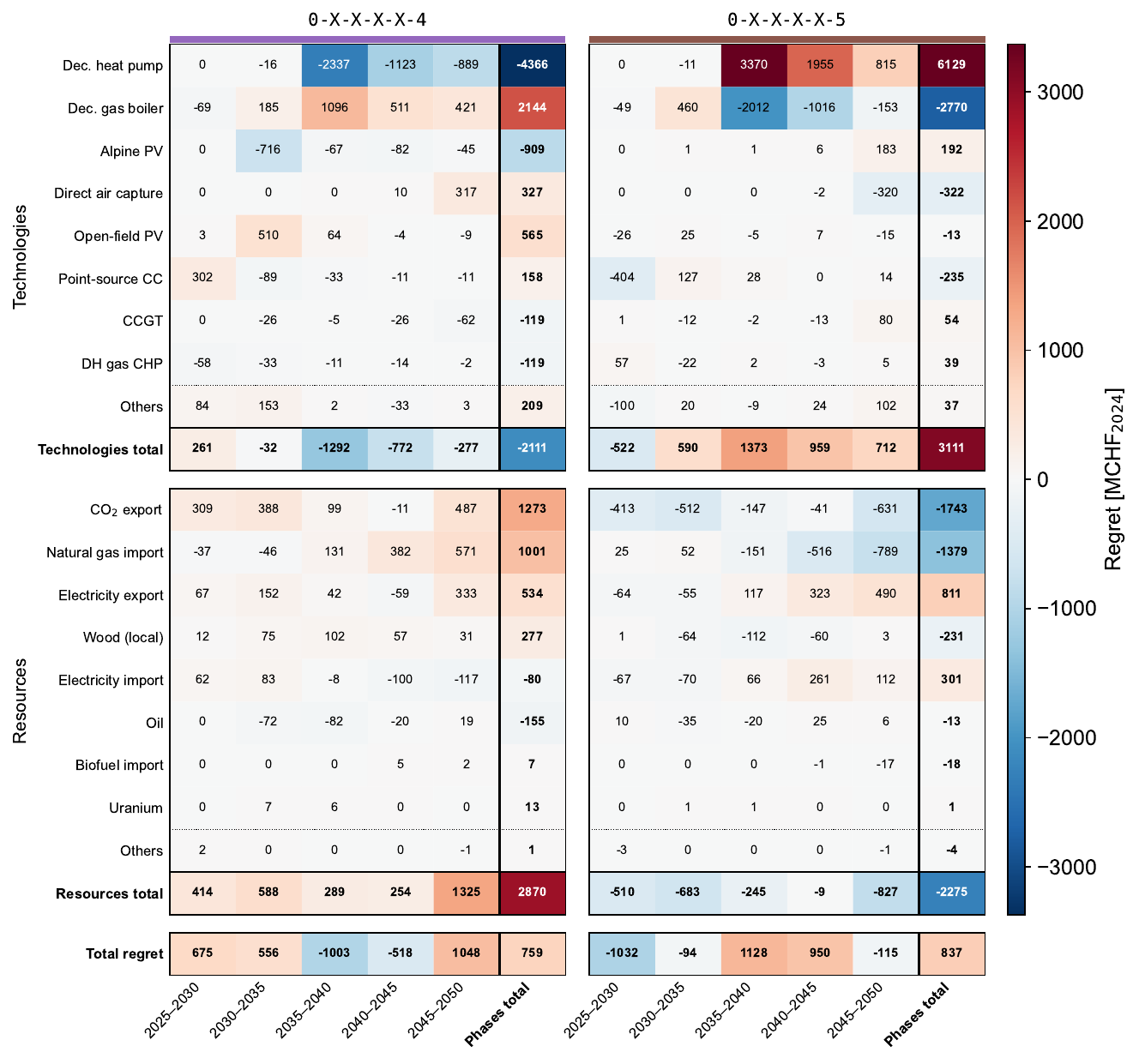}
\caption{Average regret of the partial strategies broken down by technology and resource and by phase, in million CHF$_{2024}$: the target states with an open path. Continued in Figs.~\ref{fig:alternatives_breakdown_2} and
\ref{fig:alternatives_breakdown_3}.}
\label{fig:alternatives_breakdown}
\end{figure}

\begin{figure}[p]
\centering
\includegraphics[width=0.9\linewidth]{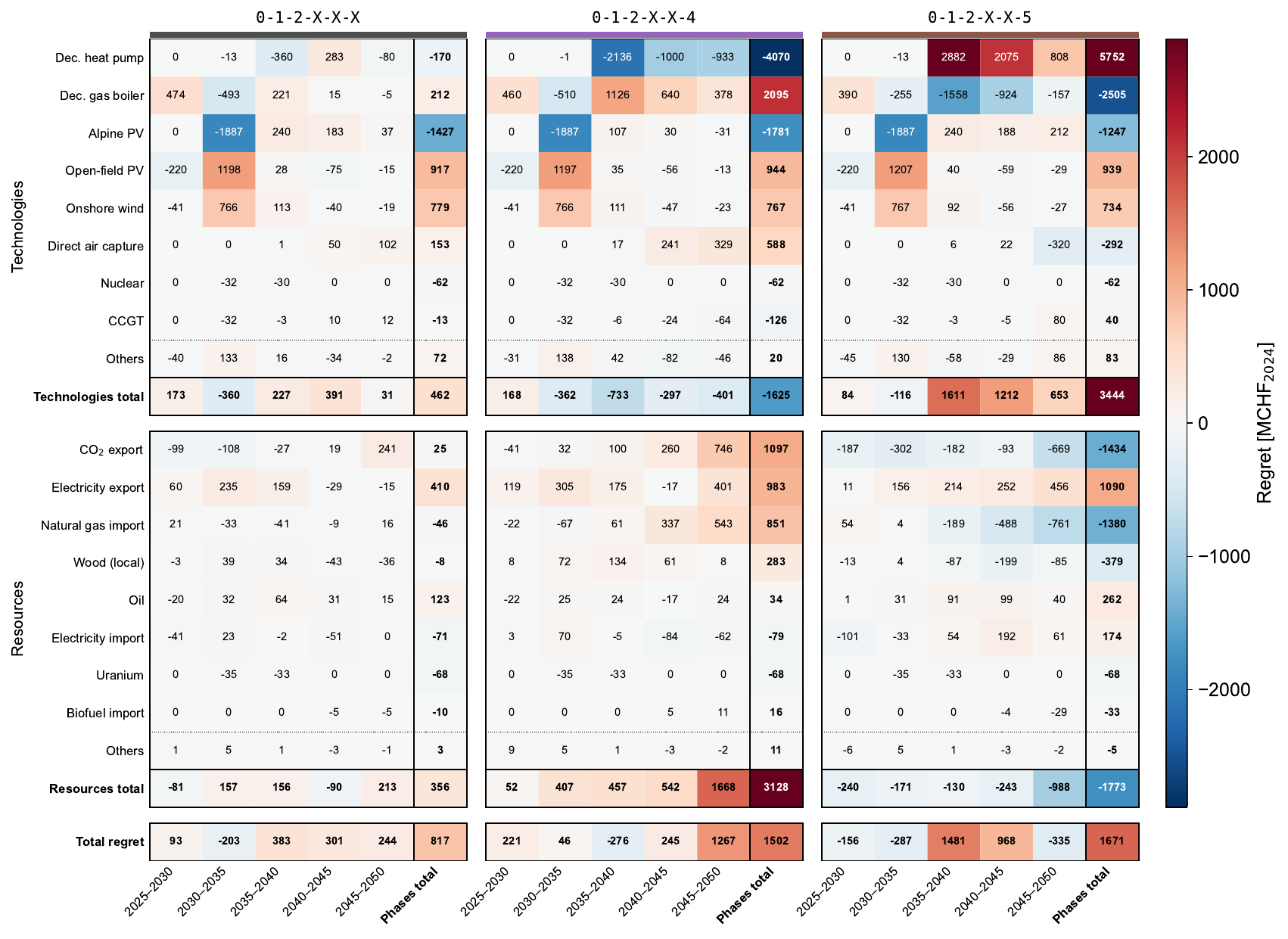}
\caption{Continued from Fig.~\ref{fig:alternatives_breakdown}: the path $\kappa_0$-$\kappa_1$-$\kappa_2$ with an open or fixed target state.}
\label{fig:alternatives_breakdown_2}
\end{figure}

\begin{figure}[p]
\centering
\includegraphics[width=0.9\linewidth]{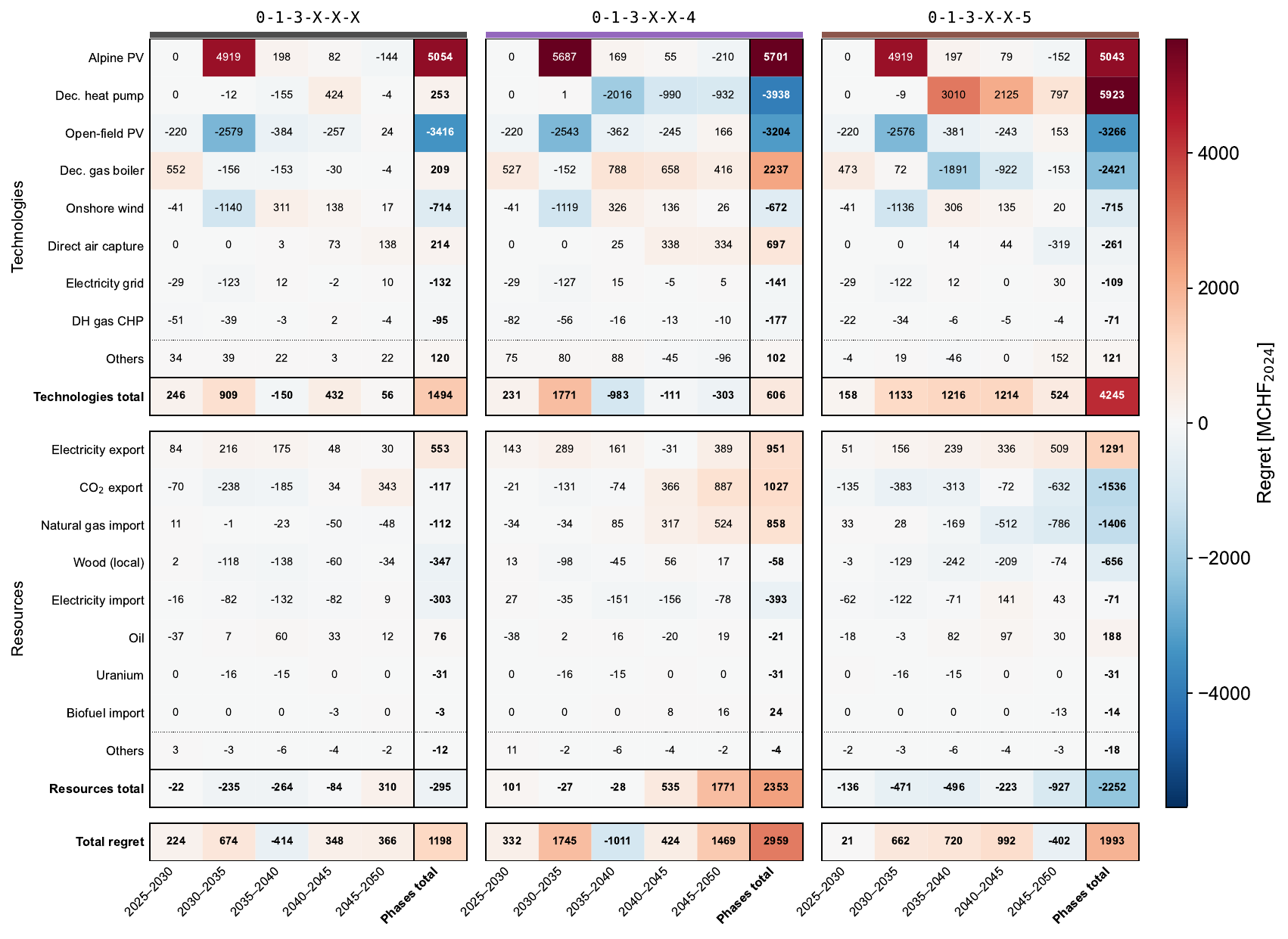}
\caption{Continued from Fig.~\ref{fig:alternatives_breakdown}: the path $\kappa_0$-$\kappa_1$-$\kappa_3$ with an open or fixed target state.}
\label{fig:alternatives_breakdown_3}
\end{figure}

\begin{figure}[p]
\centering
\includegraphics[width=0.9\linewidth]{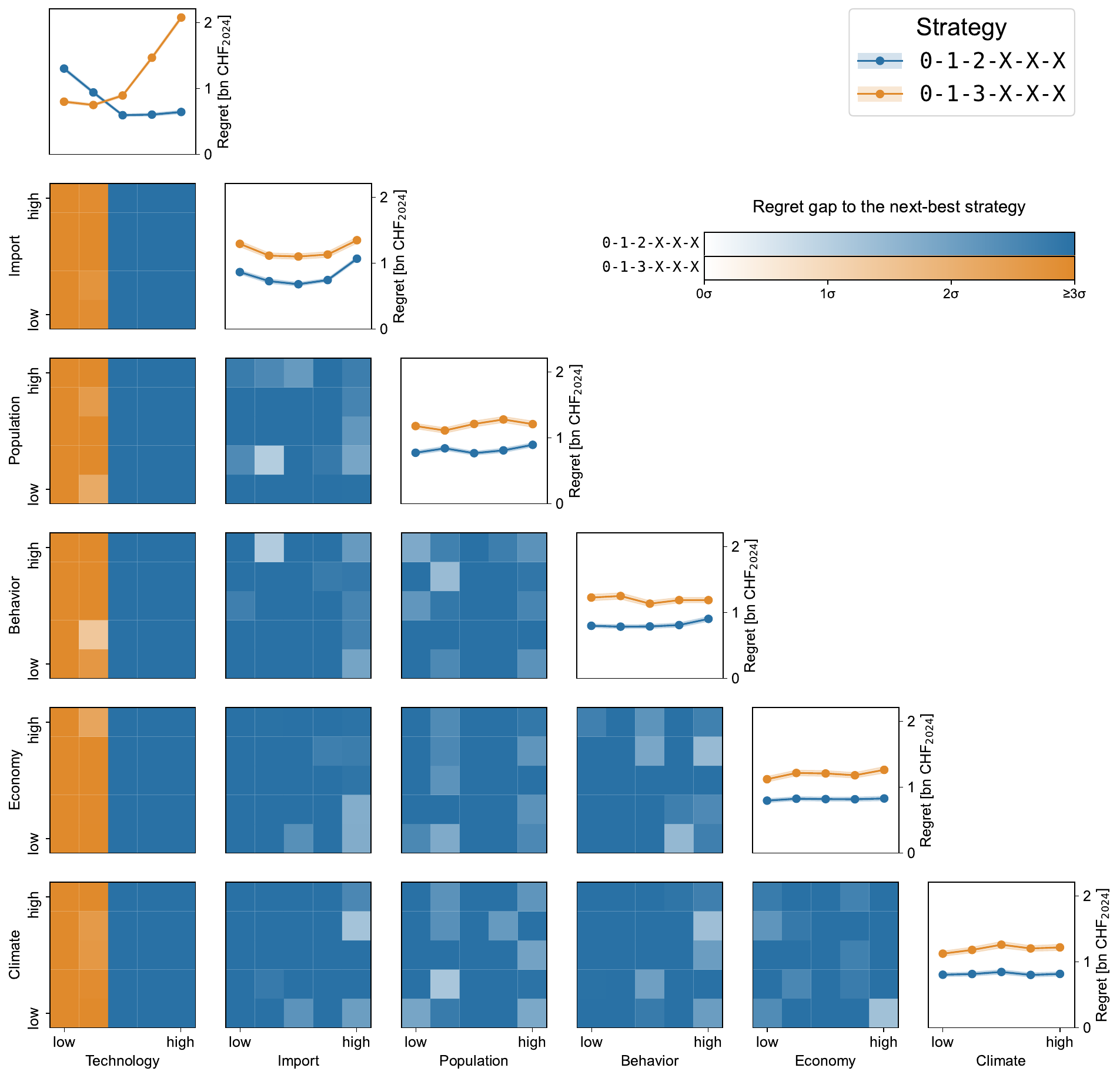}
\caption{Decision map of the two paths up to 2035 with an open target state. Off-diagonal panels show which strategy has the lowest average regret in each bin of two macro drivers, shaded by the gap to the next-best strategy; diagonal panels show the average regret over the bins of one driver.}
\label{fig:dm_path}
\end{figure}

\begin{figure}[p]
\centering
\includegraphics[width=0.9\linewidth]{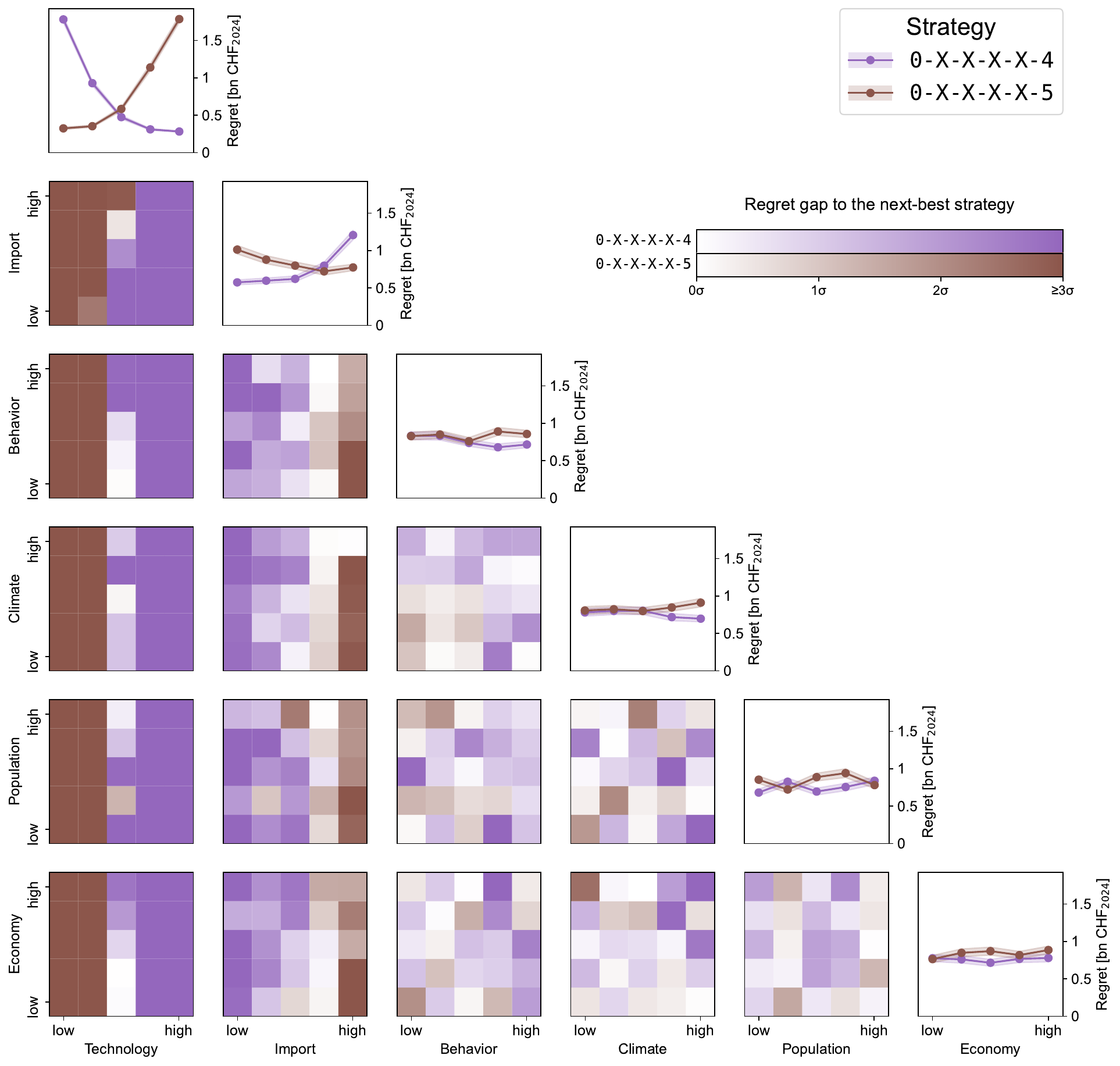}
\caption{Decision map of the two target states with an open path, as in Fig.~\ref{fig:dm_path}.}
\label{fig:dm_target}
\end{figure}

\begin{figure}[p]
\centering
\includegraphics[width=0.9\linewidth]{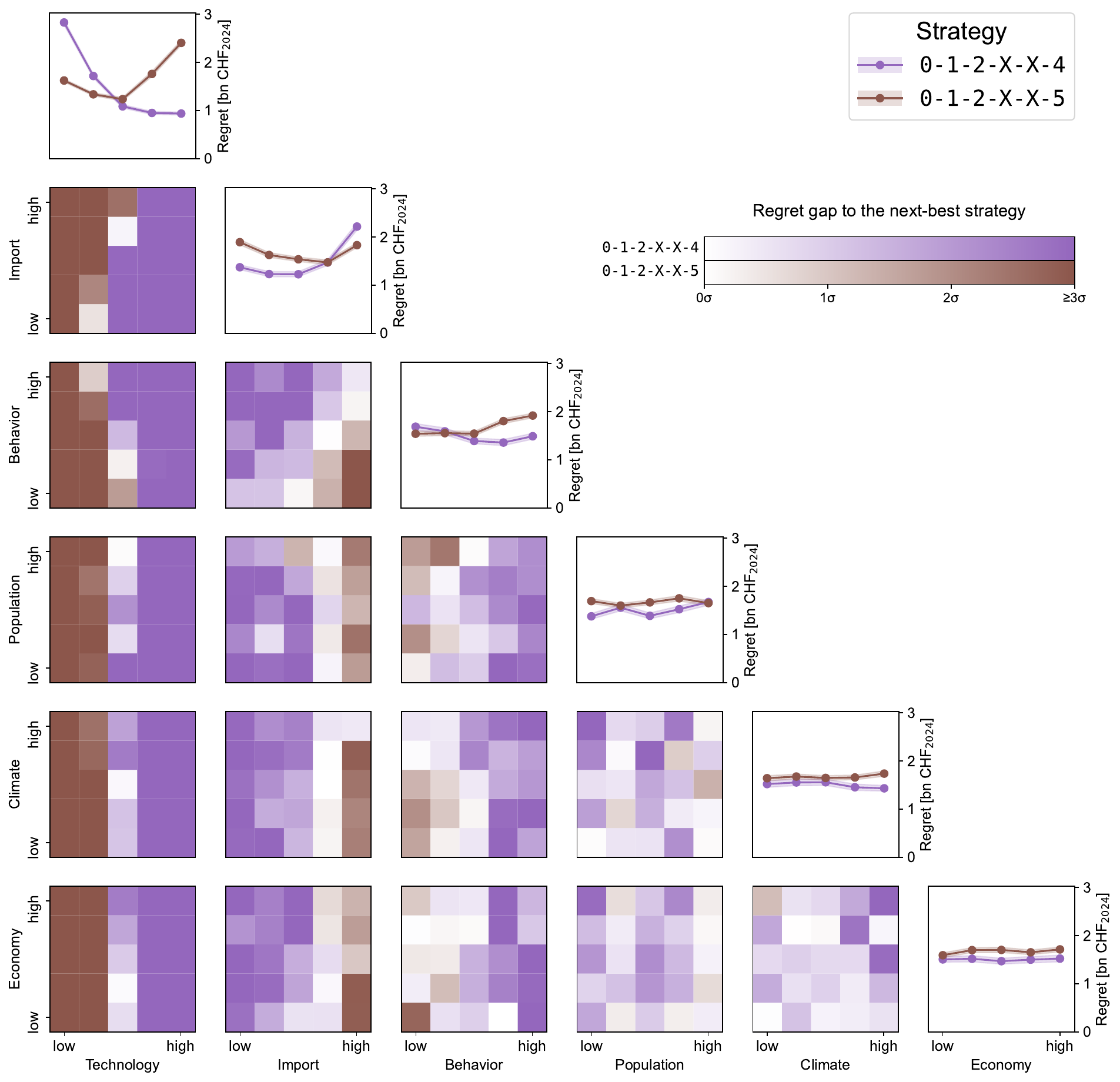}
\caption{Decision map of the two target states after the path $\kappa_0$-$\kappa_1$-$\kappa_2$, as in
Fig.~\ref{fig:dm_path}.}
\label{fig:dm_fixed_2}
\end{figure}

\begin{figure}[p]
\centering
\includegraphics[width=0.9\linewidth]{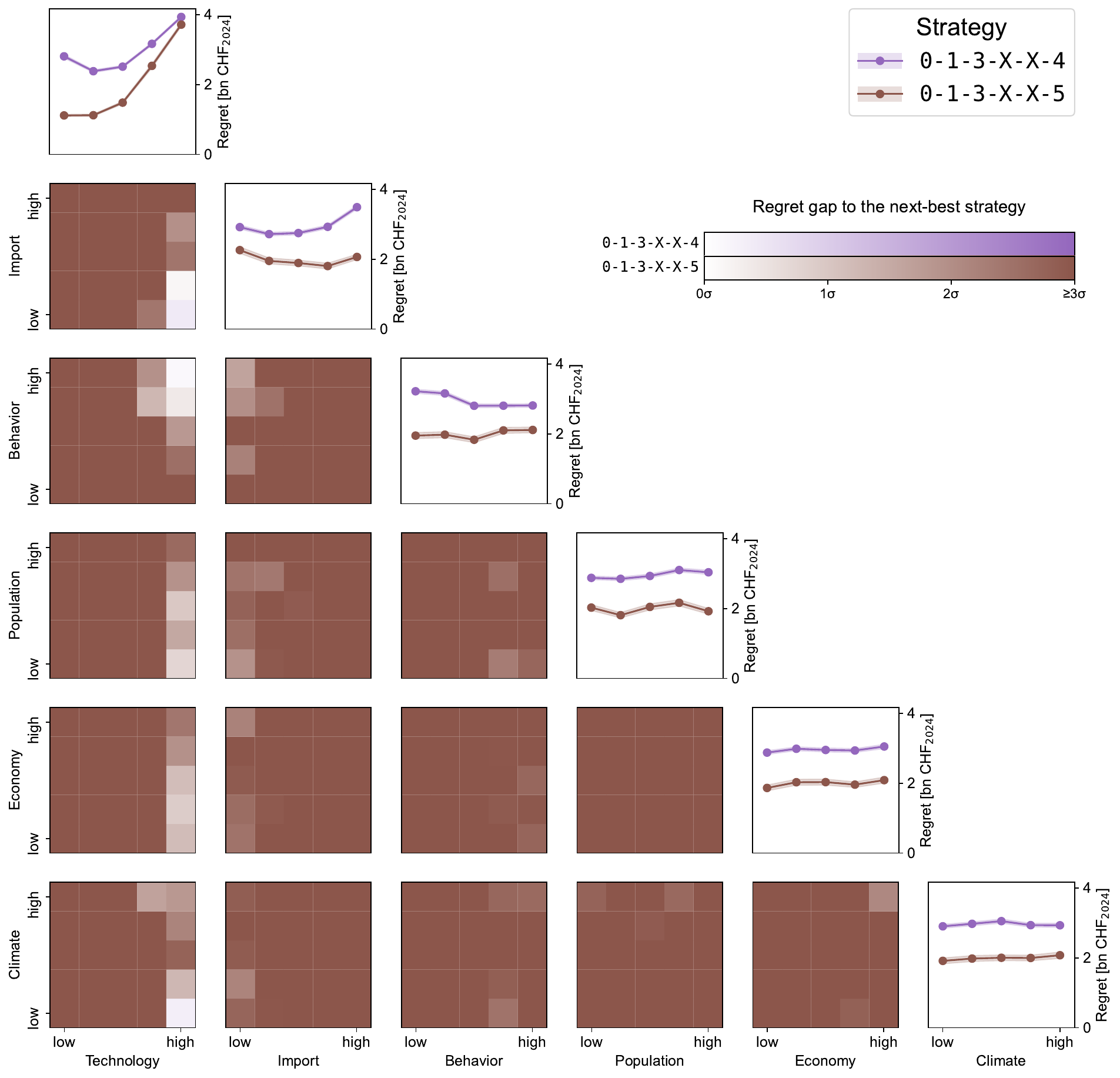}
\caption{Decision map of the two target states after the path $\kappa_0$-$\kappa_1$-$\kappa_3$, as in
Fig.~\ref{fig:dm_path}.}
\label{fig:dm_fixed_3}
\end{figure}

\clearpage

\sinote{sensitivity}{Sensitivity of decentralized heating around the central scenario}

Whether decentralized heat is supplied by gas boilers or heat pumps distinguishes the gas-based states $\kappa_2$ and $\kappa_4$ from the electrified states $\kappa_3$ and $\kappa_5$. To test the robustness of the gas-based solution, we vary one input at a time around the central scenario, in which all macro drivers are at their median ($u_d = 0.5$), with the weather years 2023, 2023, 2022, 2023, 2020 and 2022 for 2025--2050. In the central scenario, gas boilers supply 67\,\% of the decentralized heat in 2035, 86\,\% in 2040 and 33\,\% in 2050, at a total transition cost of 500\,billion CHF$_{2024}$. We vary four inputs, each in all years:
\begin{itemize}
\item the cost of CO$_2$ export, i.e., transport and storage abroad, from 100 (central) to 400\,CHF$_{2024}$/t;
\item an additional fixed cost for decentralized gas boilers of 10 to 80\,CHF$_{2024}$ per kW per year, to approximate the cost of the gas distribution grid, which the model does not represent. At 2600 full-load hours and 90\,\% energy efficiency, 10\,CHF$_{2024}$/kW/year corresponds to about 3.4\,CHF$_{2024}$/MWh of gas;
\item the investment cost of decentralized heat pumps, scaled by 90\,\% to 50\,\%;
\item a cap on the CO$_2$ export of 15 to 3\,Mt/year in all years.
\end{itemize}

\begin{figure}[h]
\centering
\includegraphics[width=0.8\linewidth]{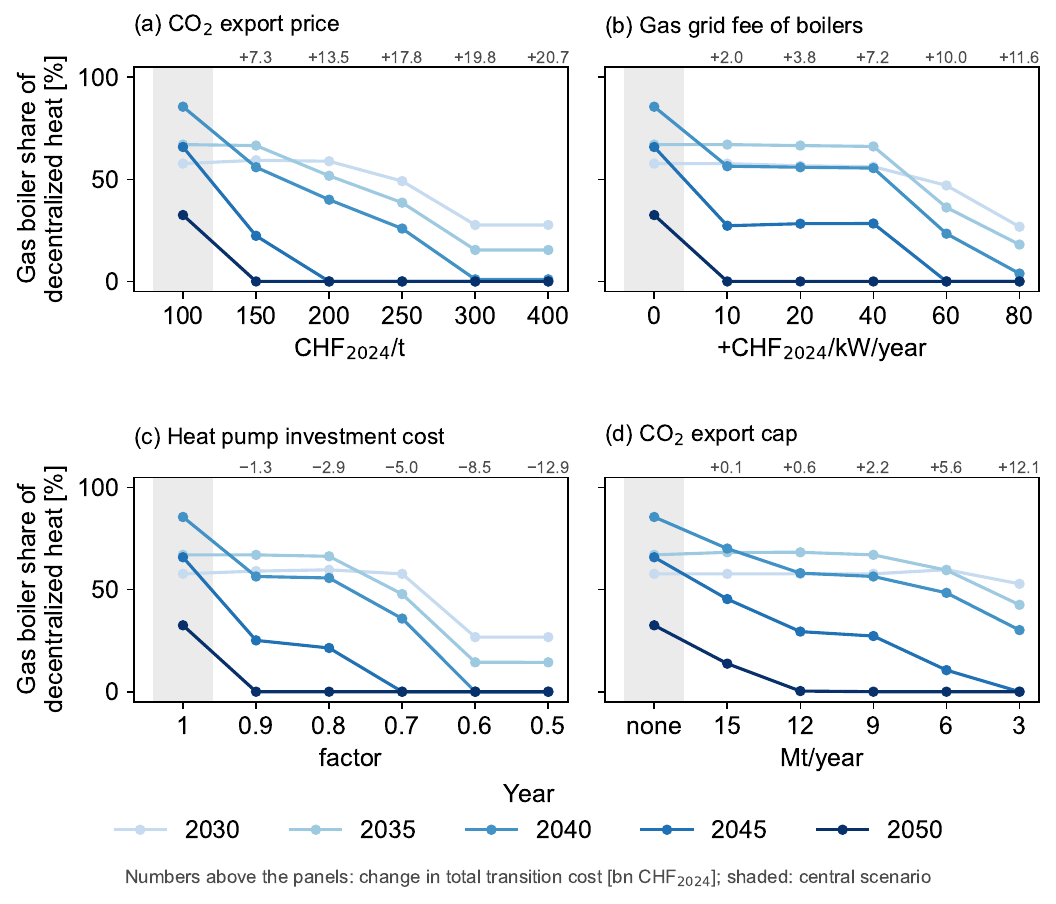}
\caption{Share of gas boilers in decentralized heat in 2030--2050 for one-at-a-time variations around the central scenario (shaded). Numbers above the panels give the change in total transition cost in billion CHF$_{2024}$.}
\label{fig:sensitivity}
\end{figure}

Gas boilers already disappear from decentralized heating in 2050 if CO$_2$ export costs are 150\,CHF$_{2024}$/t, if gas boilers pay an additional 10\,CHF$_{2024}$/kW/year, or if heat pumps are 10\,\% cheaper. Limiting the export of CO$_2$ to 12\,Mt/year or lower also removes gas boilers from the 2050 system (Fig.~\ref{fig:sensitivity}). Adapting the parameters also influences the shares of gas boilers earlier in the transition. For example, increasing the CO$_2$ export price to 200\,CHF$_{2024}$/t removes gas boilers from decentralized heating from 2045 on, and 300\,CHF$_{2024}$/t removes them already in 2040.
Combining several changes, e.g., gas grid fees together with a CO$_2$ export cap, could reduce gas boilers in decentralized heating even further. Thus, the gas-based solution in decentralized heating is sensitive to the input assumptions. Already, small changes in parameters can shift the system toward heat pumps.

\clearpage

\printbibliography[heading=bibintoc,title={Supplemental References}]

%% file: tables/state_means.tex
\begin{tabular}{l r r r r r r l}
\toprule
 & $\kappa_{0}$ & $\kappa_{1}$ & $\kappa_{2}$ & $\kappa_{3}$ & $\kappa_{4}$ & $\kappa_{5}$ & Unit \\
\midrule
Years & 2025 & 2030 & 2030--50 & 2035--45 & 2035--50 & 2040--50 & \\
\#\,occurrences & 1000 & 987 & 1607 & 894 & 733 & 779 & \\
\addlinespace
\multicolumn{8}{l}{\textbf{Electricity}} \\
Rooftop PV & 8.2 & 6.5 & 3.5 & 3.7 & 0.78 & 0.63 & GW \\
Open-field PV & 0 & 0.28 & 20.0 & 10.3 & 24.3 & 23.9 & GW \\
Alpine PV & 0 & 0 & 0.19 & 9.2 & 5.6 & 8.7 & GW \\
Onshore wind & 0.10 & 0.30 & 4.0 & 2.5 & 4.1 & 4.1 & GW \\
Nuclear & 3.0 & 2.6 & 0.28 & 0.35 & 0.02 & 0 & GW \\
CCGT & 0.33 & 0.26 & 0.12 & 0.23 & 0.50 & 1.4 & GW \\
\addlinespace
\multicolumn{8}{l}{\textbf{Building}} \\
Dec. heat pump & 24 & 16 & 5.9 & 22 & 27 & 83 & \% \\
Dec. gas boiler & 28 & 56 & 86 & 70 & 72 & 17 & \% \\
Dec. oil boiler & 37 & 18 & 6.2 & 5.7 & 0.3 & 0 & \% \\
DH heat pump & 0 & 39 & 69 & 73 & 75 & 76 & \% \\
DH waste CHP & 36 & 21 & 17 & 16 & 16 & 16 & \% \\
\addlinespace
\multicolumn{8}{l}{\textbf{Transport}} \\
Battery-electric car & 4.2 & 53 & 100 & 100 & 100 & 100 & \% \\
Battery-electric truck & 1.4 & 32 & 60 & 60 & 60 & 60 & \% \\
\addlinespace
\multicolumn{8}{l}{\textbf{Industry}} \\
Ind. waste boiler & 24 & 58 & 58 & 61 & 59 & 60 & \% \\
Ind. gas boiler & 50 & 34 & 25 & 26 & 25 & 26 & \% \\
High-T heat pump & 0 & 0.9 & 17 & 13 & 15 & 14 & \% \\
\addlinespace
\multicolumn{8}{l}{\textbf{Fuels \& feedstocks}} \\
Pyrolysis to oil & 0 & 0 & 0.38 & 0.28 & 0.58 & 0.61 & GW \\
Point-source CC & 0 & 0.44 & 1.7 & 1.6 & 1.8 & 1.9 & kt\,CO$_2$/h \\
Direct air capture & 0 & 0 & 0.01 & 0 & 1.1 & 0.31 & kt\,CO$_2$/h \\
\bottomrule
\end{tabular}

%% file: tables/sequences.tex
\begin{tabular}{l c c c c c c}
\toprule
Share [\%] & 2025 & 2030 & 2035 & 2040 & 2045 & 2050 \\
\midrule
39 & $\kappa_{0}$ & $\kappa_{1}$ & $\kappa_{2}$ & $\kappa_{2}$ & $\kappa_{2}$ & $\kappa_{4}$ \\
17 & $\kappa_{0}$ & $\kappa_{1}$ & $\kappa_{3}$ & $\kappa_{3}$ & $\kappa_{5}$ & $\kappa_{5}$ \\
11 & $\kappa_{0}$ & $\kappa_{1}$ & $\kappa_{3}$ & $\kappa_{3}$ & $\kappa_{3}$ & $\kappa_{5}$ \\
7 & $\kappa_{0}$ & $\kappa_{1}$ & $\kappa_{2}$ & $\kappa_{2}$ & $\kappa_{4}$ & $\kappa_{5}$ \\
5 & $\kappa_{0}$ & $\kappa_{1}$ & $\kappa_{3}$ & $\kappa_{3}$ & $\kappa_{4}$ & $\kappa_{5}$ \\
\bottomrule
\end{tabular}

%% file: tables/regret_metrics.tex
\begin{tabular}{l r r r r r r}
\toprule
Strategy & Minimum & Median & Average & VaR$_{90}$ & CVaR$_{90}$ & Maximum \\
\midrule
\emph{Gas \& CO$_2$ removal} & \textbf{0.46} & \textbf{1.89} & \textbf{2.43} & 4.82 & 6.30 & 13.08 \\
\emph{Early electrification} & 0.78 & 2.66 & 3.21 & 5.77 & 6.62 & \textbf{8.60} \\
\emph{Late switch} & 0.73 & 2.80 & 2.97 & \textbf{4.49} & \textbf{5.46} & 11.60 \\
\bottomrule
\end{tabular}

%% file: tables/alternatives_design.tex
\begin{tabular}{l c c c}
\toprule
& \multicolumn{3}{c}{Target state in 2050} \\
\cmidrule{2-4}
Path up to 2035 & X & $\kappa_4$ & $\kappa_5$ \\
\midrule
$\kappa_0$-X-X & \texttt{0-X-X-X-X-X} & \texttt{0-X-X-X-X-4} & \texttt{0-X-X-X-X-5} \\
$\kappa_0$-$\kappa_1$-$\kappa_2$ & \texttt{0-1-2-X-X-X} & \texttt{0-1-2-X-X-4} & \texttt{0-1-2-X-X-5} \\
$\kappa_0$-$\kappa_1$-$\kappa_3$ & \texttt{0-1-3-X-X-X} & \texttt{0-1-3-X-X-4} & \texttt{0-1-3-X-X-5} \\
\bottomrule
\end{tabular}

%% file: tables/currency_factors.tex
\begin{tabular}{l c l}
\toprule
Price basis & Factor to CHF$_{2024}$ & Used for \\
\midrule
EUR$_{2015}$ & 1.1751 & EnergyScope Pathway costs and prices; electricity price 2015 \\
CHF$_{2017}$ & 1.0681 & CROSS renewable H$_2$ import price; hydro dam \\
EUR$_{2020}$ & 1.1228 & DEA buses, trucks and Fischer--Tropsch fuel; electricity price 2020 \\
CHF$_{2020}$ & 1.0488 & CROSS investment costs \\
USD$_{2022}$ & 0.9900 & CO$_2$ compensation (direct air capture) \\
USD$_{2023}$ & 0.9167 & CROSS oil and gas import prices \\
EUR$_{2024}$ & 0.9530 & electricity price 2024 \\
CHF$_{2024}$ & 1.0000 & Nexus-e electricity prices (assumed) \\
EUR$_{2025}$ & 0.9297 & DEA direct electric heating (assuming 2.5\% inflation from EUR$_{2024}$) \\
CHF$_{2025}$ & 1.0000 & passenger train, hydro reservoir (as CHF$_{2024}$) \\
\bottomrule
\end{tabular}

%% file: tables/demands.tex
\begin{tabular}{l l r r r r r r }
\toprule
End-use demand & Unit & 2025 & 2030 & 2035 & 2040 & 2045 & 2050 \\
\midrule
Electricity, households & GWh & 11{,}984 & 11{,}697 & 11{,}377 & 11{,}074 & 10{,}829 & 10{,}635 \\
Electricity, services & GWh & 14{,}151 & 14{,}008 & 13{,}848 & 13{,}698 & 13{,}533 & 13{,}333 \\
Electricity, industry & GWh & 13{,}312 & 13{,}175 & 13{,}043 & 12{,}941 & 12{,}845 & 12{,}734 \\
High-temperature heat, services & GWh & 451 & 468 & 491 & 515 & 547 & 578 \\
High-temperature heat, industry & GWh & 11{,}452 & 11{,}110 & 10{,}960 & 10{,}868 & 10{,}963 & 11{,}073 \\
Medium-temperature heat, services & GWh & 182 & 189 & 189 & 189 & 181 & 172 \\
Medium-temperature heat, industry & GWh & 4{,}621 & 4{,}483 & 4{,}222 & 3{,}992 & 3{,}636 & 3{,}298 \\
Space heating, households & GWh & 36{,}963 & 36{,}860 & 36{,}596 & 36{,}036 & 35{,}249 & 34{,}525 \\
Space heating, services & GWh & 16{,}322 & 16{,}346 & 16{,}297 & 16{,}125 & 15{,}870 & 15{,}678 \\
Space heating, industry & GWh & 2{,}885 & 2{,}692 & 2{,}534 & 2{,}395 & 2{,}270 & 2{,}178 \\
Hot water, households & GWh & 7{,}053 & 7{,}016 & 6{,}925 & 6{,}811 & 6{,}703 & 6{,}596 \\
Hot water, services & GWh & 2{,}394 & 2{,}266 & 2{,}156 & 2{,}067 & 1{,}993 & 1{,}929 \\
Hot water, industry & GWh & 510 & 485 & 463 & 445 & 431 & 418 \\
Passenger mobility & Mpkm & 116{,}199 & 118{,}448 & 119{,}594 & 120{,}048 & 119{,}695 & 118{,}947 \\
Freight mobility & Mtkm & 26{,}230 & 27{,}912 & 29{,}293 & 30{,}538 & 31{,}817 & 32{,}974 \\
Aviation & GWh & 19{,}945 & 20{,}339 & 20{,}741 & 21{,}151 & 21{,}570 & 21{,}996 \\
Non-energy demand & GWh & 2{,}118 & 2{,}118 & 2{,}118 & 2{,}118 & 2{,}118 & 2{,}118 \\
\bottomrule
\end{tabular}

%% file: tables/resources.tex
\setlength{\tabcolsep}{3.5pt}
\begin{threeparttable}
\begin{tabular}{l l l r r r r r r }
\toprule
Resource & Parameter & Unit & 2025 & 2030 & 2035 & 2040 & 2045 & 2050 \\
\midrule
Ammonia import & gwp\_op\tnote{a} & kgCO$_2$/kWh & 0.285 & 0.285 & 0.285 & 0.285 & 0.285 & 0.285 \\
CO$_2$ (agriculture) & must\_use\tnote{b} & ktCO$_2$/yr & 6{,}086 & 6{,}300 & 5{,}905 & 5{,}510 & 5{,}115 & 4{,}720 \\
CO$_2$ (cement) & avail\tnote{c} & ktCO$_2$/yr & 1{,}673 & 1{,}673 & 1{,}673 & 1{,}673 & 1{,}673 & 1{,}673 \\
 & must\_use\tnote{c} & ktCO$_2$/yr & 1{,}673 & 1{,}673 & 1{,}673 & 1{,}673 & 1{,}673 & 1{,}673 \\
CO$_2$ compensation & c\_op\tnote{d} & CHF$_{2024}$/tCO$_2$ & 1{,}173 & 1{,}012 & 852 & 691 & 531 & 370 \\
CO$_2$ export & c\_op\tnote{e} & CHF$_{2024}$/tCO$_2$ & 100 & 100 & 100 & 100 & 100 & 100 \\
Coal & gwp\_op\tnote{a} & kgCO$_2$/kWh & 0.401 & 0.401 & 0.401 & 0.401 & 0.401 & 0.401 \\
Electricity import / export & avail\tnote{f} & GWh/yr & $\infty$ & $\infty$ & $\infty$ & $\infty$ & $\infty$ & $\infty$ \\
 & c\_op\tnote{g} & CHF$_{2024}$/MWh & 70.2 & 59.1 & 48.9 & 38.7 & 42.2 & 45.8 \\
Hydro dam & avail\tnote{h} & GWh/yr & 18{,}201 & 18{,}201 & 18{,}201 & 18{,}201 & 18{,}201 & 18{,}201 \\
Methanol import & gwp\_op\tnote{a} & kgCO$_2$/kWh & 0.350 & 0.350 & 0.350 & 0.350 & 0.350 & 0.350 \\
Natural gas import & c\_op\tnote{i} & CHF$_{2024}$/MWh & 32.4 & 18.8 & 17.5 & 16.3 & 16.3 & 16.3 \\
 & gwp\_op\tnote{j} & kgCO$_2$/kWh & 0.200 & 0.200 & 0.200 & 0.200 & 0.200 & 0.200 \\
Oil & c\_op\tnote{i} & CHF$_{2024}$/MWh & 42.8 & 39.0 & 36.5 & 34.1 & 32.7 & 31.4 \\
 & gwp\_op\tnote{j} & kgCO$_2$/kWh & 0.265 & 0.265 & 0.265 & 0.265 & 0.265 & 0.265 \\
Renewable H$_2$ import & c\_op\tnote{i} & CHF$_{2024}$/MWh & 341 & 319 & 297 & 278 & 260 & 241 \\
Run-of-river hydro & avail\tnote{k} & GWh/yr & 17{,}458 & 17{,}458 & 17{,}458 & 17{,}458 & 17{,}458 & 17{,}458 \\
Waste & avail\tnote{l} & GWh/yr & 15{,}313 & 16{,}452 & 17{,}111 & 17{,}741 & 18{,}350 & 18{,}914 \\
 & gwp\_op\tnote{m} & kgCO$_2$/kWh & 0.165 & 0.165 & 0.165 & 0.165 & 0.165 & 0.165 \\
 & must\_use\tnote{l} & GWh/yr & 15{,}313 & 16{,}452 & 17{,}111 & 17{,}741 & 18{,}350 & 18{,}914 \\
Waste import & c\_op\tnote{n} & CHF$_{2024}$/MWh & 51.3 & 52.8 & 54.2 & 55.7 & 57.2 & 58.7 \\
 & gwp\_op\tnote{m} & kgCO$_2$/kWh & 0.165 & 0.165 & 0.165 & 0.165 & 0.165 & 0.165 \\
Wet biomass & avail\tnote{l} & GWh/yr & 8{,}316 & 11{,}012 & 11{,}258 & 11{,}501 & 11{,}754 & 12{,}013 \\
Wood (local) & avail\tnote{l} & GWh/yr & 13{,}316 & 14{,}952 & 15{,}063 & 15{,}151 & 15{,}230 & 15{,}300 \\
\bottomrule
\end{tabular}
\begin{tablenotes}\footnotesize
\item[a] Combustion emissions of EnergyScope Pathway, counted because these carriers are burned in Switzerland.
\item[b] Agricultural residual emissions of the CROSS scenarios \cite{marcucci_cross_drivers_2025}.
\item[c] Mean geogenic process emissions of the Swiss cement plants in 2015--2024 (Cemsuisse \cite{cemsuisse_jahresbericht_2026}), held constant.
\item[d] Solid-sorbent direct air capture with storage, learning curve of Sievert et al. \cite{sievert_considering_2024} reaching a cumulative capacity of 1\,Gt in 2050.
\item[e] Transport and storage price of CO$_2$ in the North Sea of the CROSS scenario protocol \cite{marcucci_cross_2025}.
\item[f] Bounded by the net transfer capacities and by an annual net transfer of 20\,TWh/yr (Note~\ref{note:technologies}).
\item[g] Interpolated linearly between the Swiss day-ahead average of Energy-Charts \cite{fraunhofer_ise_day-ahead_nodate} in 2024 and the mean of the hourly Nexus-e price \cite{gjorgiev_nexus-e_2022} in 2030, 2040 and 2050.
\item[h] Mean reservoir inflow 2016--2023, reconstructed from the filling level and the reservoir generation of Energy-Charts \cite{fraunhofer_ise_storage_nodate}.
\item[i] CROSS import prices \cite{sweet_cross_import_2025}, Reference variant, converted to CHF$_{2024}$.
\item[j] Carbon intensity of the imported fuel \cite{wiest_low-regret_2026}.
\item[k] Mean full-load hours 2016--2023 of Energy-Charts \cite{fraunhofer_ise_net_nodate, fraunhofer_ise_installed_nodate} times the installed 4.28\,GW.
\item[l] CROSS biomass potentials \cite{sweet_cross_biomass_2025}, reference scenario, with forest wood from its scenario (1), continued stock increase.
\item[m] Half of the 0.33\,kgCO$_2$/kWh of municipal waste is biogenic \cite{wiest_low-regret_2026}.
\item[n] Own assumption: twice the domestic waste price of EnergyScope Pathway.
\end{tablenotes}
\end{threeparttable}

%% file: tables/cross_mapping.tex
\begin{threeparttable}
\begin{tabular}{l l}
\toprule
Model technology & CROSS technology and size \\
\midrule
Run-of-river hydro & Hydropower run-of-river, $<$ 10MW \\
Onshore wind & Wind\tnote{a} \\
Rooftop PV & Solar PV, 30 kW\tnote{b} \\
Open-field PV & Solar PV, 1000 kW \\
Alpine PV & Solar PV, 1000 kW \\
CCGT & Gas turbine, 100-500 MW \\
Dec. heat pump & Air source heat pump, 10 kWth \\
DH heat pump & Ground source heat pump, 200 - 500 kWth \\
Dec. wood boiler & Wood pellets boiler, 20 kW \\
DH wood boiler & Wood chips boiler, 500 kW \\
Ind. wood boiler & Wood chips boiler, 500 kW \\
Dec. gas boiler & Gas boiler, 20 kW \\
DH gas boiler & Gas boiler, 200 - 500 kW \\
Ind. gas boiler & Gas boiler, 200 - 500 kW \\
Dec. oil boiler & Oil boiler, 20 kW \\
DH oil boiler & Oil boiler, 200 - 500 kW \\
Ind. oil boiler & Oil boiler, 200 - 500 kW \\
Dec. solar thermal & Solar thermal, 10 m2 \\
DH solar thermal & Solar thermal, 200 m2 \\
District heating network & District heating, 200 - 500 kW \\
Li-ion battery & Li-Ion  battery, Large scale \\
Electrolysis & Electrolyser, 100 MWAC \\
\bottomrule
\end{tabular}
\begin{tablenotes}\footnotesize
\item[a] No size class in the CROSS file.
\item[b] Close to the capacity-weighted Swiss 2024 additions.
\end{tablenotes}
\end{threeparttable}

%% file: tables/technologies_electricity.tex
\setlength{\tabcolsep}{3.5pt}
\begin{threeparttable}
\begin{tabular}{l l l r r r r r r }
\toprule
Technology & Parameter & Unit & 2025 & 2030 & 2035 & 2040 & 2045 & 2050 \\
\midrule
Alpine PV & c\_inv\tnote{a} & CHF$_{2024}$/kW & 1{,}463 & 1{,}310 & 1{,}156 & 1{,}110 & 1{,}065 & 1{,}019 \\
 & c\_maint\tnote{b} & CHF$_{2024}$/kW/yr & 14.5 & 12.7 & 11.7 & 10.6 & 9.6 & 8.6 \\
 & f\_max\tnote{c} & GW & $\infty$ & 9.34 & 9.34 & 9.34 & 9.34 & 9.34 \\
 & lifetime\tnote{b} & yr & 25 & 25 & 25 & 25 & 25 & 25 \\
CCGT & c\_inv\tnote{d} & CHF$_{2024}$/kW & 1{,}206 & 1{,}185 & 1{,}164 & 1{,}157 & 1{,}149 & 1{,}142 \\
Hydro dam & c\_inv\tnote{e} & CHF$_{2024}$/kW & 6{,}409 & 6{,}409 & 6{,}409 & 6{,}409 & 6{,}409 & 6{,}409 \\
 & c\_maint\tnote{f} & CHF$_{2024}$/kW/yr & 32.0 & 32.0 & 32.0 & 32.0 & 32.0 & 32.0 \\
 & f\_max\tnote{g} & GW & $\infty$ & 8.26 & 8.26 & 8.26 & 8.26 & 8.26 \\
 & lifetime\tnote{h} & yr & 40 & 40 & 40 & 40 & 40 & 40 \\
Hydro reservoir & c\_inv\tnote{i} & CHF$_{2024}$/kWh & 1.28 & 1.28 & 1.28 & 1.28 & 1.28 & 1.28 \\
 & c\_maint\tnote{j} & CHF$_{2024}$/kWh/yr & 0 & 0 & 0 & 0 & 0 & 0 \\
 & f\_max\tnote{k} & GWh & $\infty$ & 6{,}980 & 7{,}460 & 7{,}940 & 8{,}420 & 8{,}900 \\
 & f\_min\tnote{k} & GWh & 0 & 6{,}500 & 6{,}500 & 6{,}500 & 6{,}500 & 6{,}500 \\
 & lifetime\tnote{l} & yr & 60 & 60 & 60 & 60 & 60 & 60 \\
Li-ion battery & c\_inv\tnote{m} & CHF$_{2024}$/kWh & 140 & 113 & 85 & 78 & 71 & 64 \\
Nuclear & c\_inv\tnote{n} & CHF$_{2024}$/kW & 8{,}000 & 8{,}000 & 8{,}000 & 8{,}000 & 8{,}000 & 8{,}000 \\
 & c\_maint\tnote{o} & CHF$_{2024}$/kW/yr & 280 & 280 & 280 & 280 & 280 & 280 \\
 & c\_p\tnote{p} & -- & 0.900 & 0.849 & 0.849 & 0.849 & 0.849 & 0.849 \\
 & f\_max\tnote{q} & GW & 2.97 & 2.60 & 2.24 & 1.23 & $\infty$ & $\infty$ \\
 & t\_constr\tnote{o} & yr & 15 & 15 & 15 & 15 & 15 & 15 \\
Onshore wind & c\_inv\tnote{d} & CHF$_{2024}$/kW & 2{,}487 & 2{,}353 & 2{,}218 & 2{,}133 & 2{,}048 & 1{,}962 \\
 & f\_max\tnote{c} & GW & $\infty$ & 4.61 & 4.61 & 4.61 & 4.61 & 4.61 \\
Open-field PV & c\_inv\tnote{d} & CHF$_{2024}$/kW & 732 & 655 & 578 & 555 & 532 & 510 \\
 & c\_maint\tnote{b} & CHF$_{2024}$/kW/yr & 14.5 & 12.7 & 11.7 & 10.6 & 9.6 & 8.6 \\
 & f\_max\tnote{r} & GW & $\infty$ & 25.9 & 25.9 & 25.9 & 25.9 & 25.9 \\
 & lifetime\tnote{b} & yr & 25 & 25 & 25 & 25 & 25 & 25 \\
Pumped hydro storage & f\_max\tnote{s} & GWh & 84.5 & 84.5 & 84.5 & 84.5 & 84.5 & 84.5 \\
Rooftop PV & c\_inv\tnote{d} & CHF$_{2024}$/kW & 1{,}562 & 1{,}408 & 1{,}255 & 1{,}124 & 992 & 861 \\
 & f\_max\tnote{c} & GW & $\infty$ & 43.8 & 43.8 & 43.8 & 43.8 & 43.8 \\
Run-of-river hydro & c\_inv\tnote{d} & CHF$_{2024}$/kW & 9{,}115 & 7{,}815 & 6{,}516 & 6{,}543 & 6{,}570 & 6{,}597 \\
 & f\_max\tnote{g} & GW & $\infty$ & 4.28 & 4.28 & 4.28 & 4.28 & 4.28 \\
\bottomrule
\end{tabular}
\begin{tablenotes}\footnotesize
\item[a] CROSS investment costs \cite{sweet_cross_investment_2026} of 1000\,kW solar PV, doubled for the alpine installation (own assumption).
\item[b] As for rooftop PV.
\item[c] Mean of the High and Low potential of the CROSS scenario protocol \cite{marcucci_cross_2025} (Tab.~12), divided by the mean full-load hours of 2016--2023.
\item[d] CROSS investment costs \cite{sweet_cross_investment_2026}, release 2026-02-24, Reference variant, converted from CHF$_{2020}$; size classes in Tab.~\ref{tab:cross_mapping}.
\item[e] 6000\,CHF$_{2017}$/kW of the SES-ETH model in the CROSS cost dataset for multiple models \cite{sweet_cross_technology_2023}.
\item[f] Own assumption: 0.5\,\%/yr of the investment, as in EnergyScope TD \cite{limpens_energyscope_2019}.
\item[g] Installed capacity in 2024 (Energy-Charts \cite{fraunhofer_ise_installed_nodate}); hydropower is not expanded.
\item[h] Own assumption, as in EnergyScope Pathway.
\item[i] Grimsel lake extension: 307\,MCHF for 240\,GWh of additional storage (KWO \cite{kraftwerke_oberhasli_ag_kwo_vergrosserung_2026}).
\item[j] Own assumption: no maintenance of the reservoir itself; the maintenance of the plant is in Hydro dam.
\item[k] Effectively used storage of 6.5\,TWh, raised by dam heightening and renewal to 8.9\,TWh in equal steps from 2030 to 2050 \cite{boes_swiss_2021}.
\item[l] Own assumption.
\item[m] CROSS investment costs \cite{sweet_cross_investment_2026} of large-scale Li-ion batteries, read per kWh of storage capacity (personal communication with CROSS).
\item[n] Medium case of the overnight capital cost of the ETH Zurich Energy Science Center white paper \cite{bardow_nuclear_2026}.
\item[o] Own assumptions: O\&M 3.5\,\%/yr of the investment, construction time 15 years.
\item[p] Own assumption for the existing plants in 2025; EnergyScope Pathway afterwards.
\item[q] The existing plants operate for at most 60 years (Beznau 1 until 2030, Beznau 2 until 2035, G\"osgen until 2040, Leibstadt until 2045) and may retire earlier; new plants are not bounded.
\item[r] Full potential of free areas along roads and motorway embankments, 30.3\,TWh/yr \cite{remund_schweizer_2019, hugi_meta-study_2025}, divided by the mean full-load hours of 2016--2023.
\item[s] Installed 3.52\,GW in 2024 (Energy-Charts \cite{fraunhofer_ise_installed_nodate}) for 24\,h; not expanded.
\end{tablenotes}
\end{threeparttable}

%% file: tables/technologies_transport.tex
\setlength{\tabcolsep}{3.5pt}
\begin{threeparttable}
\begin{tabular}{l l l r r r r r r }
\toprule
Technology & Parameter & Unit & 2025 & 2030 & 2035 & 2040 & 2045 & 2050 \\
\midrule
Battery-electric bus & c\_inv\tnote{a} & CHF$_{2024}$/pkm/h & 2{,}241 & 1{,}421 & 1{,}289 & 1{,}158 & 1{,}055 & 951 \\
 & c\_maint\tnote{a} & CHF$_{2024}$/pkm/h/yr & 39.1 & 36.4 & 34.2 & 31.9 & 30.2 & 28.4 \\
 & c\_p\tnote{a} & -- & 0.297 & 0.297 & 0.297 & 0.297 & 0.297 & 0.297 \\
 & lifetime\tnote{a} & yr & 12 & 12 & 12 & 12 & 12 & 12 \\
Battery-electric car & c\_p\tnote{b} & -- & 0.0418 & 0.0418 & 0.0418 & 0.0418 & 0.0418 & 0.0418 \\
Battery-electric truck & c\_inv\tnote{c} & CHF$_{2024}$/tkm/h & 517 & 308 & 300 & 293 & 287 & 280 \\
 & c\_maint\tnote{c} & CHF$_{2024}$/tkm/h/yr & 10.7 & 10.4 & 10.2 & 10.0 & 9.7 & 9.5 \\
 & c\_p\tnote{c} & -- & 0.200 & 0.200 & 0.200 & 0.200 & 0.200 & 0.200 \\
 & lifetime\tnote{c} & yr & 13.8 & 13.8 & 13.8 & 13.8 & 13.8 & 13.8 \\
Conventional bus (oil) & c\_inv\tnote{a} & CHF$_{2024}$/pkm/h & 1{,}035 & 963 & 904 & 845 & 799 & 753 \\
 & c\_maint\tnote{a} & CHF$_{2024}$/pkm/h/yr & 44.0 & 40.9 & 38.4 & 35.9 & 34.0 & 32.0 \\
 & c\_p\tnote{a} & -- & 0.297 & 0.297 & 0.297 & 0.297 & 0.297 & 0.297 \\
 & lifetime\tnote{a} & yr & 12 & 12 & 12 & 12 & 12 & 12 \\
Conventional car (oil) & c\_p\tnote{b} & -- & 0.0418 & 0.0418 & 0.0418 & 0.0418 & 0.0418 & 0.0418 \\
Conventional truck (oil) & c\_inv\tnote{c} & CHF$_{2024}$/tkm/h & 470 & 457 & 446 & 434 & 424 & 413 \\
 & c\_maint\tnote{c} & CHF$_{2024}$/tkm/h/yr & 17.0 & 16.5 & 16.1 & 15.7 & 15.3 & 14.9 \\
 & c\_p\tnote{c} & -- & 0.200 & 0.200 & 0.200 & 0.200 & 0.200 & 0.200 \\
 & lifetime\tnote{c} & yr & 13.8 & 13.8 & 13.8 & 13.8 & 13.8 & 13.8 \\
Electrolysis & c\_inv\tnote{d} & CHF$_{2024}$/kW & 2{,}717 & 1{,}969 & 1{,}728 & 1{,}487 & 1{,}245 & 1{,}004 \\
Fischer-Tropsch fuel & c\_inv\tnote{e} & CHF$_{2024}$/kW & 2{,}209 & 1{,}910 & 1{,}612 & 1{,}313 & 1{,}194 & 1{,}075 \\
 & c\_maint\tnote{e} & CHF$_{2024}$/kW/yr & 66.3 & 57.3 & 48.4 & 39.4 & 35.8 & 32.2 \\
 & c\_p\tnote{e} & -- & 0.942 & 0.942 & 0.942 & 0.942 & 0.942 & 0.942 \\
 & lifetime\tnote{e} & yr & 25 & 25 & 25 & 25 & 25 & 25 \\
 & t\_constr\tnote{e} & yr & 2 & 2 & 2 & 2 & 2 & 2 \\
Fuel-cell bus & c\_inv\tnote{a} & CHF$_{2024}$/pkm/h & 2{,}446 & 2{,}064 & 1{,}751 & 1{,}439 & 1{,}195 & 951 \\
 & c\_maint\tnote{a} & CHF$_{2024}$/pkm/h/yr & 39.1 & 36.4 & 34.2 & 31.9 & 30.2 & 28.4 \\
 & c\_p\tnote{a} & -- & 0.297 & 0.297 & 0.297 & 0.297 & 0.297 & 0.297 \\
 & lifetime\tnote{a} & yr & 12 & 12 & 12 & 12 & 12 & 12 \\
Fuel-cell car & c\_p\tnote{b} & -- & 0.0418 & 0.0418 & 0.0418 & 0.0418 & 0.0418 & 0.0418 \\
Fuel-cell truck & c\_inv\tnote{c} & CHF$_{2024}$/tkm/h & 555 & 504 & 456 & 408 & 365 & 321 \\
 & c\_maint\tnote{c} & CHF$_{2024}$/tkm/h/yr & 10.6 & 10.3 & 10.1 & 9.8 & 9.6 & 9.4 \\
 & c\_p\tnote{c} & -- & 0.200 & 0.200 & 0.200 & 0.200 & 0.200 & 0.200 \\
 & lifetime\tnote{c} & yr & 13.8 & 13.8 & 13.8 & 13.8 & 13.8 & 13.8 \\
Hybrid-electric car & c\_p\tnote{b} & -- & 0.0418 & 0.0418 & 0.0418 & 0.0418 & 0.0418 & 0.0418 \\
Methanol car & c\_p\tnote{b} & -- & 0.0418 & 0.0418 & 0.0418 & 0.0418 & 0.0418 & 0.0418 \\
Natural-gas car & c\_p\tnote{b} & -- & 0.0418 & 0.0418 & 0.0418 & 0.0418 & 0.0418 & 0.0418 \\
Passenger train & c\_inv\tnote{f} & CHF$_{2024}$/pkm/h & 1{,}010 & 1{,}010 & 1{,}010 & 1{,}010 & 1{,}010 & 1{,}010 \\
 & c\_maint\tnote{f} & CHF$_{2024}$/pkm/h/yr & 20.2 & 20.2 & 20.2 & 20.2 & 20.2 & 20.2 \\
 & c\_p\tnote{f} & -- & 0.309 & 0.309 & 0.309 & 0.309 & 0.309 & 0.309 \\
 & fmax\_perc\tnote{g} & -- & 0.833 & 0.833 & 0.833 & 0.833 & 0.833 & 0.833 \\
 & fmin\_perc\tnote{g} & -- & 0.833 & 0.833 & 0.833 & 0.833 & 0.833 & 0.833 \\
Plug-in hybrid car & c\_p\tnote{b} & -- & 0.0418 & 0.0418 & 0.0418 & 0.0418 & 0.0418 & 0.0418 \\
Tramway / trolleybus & c\_p\tnote{h} & -- & 0.309 & 0.309 & 0.309 & 0.309 & 0.309 & 0.309 \\
 & fmax\_perc\tnote{g} & -- & 0.0398 & 0.0398 & 0.0398 & 0.0398 & 0.0398 & 0.0398 \\
 & fmin\_perc\tnote{g} & -- & 0.0398 & 0.0398 & 0.0398 & 0.0398 & 0.0398 & 0.0398 \\
\bottomrule
\end{tabular}
\begin{tablenotes}\footnotesize
\item[a] Danish Energy Agency \cite{danish_energy_agency_technology_nodate}, city bus (class B1), city route; capacity factor 0.297 (own assumption).
\item[b] Share of the day a car is driven: twice 30.1\,min of commuting per day in 2023 (FSO \cite{bundesamt_fur_statistik_bfs_erwerbstatige_2026}).
\item[c] Danish Energy Agency \cite{danish_energy_agency_technology_nodate}, heavy truck up to 56\,t (class L2), urban delivery; capacity factor 0.2 (own assumption).
\item[d] CROSS investment costs \cite{sweet_cross_investment_2026} of a 100\,MW electrolyzer per kW$_{\mathrm{AC}}$ of input, converted to per kW of H$_2$ output with the electricity input per unit of H$_2$ (1.118 in 2020, 1.097 in 2025, 1.076 in 2030, constant afterwards).
\item[e] Danish Energy Agency \cite{danish_energy_agency_technology_nodate-1}, renewable fuels, hydrogen to jet fuel; fixed O\&M of 3\,\%/yr of the investment (own assumption).
\item[f] SBB order of 116 double-deck trains for CHF\,2\,billion \cite{siemens_mobility_ag_siemens_2026} and SBB reporting 2024, per peak passenger capacity; O\&M 2\,\%/yr of the investment (own assumption).
\item[g] Share of rail (83.3\,\%) and tram (4.0\,\%) in public transport in 2025 (CROSS passenger demand \cite{sweet_cross_end-use_2025}), held constant.
\item[h] 731 trams in 2020 (FSO \cite{bundesamt_fur_statistik_bfs_offentlicher_2026}, table T4) with 50 passengers on average at 8\,km/h (own assumption).
\end{tablenotes}
\end{threeparttable}

%% file: tables/technologies_heat.tex
\setlength{\tabcolsep}{3.5pt}
\begin{threeparttable}
\begin{tabular}{l l l r r r r r r }
\toprule
Technology & Parameter & Unit & 2025 & 2030 & 2035 & 2040 & 2045 & 2050 \\
\midrule
Dec. direct elec. & c\_inv\tnote{a} & CHF$_{2024}$/kW & 954 & 930 & 907 & 885 & 863 & 842 \\
 & c\_maint\tnote{a} & CHF$_{2024}$/kW/yr & 8.51 & 8.30 & 7.90 & 7.89 & 7.70 & 7.51 \\
 & c\_p\tnote{a} & -- & 1 & 1 & 1 & 1 & 1 & 1 \\
 & lifetime\tnote{a} & yr & 30 & 30 & 30 & 30 & 30 & 30 \\
Dec. gas boiler & c\_inv\tnote{b} & CHF$_{2024}$/kW & 1{,}384 & 1{,}384 & 1{,}384 & 1{,}384 & 1{,}384 & 1{,}384 \\
Dec. heat pump & c\_inv\tnote{b} & CHF$_{2024}$/kW & 3{,}142 & 3{,}032 & 2{,}907 & 2{,}781 & 2{,}735 & 2{,}689 \\
Dec. oil boiler & c\_inv\tnote{b} & CHF$_{2024}$/kW & 1{,}783 & 1{,}783 & 1{,}783 & 1{,}783 & 1{,}783 & 1{,}783 \\
Dec. solar thermal & c\_inv\tnote{b} & CHF$_{2024}$/kW & 2{,}014 & 1{,}909 & 1{,}769 & 1{,}629 & 1{,}606 & 1{,}583 \\
Dec. wood boiler & c\_inv\tnote{b} & CHF$_{2024}$/kW & 2{,}150 & 2{,}150 & 2{,}150 & 2{,}150 & 2{,}150 & 2{,}150 \\
DH gas boiler & c\_inv\tnote{b} & CHF$_{2024}$/kW & 252 & 252 & 252 & 252 & 252 & 252 \\
DH heat pump & c\_inv\tnote{b} & CHF$_{2024}$/kW & 2{,}498 & 2{,}280 & 2{,}149 & 2{,}018 & 1{,}962 & 1{,}906 \\
DH oil boiler & c\_inv\tnote{b} & CHF$_{2024}$/kW & 388 & 388 & 388 & 388 & 388 & 388 \\
DH solar thermal & c\_inv\tnote{b} & CHF$_{2024}$/kW & 892 & 831 & 810 & 790 & 776 & 762 \\
DH wood boiler & c\_inv\tnote{b} & CHF$_{2024}$/kW & 451 & 451 & 451 & 451 & 451 & 451 \\
Ind. gas boiler & c\_inv\tnote{b} & CHF$_{2024}$/kW & 252 & 252 & 252 & 252 & 252 & 252 \\
 & fmin\_perc\tnote{c} & -- & 0.000 & 0.300 & 0.300 & 0.300 & 0.300 & 0.300 \\
Ind. oil boiler & c\_inv\tnote{b} & CHF$_{2024}$/kW & 388 & 388 & 388 & 388 & 388 & 388 \\
Ind. wood boiler & c\_inv\tnote{b} & CHF$_{2024}$/kW & 451 & 451 & 451 & 451 & 451 & 451 \\
\bottomrule
\end{tabular}
\begin{tablenotes}\footnotesize
\item[a] Danish Energy Agency \cite{danish_energy_agency_technology_nodate-2}, heating installations, electric heating of a new single-family house.
\item[b] CROSS investment costs \cite{sweet_cross_investment_2026}, release 2026-02-24, Reference variant, converted from CHF$_{2020}$; size classes in Tab.~\ref{tab:cross_mapping}.
\item[c] At least 30\,\% of industrial heat from gas, after the 29\,\% of gaseous carriers in 2050 of the Energieperspektiven 2050+, scenario ZERO Basis \cite{prognos_ag_energieperspektiven_2021}.
\end{tablenotes}
\end{threeparttable}

%% file: tables/timeseries_weights.tex
\begin{tabular}{l r}
\toprule
Time series & Weight \\
\midrule
Electricity demand & 1 \\
Space heating, households & 0.6 \\
Space heating, services and industry & 0.2 \\
Hot water & 0.2 \\
High-temperature heat & 1 \\
Passenger mobility & 1 \\
Reservoir inflow & 0.5 \\
Capacity factor, Run-of-river hydro & 0.5 \\
Capacity factor, Rooftop PV & 0.33 \\
Capacity factor, Open-field PV & 0.33 \\
Capacity factor, Alpine PV & 0.33 \\
Capacity factor, Onshore wind & 1 \\
Capacity factor, Dec. solar thermal & 0.5 \\
Capacity factor, DH solar thermal & 0.5 \\
\bottomrule
\end{tabular}

%% file: tables/peak_factor.tex
\begin{tabular}{c l c c c c}
\toprule
Weather & \multicolumn{2}{c}{Annual peak} & Annual peak & Typical-day peak & $\%_{\mathrm{Peak,sh}}$ \\
\cmidrule(lr){2-3}
year & day & hour & [GW] & [GW] & [-] \\
\midrule
2016 & 18 Jan & 06:00 & 30.5 & 24.2 & 1.26 \\
2017 & 6 Jan & 06:00 & 33.6 & 22.5 & 1.50 \\
2018 & 27 Feb & 06:00 & 39.0 & 23.5 & 1.66 \\
2019 & 22 Jan & 06:00 & 30.1 & 21.4 & 1.41 \\
2020 & 26 Dec & 06:00 & 27.4 & 25.4 & 1.08 \\
2021 & 13 Feb & 06:00 & 31.5 & 20.3 & 1.55 \\
2022 & 12 Dec & 06:00 & 33.7 & 22.0 & 1.53 \\
2023 & 3 Dec & 06:00 & 32.8 & 24.2 & 1.36 \\
\bottomrule
\end{tabular}

%% file: tables/init_2025_targets.tex
\begin{tabular}{l r l}
\toprule
Quantity & Value 2025 & Unit \\
\midrule
Electricity production, Nuclear & 22{,}983 & GWh \\
Electricity production, Onshore wind & 171 & GWh \\
Electricity production, Rooftop PV & 5{,}961 & GWh \\
Hydro production (run-of-river and dam), at least & 95\,\% & of the inflow \\
Net biofuel import & 2{,}432 & GWh \\
Share of rail in freight & 38.0\,\% &  \\
Share of district heating in low-temperature heat & 9.9\,\% &  \\
Conventional car (oil) & 85.9\,\% & of its subsector \\
Hybrid-electric car & 7.5\,\% & of its subsector \\
Plug-in hybrid car & 2.1\,\% & of its subsector \\
Natural-gas car & 0.3\,\% & of its subsector \\
Battery-electric car & 4.2\,\% & of its subsector \\
Conventional bus (oil) & 12.2\,\% & of its subsector \\
Battery-electric bus & 0.4\,\% & of its subsector \\
Battery-electric truck & 2.3\,\% & of its subsector \\
Conventional truck (oil) & 97.7\,\% & of its subsector \\
Dec. oil boiler & 37.1\,\% & of its subsector \\
Dec. gas boiler & 27.8\,\% & of its subsector \\
Dec. wood boiler & 7.1\,\% & of its subsector \\
Dec. direct elec. & 3.2\,\% & of its subsector \\
Dec. heat pump & 24.3\,\% & of its subsector \\
Dec. solar thermal & 0.2\,\% & of its subsector \\
DH wood boiler & 32.3\,\% & of its subsector \\
DH gas CHP & 20.7\,\% & of its subsector \\
DH gas CHP, at most & 50.0\,\% & of its subsector \\
DH oil boiler & 2.1\,\% & of its subsector \\
DH wet biomass CHP & 8.5\,\% & of its subsector \\
DH waste CHP & 31.8\,\% & of its subsector \\
Ind. gas boiler & 50.5\,\% & of its subsector \\
Ind. waste boiler & 23.9\,\% & of its subsector \\
Ind. oil boiler & 10.3\,\% & of its subsector \\
Ind. wood boiler & 10.2\,\% & of its subsector \\
Ind. coal boiler & 5.1\,\% & of its subsector \\
\bottomrule
\end{tabular}

%% file: tables/init_2025_fleet.tex
\begin{tabular}{l r l @{\hspace{2em}} l r l}
\toprule
Technology & Capacity & Unit & Technology & Capacity & Unit \\
\midrule
Nuclear & 2.97 & GW & Conventional car (oil) & 208 & Mpkm/h \\
CCGT & 0.33 & GW & Natural-gas car & 0.711 & Mpkm/h \\
Rooftop PV & 8.17 & GW & Hybrid-electric car & 18.2 & Mpkm/h \\
Onshore wind & 0.1 & GW & Plug-in hybrid car & 8.39 & Mpkm/h \\
Run-of-river hydro & 4.28 & GW & Battery-electric car & 10.2 & Mpkm/h \\
Hydro dam & 8.26 & GW & Freight train & 3.32 & Mtkm/h \\
Ind. gas boiler & 1.02 & GW & Aircraft & 2.28 & GW \\
Ind. wood boiler & 0.219 & GW & District heating network & 1.3 & GW \\
Ind. oil boiler & 0.208 & GW & High- to medium-temperature heat & 0.848 & GW \\
Ind. coal boiler & 0.108 & GW & Oil-to-HVC & 0.207 & GW \\
Ind. waste boiler & 0.761 & GW & Li-ion battery & 1 & GWh \\
DH gas CHP & 0.235 & GW & Battery of electric cars & 9.48 & GWh \\
DH waste CHP & 0.395 & GW & Battery of plug-in hybrid cars & 1.55 & GWh \\
DH wet biomass CHP & 0.0893 & GW & Pumped hydro storage & 84.5 & GWh \\
DH wood boiler & 0.536 & GW & Hydro reservoir & 6{,}500 & GWh \\
DH oil boiler & 0.0377 & GW & TS Dec. heat pump & 7.75 & GWh \\
DH deep geothermal & 0.0061 & GW & TS DH seasonal & 327 & GWh \\
Dec. heat pump & 5.06 & GW & TS High-T & 2.54 & GWh \\
Dec. gas boiler & 8.04 & GW & Oil storage & 37{,}000 & GWh \\
Dec. wood boiler & 2.05 & GW & Conventional bus (oil) & 1.3 & Mpkm/h \\
Dec. oil boiler & 10.7 & GW & Battery-electric bus & 0.0474 & Mpkm/h \\
Dec. solar thermal & 0.244 & GW & Battery-electric truck & 0.211 & Mtkm/h \\
Dec. direct elec. & 1.02 & GW & Battery of electric buses & 0.117 & GWh \\
Tramway / trolleybus & 0.403 & Mpkm/h & Battery of electric trucks & 0.179 & GWh \\
Passenger train & 8.42 & Mpkm/h & Conventional truck (oil) & 9.07 & Mtkm/h \\
\bottomrule
\end{tabular}

%% file: tables/macro_parameters.tex
\begin{tabular}{l l l r r r r r r }
\toprule
Macro parameter & Driver & Scenario & 2025 & 2030 & 2035 & 2040 & 2045 & 2050 \\
\midrule
Population size [million] & population & Ref & 9.11 & 9.44 & 9.69 & 9.88 & 10.06 & 10.21 \\
 &  & Low &  & 9.20 & 9.25 & 9.26 & 9.25 & 9.22 \\
 &  & High &  & 9.69 & 10.13 & 10.50 & 10.86 & 11.21 \\
\midrule
Households per capita [-] & behavior & Ref & 0.450 & 0.453 & 0.459 & 0.463 & 0.466 & 0.469 \\
 &  & Low &  & 0.449 & 0.449 & 0.449 & 0.448 & 0.446 \\
 &  & High &  & 0.458 & 0.468 & 0.477 & 0.485 & 0.493 \\
\midrule
Share of working population [-] & economy & Ref & 0.493 & 0.491 & 0.491 & 0.491 & 0.490 & 0.486 \\
 &  & Low &  & 0.486 & 0.481 & 0.476 & 0.470 & 0.461 \\
 &  & High &  & 0.496 & 0.500 & 0.506 & 0.509 & 0.510 \\
\midrule
GDP per worker [kCHF$_{2017}$] & economy & Ref & 177.9 & 188.8 & 200.3 & 212.7 & 225.9 & 240.0 \\
 &  & Low &  & 185.0 & 192.3 & 199.9 & 207.8 & 216.0 \\
 &  & High &  & 192.5 & 208.3 & 225.5 & 244.0 & 264.0 \\
\bottomrule
\end{tabular}

%% file: tables/parameter_scaling.tex
\begin{tabular}{l l l r r r r r r }
\toprule
Model input & Unit & Scaled with & 2025 & 2030 & 2035 & 2040 & 2045 & 2050 \\
\midrule
Electricity, households & MWh/capita & Population & 1.32 & 1.24 & 1.17 & 1.12 & 1.08 & 1.04 \\
Electricity, services & kWh/CHF$_{2017}$ & GDP & 0.0177 & 0.016 & 0.0145 & 0.0133 & 0.0122 & 0.0112 \\
Electricity, industry & kWh/CHF$_{2017}$ & GDP & 0.0167 & 0.0151 & 0.0137 & 0.0125 & 0.0115 & 0.0107 \\
Space heating, households & MWh/household & Households & 9.02 & 8.61 & 8.24 & 7.87 & 7.51 & 7.2 \\
Space heating, services & MWh/capita & Population & 1.79 & 1.73 & 1.68 & 1.63 & 1.58 & 1.54 \\
Space heating, industry & kWh/CHF$_{2017}$ & GDP & 0.00361 & 0.00308 & 0.00266 & 0.00232 & 0.00204 & 0.00183 \\
Hot water, households & MWh/capita & Population & 0.774 & 0.743 & 0.715 & 0.689 & 0.666 & 0.646 \\
Hot water, services & MWh/capita & Population & 0.263 & 0.24 & 0.223 & 0.209 & 0.198 & 0.189 \\
Hot water, industry & kWh/CHF$_{2017}$ & GDP & 0.000638 & 0.000554 & 0.000486 & 0.000431 & 0.000387 & 0.000351 \\
Medium-temperature heat, services & kWh/CHF$_{2017}$ & GDP & 0.000228 & 0.000216 & 0.000199 & 0.000183 & 0.000163 & 0.000145 \\
Medium-temperature heat, industry & kWh/CHF$_{2017}$ & GDP & 0.00578 & 0.00512 & 0.00443 & 0.00387 & 0.00327 & 0.00277 \\
High-temperature heat, services & kWh/CHF$_{2017}$ & GDP & 0.000564 & 0.000535 & 0.000515 & 0.000499 & 0.000491 & 0.000485 \\
High-temperature heat, industry & kWh/CHF$_{2017}$ & GDP & 0.0143 & 0.0127 & 0.0115 & 0.0105 & 0.00985 & 0.0093 \\
Passenger mobility & kpkm/capita & Population & 12.8 & 12.5 & 12.3 & 12.1 & 11.9 & 11.6 \\
Freight mobility & tkm/CHF$_{2017}$ & GDP & 0.0328 & 0.0319 & 0.0308 & 0.0296 & 0.0286 & 0.0277 \\
Aviation & MWh/capita & Population & 2.19 & 2.15 & 2.14 & 2.14 & 2.14 & 2.15 \\
\midrule
Wood (local): waste wood & MWh/capita & Population & 0.406 & 0.452 & 0.452 & 0.452 & 0.452 & 0.452 \\
Wet biomass: collected organic waste & MWh/capita & Population & 0.135 & 0.151 & 0.169 & 0.187 & 0.206 & 0.226 \\
Wet biomass: sewage sludge & MWh/capita & Population & 0.162 & 0.162 & 0.162 & 0.162 & 0.162 & 0.162 \\
Wet biomass: agricultural byproducts & MWh/capita & Population & 0.0544 & 0.0775 & 0.0755 & 0.0741 & 0.0728 & 0.0717 \\
Wet biomass: manure & MWh/capita & Population & 0.561 & 0.775 & 0.755 & 0.74 & 0.728 & 0.717 \\
Waste: municipal waste & MWh/capita & Population & 0.923 & 0.895 & 0.868 & 0.84 & 0.813 & 0.786 \\
Waste: other waste fraction & MWh/capita & Population & 0.758 & 0.847 & 0.898 & 0.955 & 1.01 & 1.07 \\
\bottomrule
\end{tabular}

%% file: tables/weather_scaling.tex
\begin{tabular}{l r r r r r r r r }
\toprule
Model input & 2016 & 2017 & 2018 & 2019 & 2020 & 2021 & 2022 & 2023 \\
\midrule
Hydro reservoir inflow & 0.982 & 0.948 & 1.039 & 1.106 & 1.011 & 1.022 & 0.856 & 1.035 \\
Run-of-river inflow & 1.024 & 0.970 & 1.013 & 1.046 & 1.040 & 0.997 & 0.902 & 1.010 \\
Space heating, households & 1.087 & 1.060 & 0.977 & 1.007 & 0.952 & 1.142 & 0.916 & 0.961 \\
Space heating, services & 1.087 & 1.060 & 0.977 & 1.007 & 0.952 & 1.142 & 0.916 & 0.961 \\
Space heating, industry & 1.087 & 1.060 & 0.977 & 1.007 & 0.952 & 1.142 & 0.916 & 0.961 \\
\bottomrule
\end{tabular}

%% file: tables/drivers_demands.tex
\begin{threeparttable}
\begin{tabular}{l l l c c c c c }
\toprule
Model input & Parameter & Driver & 2030 & 2035 & 2040 & 2045 & 2050 \\
\midrule
Space heating, households & demand & climate\tnote{a} & 1.01/1.00 & 1.01/1.00 & 1.02/0.99 & 1.03/0.98 & 1.04/0.97 \\
Space heating, services & demand & climate\tnote{a} & 1.01/1.00 & 1.01/1.00 & 1.02/0.99 & 1.03/0.98 & 1.04/0.97 \\
Space heating, industry & demand & climate\tnote{a} & 1.01/1.00 & 1.01/1.00 & 1.02/0.99 & 1.03/0.98 & 1.04/0.97 \\
Electricity, households & demand & behavior\tnote{b} & 1.02/0.98 & 1.04/0.96 & 1.06/0.94 & 1.08/0.92 & 1.10/0.90 \\
Hot water, households & demand & behavior\tnote{b} & 1.02/0.98 & 1.04/0.96 & 1.06/0.94 & 1.08/0.92 & 1.10/0.90 \\
Space heating, households & demand & behavior\tnote{b} & 1.02/0.98 & 1.04/0.96 & 1.06/0.94 & 1.08/0.92 & 1.10/0.90 \\
Passenger mobility & demand & behavior\tnote{b} & 1.02/0.98 & 1.04/0.96 & 1.06/0.94 & 1.08/0.92 & 1.10/0.90 \\
Freight mobility & demand & behavior\tnote{b} & 1.02/0.98 & 1.04/0.96 & 1.06/0.94 & 1.08/0.92 & 1.10/0.90 \\
Aviation & demand & behavior\tnote{c} & 1.06/0.94 & 1.12/0.88 & 1.18/0.82 & 1.24/0.76 & 1.30/0.70 \\
Minimum share of public mobility & share & behavior\tnote{d} & 0.94/1.06 & 0.88/1.12 & 0.82/1.18 & 0.76/1.24 & 0.70/1.30 \\
Maximum share of public mobility & share & behavior\tnote{d} & 0.94/1.06 & 0.88/1.12 & 0.82/1.18 & 0.76/1.24 & 0.70/1.30 \\
\bottomrule
\end{tabular}
\begin{tablenotes}\footnotesize
\item[a] Space heating demand of CROSS \cite{sweet_cross_end-use_2025} under RCP2.6 and RCP8.5 relative to RCP4.5.
\item[b] Own assumption: $\pm 2$\,\% per five years; eco-friendly behavior lowers the demand.
\item[c] Own assumption: $\pm 6$\,\% per five years, three times the change of the other mobility demands.
\item[d] Own assumption: $\pm 6$\,\% per five years; eco-friendly behavior raises the share of public transport.
\end{tablenotes}
\end{threeparttable}

%% file: tables/drivers_resources.tex
\begin{threeparttable}
\begin{tabular}{l l l c c c c c }
\toprule
Model input & Parameter & Driver & 2030 & 2035 & 2040 & 2045 & 2050 \\
\midrule
Wet biomass: manure & availability & behavior\tnote{a} & 1.06/0.94 & 1.12/0.88 & 1.18/0.82 & 1.24/0.76 & 1.30/0.70 \\
Agricultural emissions & quantity & behavior\tnote{a} & 1.06/0.94 & 1.12/0.88 & 1.18/0.82 & 1.24/0.76 & 1.30/0.70 \\
Oil & price & import\tnote{b} & 0.58/1.10 & 0.53/1.16 & 0.48/1.22 & 0.45/1.26 & 0.43/1.29 \\
Biofuel import & price & import\tnote{b} & 0.58/1.10 & 0.53/1.16 & 0.48/1.22 & 0.45/1.26 & 0.43/1.29 \\
Coal & price & import\tnote{b} & 0.58/1.10 & 0.53/1.16 & 0.48/1.22 & 0.45/1.26 & 0.43/1.29 \\
Uranium & price & import\tnote{b} & 0.58/1.10 & 0.53/1.16 & 0.48/1.22 & 0.45/1.26 & 0.43/1.29 \\
Natural gas import & price & import\tnote{b} & 0.73/1.08 & 0.76/1.26 & 0.79/1.46 & 0.78/1.47 & 0.77/1.48 \\
H$_2$ import & price & import\tnote{b} & 0.73/1.08 & 0.76/1.26 & 0.79/1.46 & 0.78/1.47 & 0.77/1.48 \\
Ammonia import & price & import\tnote{b} & 0.73/1.08 & 0.76/1.26 & 0.79/1.46 & 0.78/1.47 & 0.77/1.48 \\
Methanol import & price & import\tnote{b} & 0.73/1.08 & 0.76/1.26 & 0.79/1.46 & 0.78/1.47 & 0.77/1.48 \\
Renewable H$_2$ import & price & import\tnote{b} & 0.70/1.40 & 0.70/1.40 & 0.70/1.40 & 0.70/1.40 & 0.70/1.40 \\
Renewable gas & price & import\tnote{b} & 0.70/1.40 & 0.70/1.40 & 0.70/1.40 & 0.70/1.40 & 0.70/1.40 \\
Renewable ammonia import & price & import\tnote{b} & 0.70/1.40 & 0.70/1.40 & 0.70/1.40 & 0.70/1.40 & 0.70/1.40 \\
Renewable methanol import & price & import\tnote{b} & 0.70/1.40 & 0.70/1.40 & 0.70/1.40 & 0.70/1.40 & 0.70/1.40 \\
CO$_2$ compensation & price & technology\tnote{c} & 0.90/1.13 & 0.89/1.18 & 0.86/1.24 & 0.82/1.35 & 0.75/1.55 \\
Electricity import & price & import\tnote{d} & 0.95/1.10 & 0.90/1.20 & 0.85/1.30 & 0.80/1.40 & 0.75/1.50 \\
Electricity export & price & import\tnote{d} & 0.95/1.10 & 0.90/1.20 & 0.85/1.30 & 0.80/1.40 & 0.75/1.50 \\
\bottomrule
\end{tabular}
\begin{tablenotes}\footnotesize
\item[a] Own assumption: eco-friendly diets change the cattle stock, and with it the agricultural emissions and the manure, by $\pm 6$\,\% per five years.
\item[b] Low and High variants of the CROSS import prices \cite{sweet_cross_import_2025} relative to the Reference; carriers that CROSS does not price follow the closest priced one (oil, gas or renewable hydrogen).
\item[c] 5th and 95th percentile of the cost of solid-sorbent direct air capture with storage of Sievert et al.\ \cite{sievert_considering_2024}, relative to the central case.
\item[d] Own assumption: $-5$ to $+10$\,\% per five years relative to the Nexus-e projection.
\end{tablenotes}
\end{threeparttable}

%% file: tables/drivers_technologies.tex
\begin{threeparttable}
\begin{tabular}{l l l c c c c c }
\toprule
Model input & Parameter & Driver & 2030 & 2035 & 2040 & 2045 & 2050 \\
\midrule
Run-of-river hydro & investment cost & technology\tnote{a} & 0.62/1.38 & 0.62/1.38 & 0.62/1.38 & 0.62/1.38 & 0.62/1.38 \\
Onshore wind & investment cost & technology\tnote{a} & 0.80/1.20 & 0.80/1.20 & 0.80/1.20 & 0.80/1.20 & 0.80/1.20 \\
Rooftop PV & investment cost & technology\tnote{a} & 0.52/1.48 & 0.52/1.48 & 0.52/1.48 & 0.52/1.48 & 0.52/1.48 \\
Open-field PV & investment cost & technology\tnote{a} & 0.41/1.59 & 0.41/1.59 & 0.41/1.59 & 0.41/1.59 & 0.41/1.59 \\
Alpine PV & investment cost & technology\tnote{a} & 0.41/1.59 & 0.41/1.59 & 0.41/1.59 & 0.41/1.59 & 0.41/1.59 \\
CCGT & investment cost & technology\tnote{a} & 0.66/1.61 & 0.66/1.61 & 0.66/1.61 & 0.66/1.61 & 0.66/1.61 \\
Dec. heat pump & investment cost & technology\tnote{a} & 0.71/1.35 & 0.71/1.35 & 0.71/1.35 & 0.71/1.35 & 0.71/1.35 \\
DH heat pump & investment cost & technology\tnote{a} & 0.70/1.40 & 0.70/1.40 & 0.70/1.40 & 0.70/1.40 & 0.70/1.40 \\
Dec. wood boiler & investment cost & technology\tnote{a} & 0.70/1.40 & 0.70/1.40 & 0.70/1.40 & 0.70/1.40 & 0.70/1.40 \\
DH wood boiler & investment cost & technology\tnote{a} & 0.70/1.42 & 0.70/1.42 & 0.70/1.42 & 0.70/1.42 & 0.70/1.42 \\
Ind. wood boiler & investment cost & technology\tnote{a} & 0.70/1.42 & 0.70/1.42 & 0.70/1.42 & 0.70/1.42 & 0.70/1.42 \\
Dec. gas boiler & investment cost & technology\tnote{a} & 0.70/1.40 & 0.70/1.40 & 0.70/1.40 & 0.70/1.40 & 0.70/1.40 \\
DH gas boiler & investment cost & technology\tnote{a} & 0.71/1.42 & 0.71/1.42 & 0.71/1.42 & 0.71/1.42 & 0.71/1.42 \\
Ind. gas boiler & investment cost & technology\tnote{a} & 0.71/1.42 & 0.71/1.42 & 0.71/1.42 & 0.71/1.42 & 0.71/1.42 \\
Dec. oil boiler & investment cost & technology\tnote{a} & 0.70/1.40 & 0.70/1.40 & 0.70/1.40 & 0.70/1.40 & 0.70/1.40 \\
DH oil boiler & investment cost & technology\tnote{a} & 0.70/1.41 & 0.70/1.41 & 0.70/1.41 & 0.70/1.41 & 0.70/1.41 \\
Ind. oil boiler & investment cost & technology\tnote{a} & 0.70/1.41 & 0.70/1.41 & 0.70/1.41 & 0.70/1.41 & 0.70/1.41 \\
Dec. solar thermal & investment cost & technology\tnote{a} & 0.70/1.40 & 0.70/1.40 & 0.70/1.40 & 0.70/1.40 & 0.70/1.40 \\
DH solar thermal & investment cost & technology\tnote{a} & 0.69/1.40 & 0.69/1.40 & 0.69/1.40 & 0.69/1.40 & 0.69/1.40 \\
District heating network & investment cost & technology\tnote{a} & 0.70/1.39 & 0.70/1.39 & 0.70/1.39 & 0.70/1.39 & 0.70/1.39 \\
Li-ion battery & investment cost & technology\tnote{a} & 0.70/1.40 & 0.70/1.41 & 0.70/1.41 & 0.70/1.41 & 0.70/1.41 \\
Electrolysis & investment cost & technology\tnote{a} & 0.70/1.40 & 0.70/1.40 & 0.70/1.40 & 0.70/1.40 & 0.70/1.40 \\
Nuclear & investment cost & technology\tnote{b} & 0.62/1.50 & 0.62/1.50 & 0.62/1.50 & 0.62/1.50 & 0.62/1.50 \\
\bottomrule
\end{tabular}
\begin{tablenotes}\footnotesize
\item[a] Low and High variants of the CROSS investment costs \cite{sweet_cross_investment_2026} relative to the Reference (size classes in Tab.~\ref{tab:cross_mapping}).
\item[b] Low and high cases of the overnight capital cost \cite{bardow_nuclear_2026}, 5000 and 12000\,CHF$_{2024}$/kW, relative to the medium case of 8000\,CHF$_{2024}$/kW.
\end{tablenotes}
\end{threeparttable}

%% file: tables/pca_loadings_all_1.tex
\begin{tabular}{l r r r l}
\toprule
Technology & PC1 (63\,\%) & PC2 (16\,\%) & PC3 (8\,\%) & Unit \\
\midrule
\multicolumn{5}{l}{\textbf{Electricity}} \\
Open-field PV & $+9.0$ & $+3.4$ & $-3.6$ & GW \\
Alpine PV & $+2.0$ & $+1.9$ & $+1.8$ & GW \\
Rooftop PV & $-2.4$ & $-0.96$ & $+0.08$ & GW \\
Onshore wind & $+1.5$ & $+0.29$ & $-0.42$ & GW \\
Nuclear & $-1.2$ & $-0.21$ & $+0.17$ & GW \\
CCGT & $+0.09$ & $+0.46$ & $+0.12$ & GW \\
CCGT (ammonia) & $+0.00$ & $+0.00$ & $+0.00$ & GW \\
Hydro dam & $+0.00$ & $-0.00$ & $+0.00$ & GW \\
Run-of-river hydro & $+0.00$ & $-0.00$ & $+0.00$ & GW \\
\addlinespace
\multicolumn{5}{l}{\textbf{Building}} \\
Dec. heat pump & $+3.6$ & $+22.9$ & $+10.9$ & pp \\
Dec. gas boiler & $+13.4$ & $-21.4$ & $-10.4$ & pp \\
Dec. oil boiler & $-12.8$ & $-0.46$ & $-1.1$ & pp \\
Dec. wood boiler & $-2.7$ & $-0.72$ & $+0.46$ & pp \\
Dec. direct elec. & $-1.3$ & $-0.27$ & $+0.11$ & pp \\
Dec. solar thermal & $-0.07$ & $-0.03$ & $+0.01$ & pp \\
DH heat pump & $+28.1$ & $-0.20$ & $+3.6$ & pp \\
DH wood boiler & $-11.5$ & $+1.3$ & $-1.8$ & pp \\
DH waste CHP & $-6.9$ & $+1.1$ & $-1.9$ & pp \\
DH gas CHP & $-6.1$ & $-2.2$ & $+0.98$ & pp \\
DH wet biomass CHP & $-2.8$ & $-0.26$ & $-0.59$ & pp \\
DH oil boiler & $-0.68$ & $+0.22$ & $-0.26$ & pp \\
DH deep geothermal & $-0.13$ & $+0.04$ & $-0.05$ & pp \\
DH gas boiler & $-0.00$ & $-0.00$ & $+0.00$ & pp \\
\bottomrule
\end{tabular}

%% file: tables/pca_loadings_all_2.tex
\begin{tabular}{l r r r l}
\toprule
Technology & PC1 (63\,\%) & PC2 (16\,\%) & PC3 (8\,\%) & Unit \\
\midrule
\multicolumn{5}{l}{\textbf{Storage}} \\
TS Dec. gas boiler & $+5.6$ & $-6.6$ & $-0.67$ & GWh \\
TS Dec. heat pump & $+0.19$ & $+5.3$ & $+2.8$ & GWh \\
Li-ion battery & $-0.03$ & $+0.88$ & $+0.24$ & GWh \\
TS Dec. oil boiler & $+0.32$ & $-0.62$ & $+0.13$ & GWh \\
TS Dec. wood boiler & $-0.13$ & $-0.44$ & $+0.41$ & GWh \\
TS High-T & $-0.43$ & $-0.05$ & $-0.06$ & GWh \\
TS Dec. direct elec. & $-0.00$ & $-0.14$ & $+0.09$ & GWh \\
TS DH short-term & $+0.00$ & $+0.00$ & $+0.00$ & GWh \\
H$_2$ storage & $+0.00$ & $+0.00$ & $+0.00$ & GWh \\
CO$_2$ storage & $+0.00$ & $+0.00$ & $+0.00$ & GWh \\
Pumped hydro storage & $+0.00$ & $-0.00$ & $+0.00$ & GWh \\
Oil storage & $-12.1$ & $+3.1$ & $-3.8$ & TWh \\
TS DH seasonal & $+1.6$ & $+0.35$ & $-0.07$ & TWh \\
Hydro reservoir & $+728$ & $+300$ & $-36.3$ & GWh \\
Gas storage & $+11.9$ & $+39.6$ & $+13.3$ & GWh \\
Methanol storage & $+5.1$ & $+17.3$ & $+5.3$ & GWh \\
Ammonia storage & $+12.8$ & $+7.0$ & $+1.4$ & GWh \\
\addlinespace
\multicolumn{5}{l}{\textbf{Transport}} \\
Battery-electric car & $+35.8$ & $-2.2$ & $+2.1$ & pp \\
Conventional car (oil) & $-32.1$ & $+2.0$ & $-1.9$ & pp \\
Hybrid-electric car & $-2.8$ & $+0.17$ & $-0.17$ & pp \\
Plug-in hybrid car & $-0.79$ & $+0.04$ & $-0.04$ & pp \\
Natural-gas car & $-0.11$ & $+0.01$ & $-0.01$ & pp \\
Conventional bus (oil) & $-2.2$ & $-3.0$ & $+0.10$ & pp \\
Battery-electric bus & $+2.2$ & $+3.0$ & $-0.10$ & pp \\
Tramway / trolleybus & $+0.00$ & $-0.00$ & $+0.00$ & pp \\
Passenger train & $+0.00$ & $-0.00$ & $+0.00$ & pp \\
Conventional truck (oil) & $-22.5$ & $+1.6$ & $-1.6$ & pp \\
Battery-electric truck & $+21.9$ & $-1.4$ & $+1.4$ & pp \\
Freight train & $+0.65$ & $-0.21$ & $+0.25$ & pp \\
Aircraft & $+0.00$ & $-0.00$ & $+0.00$ & pp \\
\addlinespace
\multicolumn{5}{l}{\textbf{Industry}} \\
Ind. waste boiler & $+11.7$ & $-3.0$ & $+5.0$ & pp \\
Ind. gas boiler & $-9.1$ & $+1.3$ & $-0.84$ & pp \\
High-T heat pump & $+6.5$ & $+0.18$ & $-2.5$ & pp \\
Ind. wood boiler & $-4.0$ & $-0.11$ & $+0.20$ & pp \\
Ind. oil boiler & $-3.4$ & $+1.1$ & $-1.3$ & pp \\
Ind. coal boiler & $-1.6$ & $+0.53$ & $-0.63$ & pp \\
\addlinespace
\multicolumn{5}{l}{\textbf{Fuels \& feedstocks}} \\
Point-source CC & $+0.71$ & $+0.12$ & $-0.06$ & kt\,CO$_2$/h \\
Direct air capture & $+0.16$ & $+0.12$ & $-0.09$ & kt\,CO$_2$/h \\
Biomethanation & $+0.16$ & $+0.01$ & $+0.05$ & GW \\
Electrolysis & $+0.02$ & $+0.05$ & $+0.01$ & GW \\
Steam methane reforming & $-0.00$ & $-0.00$ & $+0.00$ & GW \\
Oil-to-HVC & $-0.07$ & $+0.01$ & $-0.02$ & GW \\
Biomass-to-HVC & $+0.06$ & $-0.03$ & $+0.01$ & GW \\
Methanol-to-HVC & $+0.01$ & $+0.01$ & $+0.00$ & GW \\
Synthetic methanolation & $+0.00$ & $+0.01$ & $+0.00$ & GW \\
Biomass-to-methanol & $+0.01$ & $+0.01$ & $+0.00$ & GW \\
Haber-Bosch synthesis & $+0.01$ & $+0.00$ & $+0.00$ & GW \\
Pyrolysis to oil & $+0.20$ & $+0.10$ & $-0.05$ & GW \\
Fischer-Tropsch fuel & $+0.01$ & $+0.03$ & $+0.01$ & GW \\
\bottomrule
\end{tabular}

%% file: tables/state_year_means_1.tex
\begin{tabular}{l r r r r r r r r r r l}
\toprule
 & $\kappa_{1}$ & $\kappa_{2}$ & $\kappa_{2}$ & $\kappa_{2}$ & $\kappa_{3}$ & $\kappa_{3}$ & $\kappa_{4}$ & $\kappa_{4}$ & $\kappa_{5}$ & $\kappa_{5}$ & \\
 & 2030 & 2035 & 2040 & 2045 & 2035 & 2040 & 2045 & 2050 & 2045 & 2050 & Unit \\
\midrule
\#\,occurrences & 987 & 592 & 576 & 423 & 389 & 367 & 180 & 478 & 259 & 519 & \\
\addlinespace
\multicolumn{12}{l}{\textbf{Electricity}} \\
Open-field PV & 0.28 & 19.2 & 20.1 & 21.0 & 9.7 & 10.5 & 22.7 & 24.8 & 22.7 & 24.6 & GW \\
Alpine PV & 0 & 0.15 & 0.13 & 0.33 & 9.1 & 9.2 & 4.4 & 5.7 & 8.4 & 8.9 & GW \\
Rooftop PV & 6.5 & 4.9 & 3.3 & 1.6 & 4.9 & 3.3 & 1.6 & 0 & 1.6 & 0.12 & GW \\
Onshore wind & 0.30 & 3.9 & 4.0 & 4.2 & 2.3 & 2.5 & 4.0 & 4.4 & 3.7 & 4.3 & GW \\
Nuclear & 2.6 & 0.70 & 0 & 0 & 0.81 & 0 & 0 & 0 & 0 & 0 & GW \\
CCGT & 0.26 & 0.21 & 0.09 & 0.03 & 0.32 & 0.16 & 0.13 & 0.64 & 0.52 & 1.9 & GW \\
CCGT (ammonia) & 0 & 0 & 0 & 0 & 0 & 0 & 0 & 0 & 0 & 0 & GW \\
Hydro dam & 8.3 & 8.3 & 8.3 & 8.3 & 8.3 & 8.3 & 8.3 & 8.3 & 8.3 & 8.3 & GW \\
Run-of-river hydro & 4.3 & 4.3 & 4.3 & 4.3 & 4.3 & 4.3 & 4.3 & 4.3 & 4.3 & 4.3 & GW \\
\addlinespace
\multicolumn{12}{l}{\textbf{Building}} \\
Dec. heat pump & 16 & 9.0 & 3.6 & 4.6 & 8.9 & 28 & 40 & 22 & 71 & 90 & \% \\
Dec. gas boiler & 56 & 69 & 96 & 95 & 73 & 72 & 60 & 78 & 29 & 10 & \% \\
Dec. oil boiler & 18 & 16 & 0 & 0 & 13 & 0 & 0 & 0 & 0 & 0 & \% \\
Dec. wood boiler & 6.6 & 3.6 & 0 & 0 & 3.6 & 0 & 0 & 0 & 0 & 0 & \% \\
Dec. direct elec. & 2.7 & 1.8 & 0 & 0 & 1.7 & 0 & 0 & 0 & 0 & 0 & \% \\
Dec. solar thermal & 0.2 & 0.1 & 0.1 & 0 & 0.1 & 0.1 & 0 & 0 & 0 & 0 & \% \\
DH heat pump & 39 & 63 & 72 & 74 & 68 & 76 & 76 & 75 & 76 & 77 & \% \\
DH wood boiler & 12 & 3.2 & 0.1 & 0 & 1.1 & 0 & 0 & 0 & 0 & 0 & \% \\
DH waste CHP & 21 & 17 & 17 & 16 & 15 & 16 & 16 & 17 & 15 & 16 & \% \\
DH gas CHP & 24 & 14 & 8.8 & 8.3 & 14 & 8.0 & 8.2 & 7.3 & 8.8 & 7.3 & \% \\
DH wet biomass CHP & 4.5 & 3.3 & 1.5 & 1.1 & 1.6 & 0.3 & 0.2 & 0.7 & 0.1 & 0 & \% \\
DH oil boiler & 0 & 0 & 0 & 0 & 0 & 0 & 0 & 0 & 0 & 0 & \% \\
DH deep geothermal & 0 & 0 & 0 & 0 & 0 & 0 & 0 & 0 & 0 & 0 & \% \\
DH gas boiler & 0 & 0 & 0 & 0 & 0 & 0 & 0 & 0 & 0 & 0 & \% \\
\bottomrule
\end{tabular}

%% file: tables/state_year_means_2.tex
\begin{tabular}{l r r r r r r r r r r l}
\toprule
 & $\kappa_{1}$ & $\kappa_{2}$ & $\kappa_{2}$ & $\kappa_{2}$ & $\kappa_{3}$ & $\kappa_{3}$ & $\kappa_{4}$ & $\kappa_{4}$ & $\kappa_{5}$ & $\kappa_{5}$ & \\
 & 2030 & 2035 & 2040 & 2045 & 2035 & 2040 & 2045 & 2050 & 2045 & 2050 & Unit \\
\midrule
\#\,occurrences & 987 & 592 & 576 & 423 & 389 & 367 & 180 & 478 & 259 & 519 & \\
\addlinespace
\multicolumn{12}{l}{\textbf{Storage}} \\
TS Dec. gas boiler & 17.0 & 20.2 & 25.3 & 22.6 & 20.7 & 18.7 & 14.3 & 17.8 & 7.5 & 5.3 & GWh \\
TS Dec. heat pump & 5.5 & 3.2 & 0.99 & 1.2 & 3.3 & 7.3 & 9.5 & 5.2 & 17.2 & 21.5 & GWh \\
Li-ion battery & 0.67 & 0.33 & 0 & 0 & 0.33 & 0 & 0 & 0.53 & 1.1 & 3.9 & GWh \\
TS Dec. oil boiler & 2.2 & 3.4 & 1.2 & 1.4 & 2.5 & 0.45 & 0.63 & 1.2 & 0.34 & 0.57 & GWh \\
TS Dec. wood boiler & 2.0 & 1.0 & 0 & 0 & 1.0 & 0 & 0 & 0 & 0 & 0 & GWh \\
TS High-T & 2.1 & 1.6 & 1.5 & 1.6 & 1.6 & 1.2 & 1.3 & 1.4 & 1.2 & 1.3 & GWh \\
TS Dec. direct elec. & 0.51 & 0.55 & 0.03 & 0.04 & 0.51 & 0.02 & 0.03 & 0.03 & 0.01 & 0.02 & GWh \\
TS DH short-term & 0 & 0 & 0 & 0 & 0 & 0 & 0 & 0 & 0 & 0 & GWh \\
H$_2$ storage & 0 & 0 & 0 & 0 & 0 & 0 & 0 & 0 & 0 & 0 & GWh \\
CO$_2$ storage & 0 & 0 & 0 & 0 & 0 & 0 & 0 & 0 & 0 & 0 & GWh \\
Pumped hydro storage & 84.5 & 84.5 & 84.5 & 84.5 & 84.5 & 84.5 & 84.5 & 84.5 & 84.5 & 84.5 & GWh \\
Oil storage & 4.3 & 3.5 & 0.34 & 0.27 & 2.8 & 0.29 & 0.30 & 0.44 & 0.36 & 0.84 & TWh \\
TS DH seasonal & 1.3 & 2.7 & 3.8 & 4.6 & 2.8 & 3.9 & 4.5 & 4.9 & 4.4 & 4.7 & TWh \\
Hydro reservoir & 6.9 & 7.5 & 7.9 & 8.4 & 7.5 & 7.9 & 8.4 & 8.9 & 8.4 & 8.9 & TWh \\
Gas storage & 0 & 0 & 0 & 0.01 & 0 & 0 & 0 & 0.30 & 4.9 & 171.6 & GWh \\
Methanol storage & 0.01 & 0.01 & 0.01 & 0.01 & 0.01 & 0.01 & 0.01 & 3.0 & 0.01 & 71.7 & GWh \\
Ammonia storage & 1.5 & 15.3 & 22.1 & 24.6 & 27.9 & 33.5 & 35.6 & 35.7 & 39.3 & 44.8 & GWh \\
\addlinespace
\multicolumn{12}{l}{\textbf{Transport}} \\
Battery-electric car & 53 & 100 & 100 & 100 & 100 & 100 & 100 & 100 & 100 & 100 & \% \\
Conventional car (oil) & 42 & 0 & 0 & 0 & 0 & 0 & 0 & 0 & 0 & 0 & \% \\
Hybrid-electric car & 3.7 & 0 & 0 & 0 & 0 & 0 & 0 & 0 & 0 & 0 & \% \\
Plug-in hybrid car & 1.1 & 0 & 0 & 0 & 0 & 0 & 0 & 0 & 0 & 0 & \% \\
Natural-gas car & 0.1 & 0 & 0 & 0 & 0 & 0 & 0 & 0 & 0 & 0 & \% \\
Conventional bus (oil) & 13 & 13 & 13 & 7.1 & 13 & 13 & 6.7 & 0.6 & 6.9 & 0.5 & \% \\
Battery-electric bus & 0.2 & 0 & 0 & 5.7 & 0 & 0 & 6.1 & 12 & 5.8 & 12 & \% \\
Tramway / trolleybus & 4.0 & 4.0 & 4.0 & 4.0 & 4.0 & 4.0 & 4.0 & 4.0 & 4.0 & 4.0 & \% \\
Passenger train & 83 & 83 & 83 & 83 & 83 & 83 & 83 & 83 & 83 & 83 & \% \\
Conventional truck (oil) & 28 & 0 & 0 & 0 & 0 & 0 & 0 & 0 & 0 & 0 & \% \\
Battery-electric truck & 32 & 60 & 60 & 60 & 60 & 60 & 60 & 60 & 60 & 60 & \% \\
Freight train & 40 & 40 & 40 & 40 & 40 & 40 & 40 & 40 & 40 & 40 & \% \\
Aircraft & 100 & 100 & 100 & 100 & 100 & 100 & 100 & 100 & 100 & 100 & \% \\
\addlinespace
\multicolumn{12}{l}{\textbf{Industry}} \\
Ind. waste boiler & 58 & 60 & 57 & 56 & 63 & 60 & 58 & 60 & 59 & 61 & \% \\
Ind. gas boiler & 34 & 26 & 24 & 24 & 27 & 26 & 25 & 26 & 25 & 26 & \% \\
High-T heat pump & 0.9 & 14 & 19 & 20 & 10 & 14 & 18 & 14 & 16 & 13 & \% \\
Ind. wood boiler & 7.1 & 0 & 0 & 0 & 0 & 0 & 0 & 0 & 0 & 0 & \% \\
Ind. oil boiler & 0 & 0 & 0 & 0 & 0 & 0 & 0 & 0 & 0 & 0 & \% \\
Ind. coal boiler & 0 & 0 & 0 & 0 & 0 & 0 & 0 & 0 & 0 & 0 & \% \\
\addlinespace
\multicolumn{12}{l}{\textbf{Fuels \& feedstocks}} \\
Point-source CC & 0.44 & 1.6 & 1.7 & 1.8 & 1.4 & 1.7 & 1.8 & 1.9 & 1.8 & 2.0 & kt\,CO$_2$/h \\
Direct air capture & 0 & 0 & 0 & 0.01 & 0 & 0 & 0 & 1.7 & 0 & 0.46 & kt\,CO$_2$/h \\
Biomethanation & 0.32 & 0.37 & 0.41 & 0.44 & 0.41 & 0.46 & 0.48 & 0.48 & 0.51 & 0.54 & GW \\
Electrolysis & 0 & 0.01 & 0.02 & 0.02 & 0.02 & 0.02 & 0.02 & 0.03 & 0.04 & 0.22 & GW \\
Steam methane reforming & 0 & 0 & 0 & 0 & 0 & 0 & 0 & 0 & 0 & 0 & GW \\
Oil-to-HVC & 0.04 & 0.01 & 0 & 0 & 0.01 & 0 & 0 & 0 & 0 & 0 & GW \\
Biomass-to-HVC & 0.16 & 0.19 & 0.21 & 0.21 & 0.19 & 0.21 & 0.21 & 0.17 & 0.21 & 0.15 & GW \\
Methanol-to-HVC & 0 & 0 & 0 & 0 & 0 & 0 & 0 & 0.04 & 0 & 0.05 & GW \\
Synthetic methanolation & 0 & 0 & 0 & 0 & 0 & 0 & 0 & 0 & 0 & 0.04 & GW \\
Biomass-to-methanol & 0.02 & 0.02 & 0.02 & 0.02 & 0.02 & 0.02 & 0.02 & 0.07 & 0.02 & 0.07 & GW \\
Haber-Bosch synthesis & 0 & 0.01 & 0.01 & 0.02 & 0.02 & 0.02 & 0.02 & 0.02 & 0.02 & 0.03 & GW \\
Pyrolysis to oil & 0 & 0.20 & 0.42 & 0.57 & 0.12 & 0.36 & 0.57 & 0.61 & 0.54 & 0.64 & GW \\
Fischer-Tropsch fuel & 0 & 0 & 0 & 0 & 0 & 0 & 0 & 0 & 0.01 & 0.11 & GW \\
\bottomrule
\end{tabular}

%% file: tables/bandwidth.tex
\begin{tabular}{l l r r r r }
\toprule
Metric & Strategy & $\alpha = 1$\,\% & $\alpha = 5$\,\% & $\alpha = 10$\,\% & $\alpha = 20$\,\% \\
\midrule
Minimum & Gas \& CO$_2$ removal & 0.9 (1) & 0.5 (1) & 0.3 (1) & 0.2 (1) \\
 & Early electrification & 1.4 (2) & 0.8 (3) & 0.6 (3) & 0.4 (3) \\
 & Late switch & 2.0 (3) & 0.7 (2) & 0.5 (2) & 0.3 (2) \\
\addlinespace
Median & Gas \& CO$_2$ removal & 3.8 (1) & 1.9 (1) & 1.4 (1) & 1.1 (1) \\
 & Early electrification & 5.0 (2) & 2.7 (2) & 2.0 (2) & 1.4 (2) \\
 & Late switch & 5.9 (3) & 2.8 (3) & 2.1 (3) & 1.5 (3) \\
\addlinespace
Average & Gas \& CO$_2$ removal & 4.7 (1) & 2.4 (1) & 1.9 (1) & 1.5 (1) \\
 & Early electrification & 5.5 (2) & 3.2 (3) & 2.5 (3) & 1.9 (3) \\
 & Late switch & 6.6 (3) & 3.0 (2) & 2.2 (2) & 1.6 (2) \\
\addlinespace
VaR$_{90}$ & Gas \& CO$_2$ removal & 8.7 (1) & 4.8 (2) & 3.8 (2) & 3.2 (2) \\
 & Early electrification & 8.8 (2) & 5.8 (3) & 4.8 (3) & 3.6 (3) \\
 & Late switch & 11.0 (3) & 4.5 (1) & 3.3 (1) & 2.5 (1) \\
\addlinespace
CVaR$_{90}$ & Gas \& CO$_2$ removal & 11.3 (2) & 6.3 (2) & 4.9 (2) & 4.0 (2) \\
 & Early electrification & 10.1 (1) & 6.6 (3) & 5.5 (3) & 4.2 (3) \\
 & Late switch & 13.3 (3) & 5.5 (1) & 3.9 (1) & 2.9 (1) \\
\addlinespace
Maximum & Gas \& CO$_2$ removal & 23.2 (2) & 13.1 (3) & 8.7 (3) & 6.4 (3) \\
 & Early electrification & 17.9 (1) & 8.6 (1) & 7.2 (1) & 5.6 (2) \\
 & Late switch & 23.2 (2) & 11.6 (2) & 7.5 (2) & 4.6 (1) \\
\bottomrule
\end{tabular}